\documentclass[twocolumn,aps,prd,floatfix,preprintnumbers,a4paper,nofootinbib,superscriptaddress,10pt]{revtex4-2}
\usepackage{lipsum}
\usepackage{multirow}
\usepackage{import}
\usepackage[titletoc,title]{appendix}

\usepackage{graphicx}				% Use pdf, png, jpg, or eps§ with pdflatex; use eps in DVI mode
\usepackage{placeins}
\usepackage{amssymb}
\usepackage{amsmath}
\usepackage[dvipsnames]{xcolor}
\usepackage{verbatim}
\usepackage{rotating}
\usepackage{comment}
\usepackage{array}

\usepackage[colorlinks]{hyperref}
\usepackage[utf8]{inputenc}
\usepackage{txfonts}
\usepackage{braket}
\usepackage{physics}
\usepackage{mathtools}

\DeclareMathOperator*{\argmin}{arg\,min}

\begin{document}

\newcommand\red[1]{{\color[rgb]{0.75,0.0,0.0} #1}}
\newcommand\green[1]{{\color[rgb]{0.0,0.60,0.08} #1}}
\newcommand\blue[1]{{\color[rgb]{0,0.20,0.65} #1}}
\newcommand\cyan[1]{{\color[HTML]{00c3ff} #1}}
\newcommand\bluey[1]{{\color[rgb]{0.11,0.20,0.4} #1}}
\newcommand\gray[1]{{\color[rgb]{0.7,0.70,0.7} #1}}
\newcommand\grey[1]{{\color[rgb]{0.7,0.70,0.7} #1}}
\newcommand\white[1]{{\color[rgb]{1,1,1} #1}}
\newcommand\darkgray[1]{{\color[rgb]{0.3,0.30,0.3} #1}}
\newcommand\orange[1]{{\color[rgb]{.86,0.24,0.08} #1}}
\newcommand\purple[1]{{\color[rgb]{0.45,0.10,0.45} #1}}
\newcommand\note[1]{\colorbox[rgb]{0.85,0.94,1}{\textcolor{black}{\textsc{\textsf{#1}}}}}
% ***************************************** %
\def\gw#1{gravitational wave#1}
% ***************************************** %
%\def\oed#1{optimal emission direction#1}
\def\oed#1{QA frame#1}
% ***************************************** %
\def\nr#1{numerical relativity
 (NR)#1\gdef\nr{NR}}
 % ***************************************** %
\def\bh#1{black-hole
 (BH)#1\gdef\bh{BH}}
 % ***************************************** %
 \def\bbh#1{binary black hole#1
  (BBH#1)\gdef\bbh{BBH}}
% ***************************************** %
% quadrupole-aligned (QA)
\def\qa#1{quadrupole-aligned
(QA)#1\gdef\qa{QA}}
% ***************************************** %
\def\pn#1{post-Newtonian
 (PN)#1\gdef\pn{PN}}
% ***************************************** %
 \def\qnm#1{Quasinormal Mode
    (QNM)#1\gdef\qnm{QNM}}
% ***************************************** %
   \def\eob#1{effective-one-body
      (EOB)#1\gdef\eob{EOB}}
% ***************************************** %
\def\imr#1{inspiral-merger-ringdown
 (IMR)#1\gdef\imr{IMR}}
 % ***************************************** %
 \def\fig#1{Fig.~(\ref{#1})}
 % ***************************************** %
 \def\cfig#1{Fig.~\ref{#1}}
 % ***************************************** %
 \newcommand{\figs}[2]{Figures~(\ref{#1}-\ref{#2})}
 % ***************************************** %
 \def\eqn#1{Eq.~(\ref{#1})}
 % ***************************************** %
 \def\ceqn#1{Eq.~\ref{#1}}
 % ***************************************** %
 \newcommand{\Eqns}[2]{Equations~(\ref{#1})-(\ref{#2})}
 % ***************************************** %
 \newcommand{\eqns}[2]{Eqs.~(\ref{#1})-(\ref{#2})}
 % ***************************************** %
 \newcommand{\ceqns}[2]{Eqs.~\ref{#1}-\ref{#2}}
 % ***************************************** %
 \def\lm{{\ell m}}
 \def\tpsi{\tilde{\psi}}
 % ***************************************** %

%% Indistinguishable macros
\def\data{d}
\def\signal{s}
\def\noise{n}
\def\model{h}
\def\modelsig{h_\text{s}}
\def\modelbf{h_\text{bf}}
\def\param{\theta}

\def\d{{\sc{PhenomD}}}
\def\indistSNR{\rho_\text{I}}
\def\thetabf{\theta_\text{bf}}
\newcommand\thetaod[1]{\theta_{\text{bf}\, |\, #1}}
\def\thetasig{\theta_\text{s}}

\definecolor{brown-ish}{RGB}{181,180,20}

\newcommand{\mdh}[1]{{\color{blue}{ Mark: #1}}}
\newcommand{\jt}[1]{{\color{violet}{ JT: #1}}}
\newcommand{\ch}[1]{{\color{RubineRed}{ Charlie: #1}}}

\newcommand{\Cardiff}{School of Physics and Astronomy, Cardiff University, Cardiff, CF24 3AA, United Kingdom}
\newcommand{\Portsmouth}{Institute of Cosmology \& Gravitation, University of Portsmouth, Portsmouth, United Kingdom}
\newcommand{\Soton}{Mathematical Sciences \& STAG Research Centre, University of Southampton, Southampton, SO17 1BJ, United Kingdom}
\newcommand{\Caltech}{Theoretical Astrophysics Group, California Institute of Technology, Pasadena, CA 91125, U.S.A.}

%\title{Indistinguishability done properly}
\title{On the use and interpretation of signal-model indistinguishability measures for gravitational-wave astronomy}

\author{Jonathan E. Thompson}
\affiliation{\Soton}
\affiliation{\Caltech}
\author{Charlie Hoy}
\affiliation{\Portsmouth}
\author{Edward Fauchon-Jones}
\affiliation{\Cardiff}
\author{Mark Hannam} 
\affiliation{\Cardiff}

\begin{abstract}
The difference (``mismatch'') between two gravitational-wave (GW) 
signals is often used to estimate the signal-to-noise ratio (SNR) at which they will be 
distinguishable in a measurement or, alternatively, when the errors in a signal
 model will lead to biased measurements. It is well known that the 
standard approach to calculate this ``indistinguishability SNR'' is too 
conservative: a model may fail the criterion at a given SNR, but not necessarily incur a 
biased measurement of any individual parameters. This problem can be solved by 
taking into account errors orthogonal to the model space (which therefore
do not induce a bias), and calculating indistinguishability SNRs for individual 
parameters, rather than the full $N$-dimensional
parameter space. We illustrate this approach with the simple
example of aligned-spin binary-black-hole signals, and calculate accurate estimates 
of the SNR at which each parameter measurement will be biased. In general biases 
occur at much higher SNRs than predicted from the standard mismatch calculation.
Which parameters are most easily biased depends sensitively on the details of 
a given waveform model, and the location in parameter space, and in some cases
the bias SNR is as high as the conservative estimate. We also illustrate how the 
parameter bias SNR can be used to robustly specify waveform accuracy requirements for future detectors.
\end{abstract}

%\date{}

\maketitle

%%%%%%%%%%%%%%%%%%%%%%%%%%%%%%%%%%%%%%%%%%%%%
\section{Introduction}
\label{sec:Intro}

Due to improvements in detector sensitivity, the observation potential of gravitational-wave (GW) 
astronomy has grown rapidly since the first direct GW detection in 
2015~\cite{LIGOScientific:2016aoc,LIGOScientific:2016dsl,LIGOScientific:2020ibl,LIGOScientific:2021usb,LIGOScientific:2021vkt}, 
and by current forecasts it will continue to do so over the next two 
decades~\cite{abbott2020prospects,Reitze:2019iox,Evans:2023euw,Punturo:2010zza,Hild:2010id,ET:2019dnz,Abac:2025saz,LISA:2017pwj}. 
Detector networks in 2015 could be expected to observe $O(10)$ black-hole binaries per year, while 
current networks should be sensitive to $O(100)$ binaries per year~\cite{abbott2020prospects}. 
At the projected sensitivity of next-generation ground-based detectors, we will observe many thousands of 
binaries per year, and will be sensitive to all black-hole mergers in the universe. 
The additional upcoming space-based GW detector LISA~\cite{Babak:2021mhe,LISA:2024hlh} will be sensitive
to massive black hole signals of similar morphology to those seen in ground-based detectors 
(but at total masses \(>10^5\,M_\odot\)). 
%Although merger rates for these sources are uncertain~\cite{Sesana:2021jfh}, we expect to observe GW signals at large 
%signal-to-noise ratios (SNRs) between 10 to 1000.

With increased sensitivity we also observe louder signals, which will allow more accurate 
measurements; for a given source, measurement accuracy scales roughly linearly with detector 
sensitivity. However, to realize higher measurement accuracies we also require sufficiently 
accurate theoretical signal models against which to compare the detector data. The accuracy 
requirements of our models, in order to make unbiased measurements, become more stringent 
with increased signal strength. 

As such, the question of quantifying model accuracy, and determining future accuracy requirements,
is an important one. In the near term we wish to know under what circumstances we can trust the source 
inference from current models, and when we must beware of systematic biases. In the longer term, 
as part of the  extensive research and development effort to prepare for LISA~\cite{Babak:2021mhe,LISA:2024hlh}, 
Einstein Telescope~\cite{Punturo:2010zza,Hild:2010id,ET:2019dnz,Abac:2025saz}, and 
Cosmic Explorer~\cite{abbott2020prospects,Reitze:2019iox,Evans:2023euw},
we also wish to know how accurate our models must be to achieve the observatories' science goals. 
These questions are made more urgent by the large resources and many-year timescale 
required to produce accurate models: large numbers of computationally expensive numerical relativity (NR) simulations, 
 and sophisticated procedures to calibrate semi-analytic phenomenological models, or to train surrogate 
 models. 

To date it has been difficult to provide useful waveform accuracy measures. For example, in NR simulations
one typically quantifies the signal's phase accuracy from the beginning of the simulation. In GW data
analysis, on the other hand, it is more usual to consider an inner product between waveforms~\cite{Finn:1992wt}, which 
involves an optimization with respect to an overall phase shift and confuses the nominal phase uncertainty 
of the waveform. Furthermore, when estimating binary source parameters, we identify the parameters at which 
our model agrees best with the data; a single NR waveform corresponds to a binary with a single set of
parameters, so we also need a way to convert its uncertainty into measurement biases. 

A series of previous works have noted that we can define a waveform uncertainty measure based on the 
inner product above (the mismatch). The mismatch between two waveforms, for example between a true signal waveform and
a waveform model, can in turn be related to the SNR at which the two waveforms will be distinguishable in 
a measurement~\cite{lindblom:2008cm,McWilliams:2010eq,Hannam:2010ky,Baird:2013dbm,Chatziioannou:2017tdw,Toubiana:2024car}. 
Unfortunately, in practice this simple mismatch requirement is typically found to be 
extremely conservative, and so of little use in making realistic estimates of required model 
accuracy~\cite{LIGOScientific:2016ebw,Purrer:2019jcp}. 

One pragmatic solution to this problem would be to consider our most accurate NR waveforms as proxies for 
true signals, and to study how well our current models recover true source properties for selected detector 
configurations and a range of signal strengths, \textit{i.e.}, a range of SNRs. In doing this we could in 
principle identify the SNR at which a model will lead to a biased measurement in each parameter of interest, 
and determine how much more accurate the models must be for future high SNR observations. A first attempt at such 
an approach was made in P\"urrer and Haster's 2019 study, Ref.~\cite{Purrer:2019jcp}. The authors 
considered two signals and found parameter biases at vastly different SNRs, depending on both the parameter 
and the signal; some parameters are biased at an SNR of $\sim$50, while others are not biased even at SNRs of 
$\sim$2500, and some of the parameters that are most susceptible to bias in one signal and not biased at all
for the other. In order to draw some general conclusions the authors estimate an approximate ``balance SNR''
for each signal (SNRs of $\sim$50 for both signals), and use this to conclude that the mismatch uncertainties of 
models for next-generation ground-based detectors must improve on those of c.2019 models by three orders of 
magnitude, and the mismatch uncertainty of NR simulations must improve by one order of magnitude. 

In this work we revisit the indistinguishability mismatch criterion, and show that, if calculated appropriately, 
it is not conservative, but in fact an accurate measure of the SNR at which a measurement will be biased. 
The criterion is too conservative in its standard form, partly because it does not account for waveform errors
that do not contribute to measurement bias (i.e., are orthogonal to the signal manifold), but mostly because
the criterion applies to an $N$-dimensional credible interval. The criterion can, however, be calculated in
such a way as to accurately predict bias SNRs for individual parameters. In the present work we illustrate 
each of these features with respect to a simple model for quadrupole-only radiation and spins aligned 
with the the orbital angular momentum (referred to as an ``aligned-spin'' binary); we will consider 
state-of-the-art generic-binary models in future work. We stress that although we are not aware of this method
being applied to binary-black-hole waveform models with current and future ground-based detectors, the 
method itself is not new; it is discussed, either implicitly or explicitly, in works from Ref.~\cite{lindblom:2008cm}
through to Ref.~\cite{Toubiana:2024car}, and the method we use to calculate the 1D parameter bias SNRs
is equivalent to that discussed in Ref.~\cite{Toubiana:2024car}.

We also note that a more appropriate measure of model uncertainty is not the mismatch, but the square
root of the mismatch. It is this quantity that scales directly with both the signal SNR, and with standard
accuracy measures in NR simulations and, under reasonable assumptions, NR computational cost. 
This seemingly minor change has important implications for future accuracy requirements: a 
two-orders-of-magnitude improvement in mismatch can be achieved with only one order of magnitude improvement in
NR accuracy, and similarly only one order of magnitude increase in computational cost. 

Although we consider only a simple proof-of-principle model in this work, we are able to make some 
broad estimates for the required improvements in model accuracy and NR simulations for next-generation
observatories. 
These sharpen the early estimates from Ref.~\cite{Purrer:2019jcp}. In particular, we note that 
the bias SNRs depend on an individual model's construction, and in principle a 
``conservative'' estimate of the bias SNR can sometimes be correct. 
This leads to a far more stringent accuracy requirement
than in Ref.~\cite{Purrer:2019jcp}: model mismatch uncertainties must be below $10^{-6}$ to be free of
bias in observations with SNRs of $\sim$1000, an improvement of four orders of magnitude over some 
current models. On the other hand, if model construction 
can be optimized to maximize individual bias SNRs,
we may require only modest improvements over the most accurate current models, 
\textit{e.g.}, {\sc NRSur7dq4}~\cite{Varma:2019csw}. 
This large uncertainty in the required level of improvement
highlights the scale of the general problem of accuracy-requirement estimates, and we hope that
future applications of the method we discuss here to current generic models will provide more refined, 
and more useful, estimates. 

In this work we consider the problem of quantifying the accuracy of waveform models for source
measurements, such that those measurements will not be contaminated by systematics. 
What we \emph{do not} consider is the related (and likely more difficult) problem of identifying
when systematics are present in a measurement. See, for example, Refs.~\cite{Cutler:2007mi} and 
\cite{Hu:2022rjq} for discussions of strategies to identify waveform systematics in observations.

The outline of this paper is as follows. 
In Sec.~\ref{sec:IndistBias} we review the approach to estimating indistinguishability
SNR, describe why it is overly conservative and detail how one may
improve upon it. In Sec.~\ref{sec:methods} we outline the details of 
the signal waveforms we use and our parameter estimation injections. 
Section~\ref{sec:EffSNR} discusses the results for indistinguishability SNR estimation
for $N$-dimensional posterior data, and Sec.~\ref{sec:BiasSNR} 
extends this analysis to SNR estimates for individual model parameters. 
We discuss the impact of SNR estimates on next-generation detectors in
Sec.~\ref{sec:acc-req} and provide concluding thoughts in Sec.~\ref{sec:conclusions}.
Throughout this manuscript we work in units of $G=c=1$.

%%%%%%%%%%%%%%%%%%%%%%%%%%%%%%%%%%%%%%%%%%%%%
\section{Model Indistinguishability and Bias} 
\label{sec:IndistBias}

We consider the data timeseries \(\data\) collected by a ground-based GW interferometer. Under the
hypothesis that a GW signal exists in the data, we write \(\data=\signal+\noise\), where \(\signal\) is
the true signal and \(\noise\) is the stationary and Gaussian-distributed noise of the detector.
For most of this work we will consider the ``zero-noise case''
where we implicitly replace quantities relating to noise with their expectation values under 
infinite noise realizations (thereby setting \(\noise\rightarrow\langle \noise\rangle =0\))~\cite{Markovic:1993cr}.

Consider a gravitational-wave model for a compact binary coalesence, 
$\model(\param)$, parameterized by a set of
intrinsic and extrinsic parameters \(\theta\in\Theta\). The set \(\Theta\) contains
intrinsic parameters such as the primary and secondary masses, \(m_1\ge m_2\), 
the primary and secondary (dimensionless) spin vectors, \(\boldsymbol{\chi}_{1,2}\),
as well as a number of extrinsic parameters. 
For the examples in this work we focus on the constrained case of compact binaries with spins strictly aligned with the 
orbital angular momentum, thereby reducing the spin degrees of freedom to two,
denoted without loss of generality simply as \(\chi_{1z}\) and \(\chi_{2z}\). 
We emphasize however that the general approach outlined 
in this work does not rely on these simplifying approximations to the signal. 
We drop explicit parameter dependence wherever convenient for ease of reading.

It is useful to define an inner product between two signals 
$\model_1=\model(\theta_1)$ and $\model_2=\model(\theta_2)$ as, 
\begin{equation}
\Braket{\model_1|\model_2} =  4 \text{Re} \int^{f_\text{max}}_{f_\text{min}}
   \frac{\tilde{h}_1 \left(f\right) \tilde{h}^{*}_2 \left(f\right)}{\tilde{S}_n\left(f\right)} \text{d}f, \label{eq:inner}
\end{equation} 
where the tilde denotes the Fourier transform, the signals are written as functions 
of frequency $f$, the detector is sensitive in the frequency range 
$f \in [f_{\rm min}, f_{\rm max} ]$, and $\tilde{S}_n\left(f\right)$ is the detector's power spectral density. 
The SNR of a GW signal $\model$ is then given by $\rho^2 = |\model|^2 = \Braket{\model|\model}$.

We refer to the \textit{indistinguishability~SNR} as
the SNR below which two GW signals will be indistinguishable in a measurement. 
This SNR has a natural connection with the ratio of parameter bias to measurement
variance (see 
Appendix~\ref{sec:Fisher}), as that ratio itself scales directly with the SNR of the 
signal. When discussing waveform 
model errors, the indistinguishability SNR indicates the SNR above which a given 
model will lead to biased parameter inference, and we can then determine, for a 
given GW observation, whether that model can be trusted to provide unbiased 
measurements.

There has been extensive discussion of variants of the indistinguishability 
SNR in the 
literature~\cite{lindblom:2008cm,McWilliams:2010eq,Hannam:2010ky,Baird:2013dbm,Chatziioannou:2017tdw,Toubiana:2024car}.
The standard calculation provides 
only a conservative estimate relevant to parameter biases.
We give examples in Sec.~\ref{sec:IndistSNR} of signals louder than a model's 
nominal indistinguishability SNR, for which the model recovers all of the source
properties with no bias. In these scenarios the conservative estimate is of little value. 

In Sec.~\ref{sec:SNRissues} we will explain the two reasons why the standard estimate is conservative, 
and how to address them. This will lead us
to define quantities that accurately estimates bias SNRs, 
either for $N$-D sets of parameters or individual parameters; the $N$-D qualifier
will be explained in due course. In the remainder of the
paper we will provide concrete examples to illustrate these points, 
and demonstrate that we can calculate a reliable bias
SNR for all measurable parameters. This approach can then be used to 
inform waveform accuracy requirements in current and 
future detectors.

%%%%%%%%%%%%%%%%%%%%%%%%%%%%%%%%%%%%%%%%%%%%%

%%%%%%%%%%%%%%%%%%%%%%%%%%%%%%%%%%%%%%%%%%%%%
\subsection{Mismatches and Distance Metrics}
\label{sec:mismatchdistance}
   
One measure of waveform model accuracy is the 
\textit{match}, defined as the noise-weighted inner product 
optimized over some subset of parameters 
$\Theta_\text{opt}\subset\Theta$~\cite{Cutler:1994ys}, and
normalized with respect to the magnitude of each waveform, 
\begin{equation} \label{eqn: match def}
   M\left(\model_1,\model_2\right) =  \max_{\Theta_\text{opt}} \, \frac{\Braket{\model_1|\model_2}} { \left|\model_1\right| \left|\model_2\right|}.
\end{equation}
The match is unity if the two waveforms are the same, up to an overall amplitude rescaling. To quantify the difference between two 
waveforms we use the \textit{mismatch},
\begin{align}\label{eqn: mismatch} 
\mathcal{M} = {}& 1 - M\left(\model_1,\model_2\right).
\end{align}
The choice of optimization parameters $\Theta_\text{opt}$ is discussed in more detail 
later in Sec.~\ref{sec:optimal-parameters}. 

We can identify the mismatch with a measure of
normalized difference between two waveforms, $\hat{d} = \sqrt{\mathcal{M}}$~\cite{Owen:1995tm}. 
This is motivated by the usual interpretation of an inner product as the square of 
a distance, and also a consideration of uncertainties 
in waveforms. 
For the former, we can rearrange the inner product between the
difference of two waveforms to find~\cite{McWilliams:2010eq,Hannam:2010ky}, 
\begin{eqnarray}
\nonumber\left| \model_1 - \model_2 \right|^2 & = & 2 \left|\model_1\right|^2 \left( 1 -\frac{\Braket{\model_1|\model_2}}{\left|\model_1\right| \left|\model_2\right|} \right), \\
\nonumber& = & 2 \rho^2  \mathcal{M}\left(\model_1,\model_2\right), \\
 & = & 2 \rho^2 \hat{d}^{\,2}\left(\model_1, \model_2\right), \label{eq:diffToDistance}
\end{eqnarray} 
assuming that both waveforms have the same SNR, $\rho^2 = |\model_1|^2 = |\model_2|^2$. We see then that $\hat{d}$ 
is proportional to the norm of the difference between the two normalized waveforms.
In addition, when written in terms of normalized signals \(\hat{\model}=\model/\left|\model\right|\) under the same assumptions as
above and rearranged, Eq.~\eqref{eq:diffToDistance} becomes,
\begin{equation}
\hat{d}^{\,2}\left(\model_1,\model_2\right)=\frac12\left|\hat{\model}_1-\hat{\model}_2\right|^2,
\end{equation}
which we discuss further in the context of the linear signal approximation in Appendix~\ref{sec:Fisher}.

To connect the normalized distance to error measures, 
we note that the mismatch between a waveform and some approximation of it
can be related, to leading order in the amplitude and phase uncertainties 
in the approximate waveform $\Delta A$ and $\Delta \phi$, 
as $\mathcal{M} \sim (\Delta A)^2$ and  $\mathcal{M} \sim (\Delta \phi)^2$. 
(See, for example, the discussion in Sec.~IV.C.1 of Ref.~\cite{Hamilton:2023qkv}.)
Since we ultimately want to relate mismatch calculations to waveform accuracy 
requirements, we prefer to use the normalized difference $\hat{d}$, as it 
is proportional to the standard uncertainty measures
of the waveform.

%%%%%%%%%%%%%%%%%%%%%%%%%%%%%%%%%%%%%%%%%%%%%
\subsection{Indistinguishability SNR}
\label{sec:IndistSNR}

\subsubsection{Standard definitions}

If we assume that the statistical likelihood behaves as a 
Gaussian in $|\model_1-\model_2|$, which is true in the high-SNR limit~\cite{Cutler:1994ys}, 
then two waveforms will be distinguishable at one
standard deviation when $|\model_1 - \model_2| > 1$, or 
$\mathcal{M} > 1/(2\rho^2)$~\cite{Flanagan:1997kp, lindblom:2008cm}. More generally, 
if we optimize the mismatch over all but $N$ parameters in a 
model, then the signals will be distinguishable with probability 
$p$ if~\cite{Baird:2013dbm}, 
\begin{equation}
\mathcal{M} > \frac{\chi_N^2(1-p)}{2\rho^2}, \label{eq:ind1}
\end{equation} 
where $\chi_N^2(1-p)$ is the chi-square value 
at probability $p$ for $N$ degrees of freedom. If we are interested in
only 1-$\sigma$, where $p = 0.657$, then $\chi_N^2(0.37) = N$, recovering 
a commonly-quoted indistinguishability criterion~\cite{Chatziioannou:2017tdw},
\begin{equation}
\mathcal{M} > \frac{N}{2\rho^2}.
\end{equation}

We see from Eq.~(\ref{eq:ind1}) that the requirement on a waveform 
model's mismatch uncertainty scales with $1/\rho^2$. 
The requirement on the normalized waveform difference, $\hat{d}$, 
therefore scales as $1/\rho$; this reflects the intuitive result
that the accuracy requirements on waveforms (e.g., the accuracy of
 their amplitude and phase), also scales with $1/\rho$;
if we detect signals twice as loud, we require waveform models twice as accurate. 

Equation~(\ref{eq:ind1}) motivates the standard way to estimate the SNR at 
which a model will yield biased parameter estimates:
we calculate the mismatch between a fiducial signal (e.g., a 
numerical-relativity waveform), and a signal model, keeping the
true intrinsic binary parameters (the masses and spins) fixed, and 
optimising over the extrinsic parameters (distance, orientation,
etc). The number of degrees of freedom is then the number of parameters 
that we have not optimized over, or which we consider
physically meaningful to measure. For example, in an aligned-spin 
system the intrinsic parameters are the total mass $M$, the mass ratio $q$,
and the two black-hole spins $\chi_{1z}$ and $\chi_{2z}$. At low SNRs it 
is not possible to measure both spins, only a mass-weighted
sum (commonly $\chi_{\rm eff} = (m_1\, \chi_{1z} + m_2\, \chi_{2z})/M$~\cite{Ajith:2009bn}), 
and so we may consider this system to have three rather
than four degrees of freedom. In many cases the appropriate number of 
degrees of freedom may be unclear. We will resolve this apparent confusion in 
the next section.

\subsubsection{Issues and Resolutions}
\label{sec:SNRissues}

As noted above, the standard application of Eq.~(\ref{eq:ind1}) leads to a conservative estimate of 
the minimum SNR at which parameter biases appear. There
are two reasons for this. 

\begin{figure}[t]
	\centering
	\includegraphics[width=0.4\textwidth]{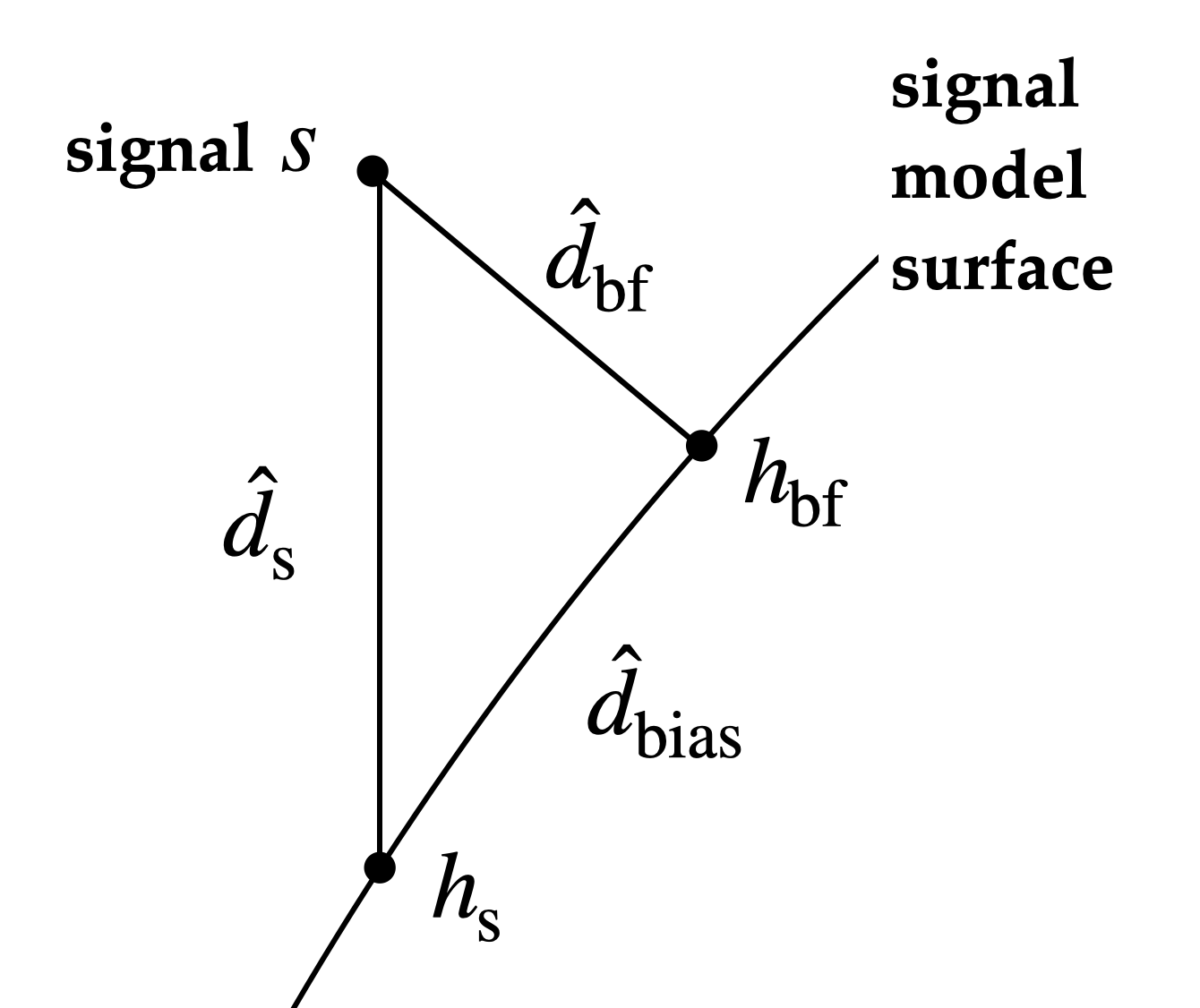}
	\caption{Illustration of the relationship between the true signal $s$, the model signal with the true
	parameters, $h_s$, and the model signal with the parameters $\thetabf$ that agrees best with 
	the true signal, $\modelbf$. We can relate the three waveforms by considering directions parallel and 
	perpendicular to the signal-model manifold, as in Eqs.~(\ref{eq:hperp})-(\ref{eq:hpara}).}
	\label{fig:manifolds}
\end{figure}

\paragraph{Uninformative Perpendicular SNR:}
The first reason is well known from discussions of waveform accuracy dating 
back to Refs.~\cite{Cutler:1994ys,Flanagan:1997kp,Cutler:2007mi,lindblom:2008cm}, and becomes clear 
when we consider the discussion above in more detail. Consider the illustration in Fig.~\ref{fig:manifolds}. 
Denote the signal waveform $\signal$ and the source parameters $\thetasig$. 
Denote the model evaluated at these true source parameters by $\modelsig$. 
The distance between $\signal$ and $\modelsig$, $\hat{d}_\text{s}$, 
is the quantity typically used in calculating the indistinguishability 
SNR. However, this is not the relevant quantity when considering 
parameter biases: we are not interested in whether $\modelsig$ can be 
distinguished from $\signal$, but whether $\modelsig$ can be distinguished
from $\modelbf$, the model evaluated at parameters that give the best agreement 
between the model and the signal, \(\thetabf\),
which are those that will be measured in a parameter 
estimation exercise. We wish to know whether the true 
parameters $\thetasig$ will lie within some credible interval (CI) around~$\thetabf$. 

It is common to treat numerical relativity waveforms as true signals, 
and calculate matches between these and a given model
calculated with the same intrinsic parameters. In general this will 
over-estimate the mismatch and lead to an indistinguishability 
SNR that is too conservative. We must instead find the model parameters 
that maximize the agreement with the NR signal, 
and then calculate the mismatch between model waveforms at these two sets of parameters. 

We are therefore interested in the distance between the true parameters
and the best-fitting model parameters that lie solely within the model manifold,
$\hat{d}_{\rm bias} = \sqrt{\mathcal{M}\left(\modelbf, \modelsig \right)}$, 
meaning that we wish to ignore the contribution to 
the signal waveform that is orthogonal to the manifold of our model. Another 
way of saying this is that the difference between
the true signal waveform and the model at the true parameters is made up of 
two contributions: one that leads to a bias in the
measured parameters, and another that does not introduce a bias, but only 
reduces the extracted SNR of the signal~\cite{Flanagan:1997kp}. (See also Sec.~II.B of 
Ref.~\cite{lindblom:2008cm}.) The appropriate indistinguishability criterion for 
\(N\) degrees of freedom is then, 
\begin{equation}
\hat{d}_{\rm bias, \,ND}^{\,2} = \mathcal{M}\left(\modelbf , \modelsig\right) > \frac{\chi_N^2(1-p)}{2\rho^2}. \label{eq:ind2}
\end{equation}

It is common to refer to $\hat{d}_\text{s}^2 = \mathcal{M}\left(\signal,\modelsig\right)$
%In much of the waveform modelling and signal-detection literature 
%[REFs] $\hat{d}_\text{s}^2 = \mathcal{M}\left(\signal,\modelsig\right)$ is referred to 
as the \emph{faithfulness} mismatch, because it is a measure of 
how well the model reproduces the signal when evaluated at the same parameters. 
The quantity $\hat{d}_\text{bf}^2 = \mathcal{M}\left(\signal,\modelbf\right)$ 
is referred to as the \emph{effectualness} mismatch, because it 
describes how effective the model is at reproducing the signal in total, 
and is the smallest mismatch
that can be achieved in a search or parameter estimation (assuming zero noise). 

The result in Eq.~\eqref{eq:ind2} has also been recently 
discussed again in Ref.~\cite{Toubiana:2024car}. 
Their Eq.~(16) takes the place of our
Eq.~(\ref{eq:ind2}) and in our notation would be,
 \begin{equation}
	\label{eq:mismatch-diff-snr}
\mathcal{M}\left(\signal, \modelsig\right) - \mathcal{M}\left(\signal ,\modelbf\right)  > \frac{\chi_N^2(1-p)}{2\rho^2}.
\end{equation} 
This is equivalent to Eq.~(\ref{eq:ind2}) because, to a good approximation, $ \hat{d}_{\rm bias}^{\,2} = \hat{d}_\text{s}^{\,2} - \hat{d}_\text{bf}^{\,2}$,
i.e., $\mathcal{M}\left(\modelbf, \modelsig\right) = \mathcal{M}\left(\signal ,\modelsig\right) - \mathcal{M}\left(\signal, \modelbf\right)$.
That the usual Pythagorean relation approximately holds 
(and we're not just being misled by the notation), can be seen by 
considering normalized waveforms and writing the signal and 
model waveforms in terms of the model at the true parameters as,
\begin{eqnarray}
\hat{\signal} & = & A\, \hat{\model}_\text{bf} + \sqrt{1-A^2}\, \hat{\model}_\perp,  \label{eq:hperp} \\
\hat{\model}_\text{s} & = & B\, \hat{\model}_\text{bf} + \sqrt{1-B^2}\, \hat{\model}_\parallel, \label{eq:hpara}
\end{eqnarray} 
where $\hat{\model}_\perp$ and $\hat{\model}_\parallel$ are both orthogonal to $\hat{\model}_\text{bf}$ and to each other. 
We note that $A = 1 - \mathcal{M}\left(\signal, \modelbf\right)$ and $B = 1 - \mathcal{M}\left(\modelbf, \modelsig\right)$, 
and so $\braket{\hat{\model}}{\hat{\model}_\text{s}} = AB \approx 1 - \mathcal{M}\left(\signal ,\modelbf\right) 
- \mathcal{M}\left(\modelbf, \modelsig\right)$ to leading order in the mismatches,
which gives us the desired result. 

\begin{figure*}[ht!]
	\centering
	\includegraphics[width=\textwidth]{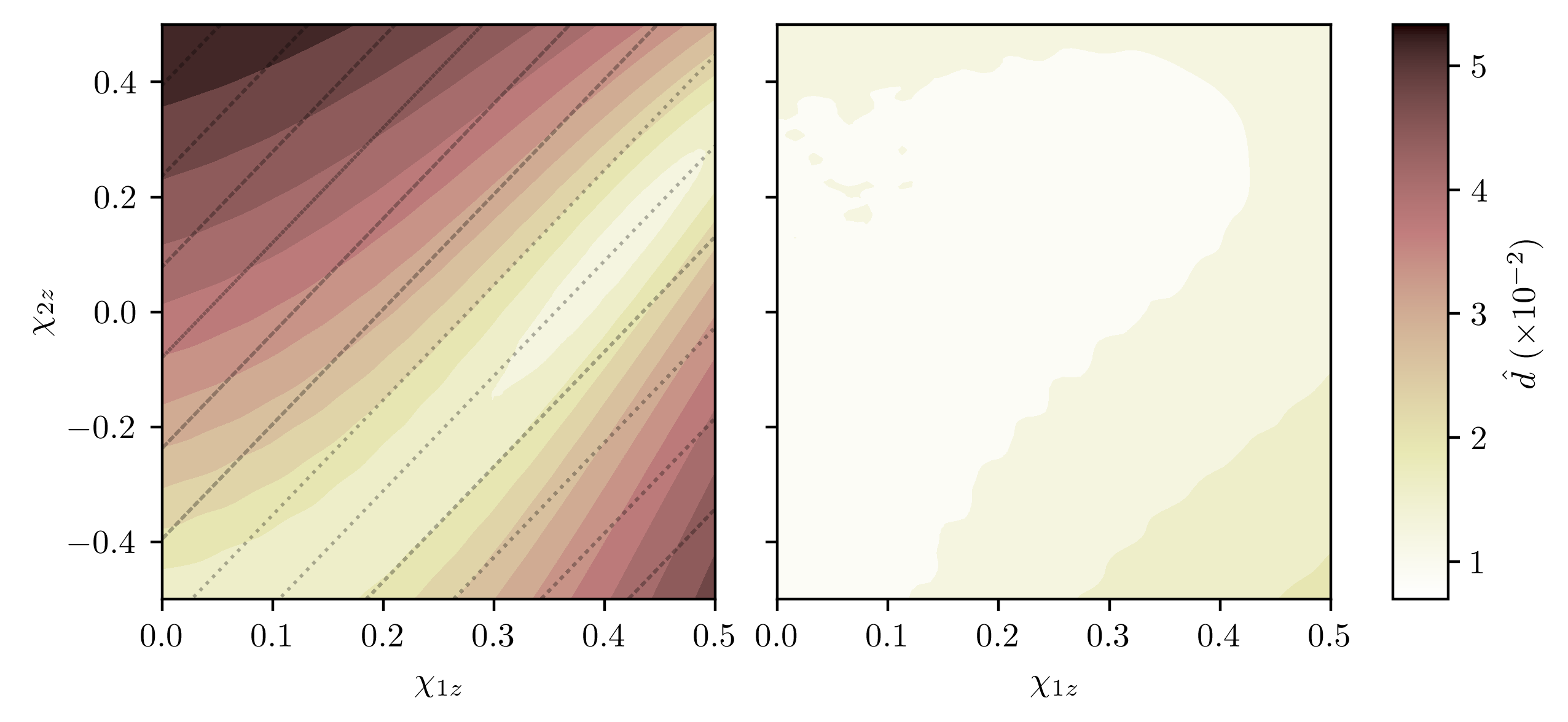}
	\caption{
	Contour plots of \(\hat{d}\) computed
	for the models \d{} and \textsc{NRHybSur3dq8}, plotted for a range of 
	\(\chi_{1z}\) and \(\chi_{2z}\) for fixed values of \((m_1,m_2)=(200,100)\)~\(M_\odot\).
	In the left panel, the faithfulness is used to compute the distance \(\hat{d}_\text{s}\), 
	while the right panel displays the distance \(\hat{d}_\text{bf}\) arising from the effectualness.
	The dotted lines show lines of constant \(\chi_\text{antisym}\), indicating
	that model accuracy varies strongly with changing~\(\chi_\text{eff}\).
	}
	\label{fig:pd_sur_snr_contours}
\end{figure*}

To illustrate the relative importance of $\hat{d}_{\rm bf}$ to $\hat{d}_{\rm s}$,
%differences between the faithfulness and effectualness of a model,
we show in Fig.~\ref{fig:pd_sur_snr_contours} 
%the different SNRs arising from the use of Eq.~\eqref{eq:ind1} using either 
$\hat{d}_{\rm s}$ in the left panel, compared to $\hat{d}_{\rm bf}$
in the right panel, between the models \d{} and \textsc{NRHybSur3dq8}. These
results are plotted for a range of \(\chi_{1z}\) and \(\chi_{2z}\) values for fixed
masses \((m_1,m_2)=(200,100)\)~\(M_\odot\).
We see that the waveform model error orthogonal to the model surface, $\hat{d}_{\rm bf}$ 
varies little over the parameter space, while $\hat{d}_{\rm s}$  
shows a clear trend of variation 
perpendicular to lines of constant
\(\chi_\text{antisym}=(m_1\chi_{1z}-m_2\chi_{2z})/M\), displayed as dotted lines
in the left panel. Seeing these results, one would hypothesize that a parameter
estimation study injecting any one of these \textsc{NRHybSur3dq8} signals and
recovering with \d{} would find comparable recovered SNRs regardless of the spin
values used for the injection. From Eq.~\eqref{eq:mismatch-diff-snr} 
one would infer that the difference in any of these
injections would be the varying levels of parameter bias seen in the parameter
estimation. We also see that $\hat{d}_{\rm bf}$ is comparable to $\hat{d}_{\rm s}$
only when $\hat{d}_{\rm s}$ is small, i.e., the bias distance that we are most interested
in, $\hat{d}_{\rm bias}$, will be well approximated by $\hat{d}_{\rm s}$ except in cases
where the indistinguishability SNR is high. Nonetheless, as we will see, we require 
$\hat{d}_{\rm bias}$ to accurately calculate bias SNRs.
 
We will refer to the simple mismatch-based indistinguishability SNR in 
Eq.~\eqref{eq:ind1} first described in Ref.~\cite{Baird:2013dbm}
as the \emph{faithfulness SNR}~\(\rho_\text{faith}\). We call the improved estimate in 
Eq.~\eqref{eq:ind2} the \emph{N-D~bias SNR}~\(\rho_\text{bias,\,ND}\). 

\begin{figure}[ht!]
	\centering
	\includegraphics[width=0.5\textwidth]{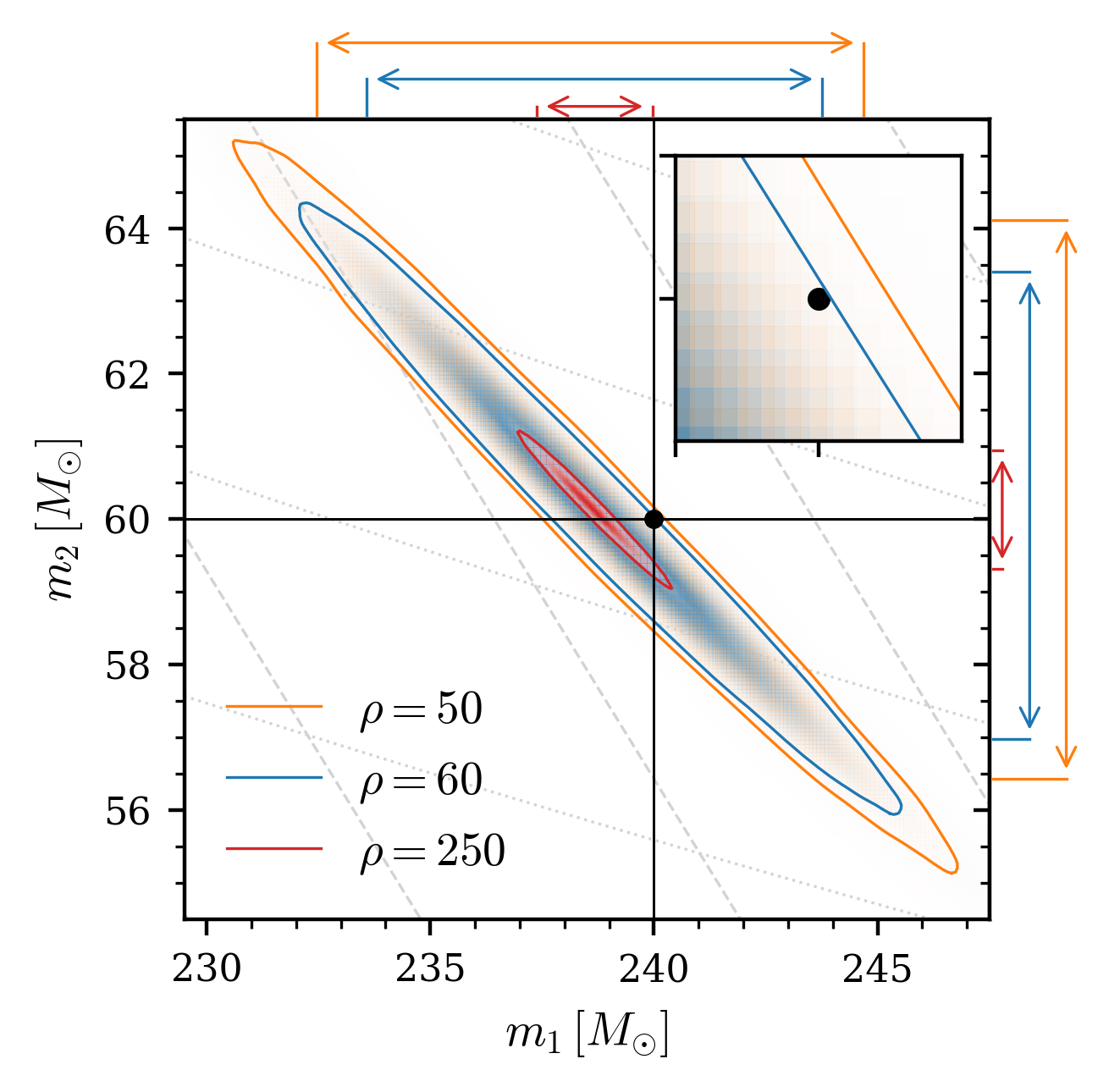}
	\caption{Measurement of the primary mass, $m_1$, and secondary mass $m_2$ for different signal-to-noise ratios $\rho$ for the case \texttt{BAM-3} described in Table~\ref{tab:simcases}. The black dot indicates the true parameters, the contours
	show the 90\% credible intervals and the horizontal/vertical lines above and to the right of the Figure show the 90\% symmetric credible intervals for the 1D marginalized posteriors. The inset shows a zoomed in portion of the posterior, focusing on the correlation between the true parameters and the credible interval at which they are biased. We show lines of constant total mass (grey dashed) and constant chirp mass (grey dotted). For this simulation, the faithfulness indistinguishability SNR is 52, the 2D bias SNR is 60, and the primary mass is estimated to be biased at $\rho\approx 250$. We see that the 2D bias SNR correctly identifies the SNR at which the 2D posterior is biased (at the 90\% credible interval) and the 1D marginalized posterior for the primary mass remains unbiased until $\rho\approx 250$.
	}
	\label{fig:m1m2_toy_q4aM075}
\end{figure}

Consider Fig.~\ref{fig:m1m2_toy_q4aM075}, which illustrates
parameter measurement for a two-dimensional toy problem where the only 
parameters in the model are $m_1$ and $m_2$.
(See Sec.~\ref{sec:signalwaveforms} for more details of this configuration.)
The figure shows the true parameters, indicated by a black dot, and the 2D 
90\%~CIs for signals at a selection of
SNRs. In this example the faithfulness 
SNR~\(\rho_\text{faith}\) from Eq.~\eqref{eq:ind1} is 52 and the 2D~bias SNR~\(\rho_\text{bias,\,2D}\) from Eq.~\eqref{eq:ind2} 
is 60. We see that the true parameters lie approximately on the 2D CI boundary for a 
signal at SNR 60, but are within the CI for a signal with SNR 50, 
consistent with the discussion above. For this example these two SNR estimates
differ only by about 10\%, but we will see cases below for which these SNR
estimates may disagree by 150\%.

\paragraph{Uninteresting Bias Directions:}
The second reason why these estimates of the indistinguishability SNR are too 
conservative arises from the fact that we are dealing with a multi-dimensional 
parameter space~\cite{Ohme:2012cba}. 
Consider the 1D CIs for $m_1$ and $m_2$ in Fig.~\ref{fig:m1m2_toy_q4aM075} separately (shown as vertical 
lines above and to the right of the figure; note 
that the 1D intervals are always narrower than the direct projection of the 2D 
intervals). Neither $m_1$ nor $m_2$ is remotely close to being biased at 
SNRs of 50 or 60. The faithfulness and 2D~bias SNRs tell us nothing 
about the potential bias of the two parameters we are interested in. 

We may expect that the 2D~bias SNR estimate may be more accurate if we consider some other 
parameterization of the masses, e.g., the total mass $M = m_1 + m_2$, or the chirp mass 
$M_{\rm chirp} = M \eta^{3/5}$ where \(\eta=m_1m_2/(m_1+m_2)^2\) is the symmetric mass
ratio. Lines of constant chirp mass and total mass are shown in 
Fig.~\ref{fig:m1m2_toy_q4aM075}, and we see that biases will not be incurred in either parameter
at SNRs 50 or 60. It may be the case that the 2D~bias SNR does apply to the chirp mass in 
low-mass binaries where the signal is dominated by the inspiral (and therefore the leading-order 
PN phasing, from which the chirp mass derives), but in general we do not expect the $N$-D~bias SNR
to apply to any single parameter of interest. We provide more detailed examples in 
Secs.~\ref{sec:EffSNR} and \ref{sec:BiasSNR} to illustrate this point.

To accurately estimate the SNR at which individual parameters 
will be biased we must calculate separately the bias SNR for each parameter
(or combination of parameters) that we are interested in. This is 
straightforward to do. We first calculate the complete set of parameters $\thetabf$ 
at which the model best agrees with the true signal, as before. We then
calculate the parameters at which the model best
agrees with the true signal, but \emph{keep the one parameter we are interested 
in fixed to its true value}. In this way, we are considering the distance between
the true and best-fit values of that one parameter along the curve connecting
these two points within the model manifold that is always closest to the true signal
in all other parameters.

We also illustrate this SNR estimate in Fig.~\ref{fig:m1m2_toy_q4aM075}. 
To find the SNR at which, say, $m_1$ is biased, we keep 
$m_1$ fixed to its true value and optimize all other parameters to find 
the parameters that give the best agreement with the true signal under
this restriction, $\thetaod{m_1}$. 
The distance between the model at $\thetabf$ and at 
$\thetaod{m_1}$, $\hat{d}\left(\modelbf, \model_{\text{bf}\,|\,m_1}\right)$,
will tell us the bias SNR for $m_1$. In this case it 
is approximately SNR~250. Since the single-parameter indistinguishability
SNRs can be related directly to the SNR at which each parameter 
will be biased, we call this the ``$m_1$ bias SNR'' and denote it by 
$\rho_{m_1}$. We note that this result is presented in a similar fashion 
in Eq.~(26) of Ref.~\cite{Toubiana:2024car},
again asserting that the Pythagorean relation holds between the distances as
described in Eq.~\eqref{eq:mismatch-diff-snr}, and our separate maximization over
all other parameters at both \(\theta_\text{s}\) and \(\theta_\text{bf}\) amounts
to their choice of maximum averaged overlap.

We see, then, that the $N$-D~bias SNR \emph{does} reliably 
predict the SNR at which a measurement will
be biased from the true parameters, but only in the 
$N$-dimensional credible interval, where $N$ is 
the number of parameters that were kept fixed in the mismatch optimization. To calculate the SNR at
which a particular parameter is biased, we must calculate the appropriate parameter bias SNR.

These statements are based on a small number of assumptions. The derivation of
Eq.~\eqref{eq:ind1} in Ref.~\cite{Baird:2013dbm} begins with an assumption of
sufficiently high SNR to allow for the Gaussian posterior scaling, which is equivalent
to assuming the linear signal approximation outlined in Appendix~\ref{sec:Fisher}.
This assumption is relaxed somewhat by using the mismatch instead of the Fisher
matrix in their calculation, though we still assume that 
the signal and model have approximately equal
SNR. We further assume that the minimization to find the best-fit parameters has
a true minimum (equivalent to the likelihood being unimodal). This is a valid
assumption for the comparable-mass black-hole-binary signals we consider here, 
though may not always hold for other sources of gravitational waves~\cite{Chua:2021aah}.
For more realistic signal models with higher multipoles, one can find reparameterizations
of the extrinsic parameters to ensure unimodal posteriors~\cite{Roulet:2022kot}.
We finally assume that the impact of the (broadly uniform) prior probability 
is negligible for this analysis, except when the best-fit parameters lie close
to a prior boundary, as discussed in Sec.~\ref{sec:prior-bounds}.

We have illustrated here that these assumptions hold for one example. 
In Sec.~\ref{sec:EffSNR} we will illustrate in more detail with two-dimensional toy 
models and full four-dimensional examples 
that Eq.~\eqref{eq:ind2} does correctly predict the 
indistinguishability SNR, so long as we calculate the normalized 
distance $\hat{d}\left(\modelbf, \modelsig\right)$; the distance
used in most applications of Eq.~\eqref{eq:ind1}, 
$\hat{d}\left(\signal, \modelsig\right)$, provides only a conservative
estimate. We then show in Sec.~\ref{sec:BiasSNR} that by optimizing the 
mismatch over all parameters but one, we can calculate the 
bias SNR for that parameter.

%%%%%%%%%%%%%%%%%%%%%%%%%%%%%%%%%%%%%%%%%%%%%
\section{Methods}
\label{sec:methods}

We now describe the specifics of the numerical set-up used to produce the results
in this paper.

%%%%%%%%%%%%%%%%%%%%%%%%%%%%%%%%%%%%%%%%%%%%%
\subsection{Signal waveforms and waveform models}
\label{sec:signalwaveforms}

\begin{table*}
	\renewcommand{\arraystretch}{1.2}
\begin{tabular}{@{} c||ccc||c|cc|cc @{}}
    \toprule
\hspace{1em}Simulation ID\hspace{1em} &	\hspace{1em}$q$\hspace{1em}	&	\hspace{1em}$\chi_{1z}$\hspace{1em}	&	\hspace{1em}$\chi_{2z}$\hspace{1em}	&	\hspace{1em}$\hat{d}_\text{s}^{\,2}$\tiny{$(\times10^{-3})$}\hspace{1em}	&	\hspace{1em}$\hat{d}_\text{bf, 4D}^{\,2}$\tiny{$(\times10^{-3})$}	&	$\hat{d}_\text{bias, 4D}^{\,2}$\tiny{$(\times10^{-3})$}	&	$\hat{d}_\text{bf, 2D}^{\,2}$\tiny{$(\times10^{-3})$}	&	$\hat{d}_\text{bias, 2D}^{\,2}$\tiny{$(\times10^{-3})$} \\ \hline
\texttt{BAM-1}	&	3	&	-0.5	&	-0.5	&	1.22	&	0.13	&	1.09	&	0.23	&	1.00 \\
\texttt{BAM-2}	&	4	&	0.25	&	0	&	0.40	&	0.14	&	0.26	&	0.34	&	0.06 \\
\texttt{BAM-3}	&	4	&	-0.75	&	0	&	0.86	&	0.23	&	0.63	&	0.25	&	0.62 \\
\texttt{BAM-4}	&	10	&	0	&	0	&	1.48	&	0.16	&	1.32	&	0.21	&	1.26 \\
\texttt{BAM-5}	&	18	&	0	&	0.4	&	3.68	&	0.29	&	3.38	&	0.51	&	3.16 \\ \hline
\texttt{SUR-1}	&	2	&	0.5	&	-0.5	&	1.75	&	0.13	&	1.62	&	0.73	&	1.02 \\
\texttt{SUR-2}	&	2	&	0.4	&	0.1	&	0.08	&	0.07	&	0.01	&	0.07	&	0.01 \\
\texttt{SUR-3}	&	2	&	0.3	&	-0.4	&	0.22	&	0.09	&	0.13	&	0.16	&	0.05 \\
\texttt{SUR-4}	&	2	&	0.05	&	0.47	&	2.49	&	0.07	&	2.40	&	0.87	&	1.59 \\
    \botrule
\end{tabular}

% \begin{tabular}{@{} c|ccc|cccc @{}}
%     \toprule
%     \hspace{1em}Simulation ID\hspace{1em} & \hspace{1em}$q$\hspace{1em} & $\chi_{1z}$ & \hspace{1em}$\chi_{2z}$\hspace{1em} & $\hat{d}^{\,2}_\text{s}$ & $\hat{d}^{\,2}_\text{bf}$ & $\hat{d}_\text{bias, 2D}$ & $\hat{d}_\text{bias, 4D}$ \\
%     \hline
%     \texttt{BAM}-1 & 3 & -0.5 & -0.5 & \hspace{1em}$1.2\times10^{-03}$\hspace{1em} & \hspace{1em}$1.3\times10^{-04}$\hspace{1em} & \hspace{1em}$3.2\times10^{-02}$\hspace{1em} & \hspace{1em}$3.3\times10^{-02}$\hspace{1em} \\
%     \texttt{BAM}-2 & 4 & 0.25 & 0 & $4.0\times10^{-04}$ & $1.4\times10^{-04}$ & $7.8\times10^{-03}$ & $1.6\times10^{-02}$ \\
%     \texttt{BAM}-3 & 4 & -0.75 & 0 & $8.6\times10^{-04}$ & $2.3\times10^{-04}$ & $2.5\times10^{-02}$ & $2.5\times10^{-02}$ \\
%     \texttt{BAM}-4 & 10 & 0 & 0 & $1.5\times10^{-03}$ & $1.6\times10^{-04}$ & $3.6\times10^{-02}$ & $3.6\times10^{-02}$ \\
%     \texttt{BAM}-5 & 18 & 0 & 0.4 & $3.7\times10^{-03}$ & $2.9\times10^{-04}$ & $5.6\times10^{-02}$ & $5.8\times10^{-02}$ \\\hline  
%     \texttt{SUR}-1 & 2 & 0.5 & -0.5 & $1.8\times10^{-03}$ & $1.3\times10^{-04}$ & $3.2\times10^{-02}$ & $4.0\times10^{-02}$ \\  
%     \texttt{SUR}-2 & 2 & 0.4 & 0.1 & $8.0\times10^{-05}$ & $6.6\times10^{-05}$ & $3.5\times10^{-03}$ & $3.7\times10^{-03}$ \\
%     \texttt{SUR}-3 & 2 & 0.3 & -0.4 & $2.2\times10^{-04}$ & $8.8\times10^{-05}$ & $7.1\times10^{-03}$ & $1.1\times10^{-02}$ \\
%     \texttt{SUR}-4 & 2 & 0.05 & 0.47 & $2.5\times10^{-03}$ & $7.2\times10^{-05}$ & $4.0\times10^{-02}$ & $4.9\times10^{-02}$ \\
%     \botrule
% \end{tabular}
\caption{Table of simulation configurations used in this work, listing the mass-ratio
and aligned dimensionless spins of each black hole, described in Sec.~\ref{sec:signalwaveforms}. 
All simulated signals are 
generated at a total mass of \(300\,M_\odot\) and a starting frequency 
\(f_\text{22}=2\)~Hz. We also present values for the faithfulness and effectualness
mismatches, $\hat{d}^{\,2}_\text{s}$ and $\hat{d}^{\,2}_\text{bf}$ respectively 
described in Sec.~\ref{sec:SNRissues}, 
between these signals and \d{} computed over a frequency range of $5$--$128$~Hz.
We finally tabulate the 2D and 4D bias distances \(\hat{d}_\text{bias}^{\,2}\) to
be used in Sec.~\ref{sec:EffSNR}. Note that the relation 
$\hat{d}_{\rm s}^2 = \hat{d}_{\rm bf}^2 + \hat{d}_{\rm bias}^2$ holds to a good approximation, 
independently of the number of degrees of freedom used in the optimisation.
}
\label{tab:simcases}
\end{table*}

In this work we predominantly use numerical-relativity waveforms as 
proxies for our signals $s$. Numerical-relativity solutions of Einstein's equations for black-hole
mergers are excellent representations of real astrophysical signals in that 
the only approximations in the NR calculations are the numerical
errors (which can in principle be reduced to any desired level with sufficient 
numerical resolution), and the calculation of the GW signal at
a finite distance from the source. (Recall that gravitational waves are formally defined at null infinity.)

We use NR waveforms produced by the BAM code, which solves the moving-puncture treatment 
of Baumgate-Shapiro-Shibata-Nakamura (BSSN) 
formulation with finite-difference methods~\cite{Bruegmann:2006ulg}.
The waveforms listed in Table~\ref{tab:simcases} were previously 
published in Ref.~\cite{Husa:2015iqa} and used to tune or verify the 
phenomenological model \d~\cite{Khan:2015jqa}. We consider only quasi-circular aligned-spin 
binaries, where the black-hole spins are aligned with the
orbital angular momentum, and the binary's orbital plane is fixed, i.e., there is 
no spin precession. We use only the dominant ($\ell=2$, $m=2$) multipole,
so that the signal's orientation and polarization can be absorbed into an 
overall amplitude factor. To generate the NR signals down to the required 
starting frequency, we use the hybrids constructed in Ref.~\cite{Kalaghatgi:2019log}
(restricting to the $\ell=2,m=\pm2$ multipoles). 

%hybridize the NR waveforms with the model \textsc{SOBNRv4}~[REFs],
%using the hybridization method outlined in Refs.~\cite{Hotokezaka:2016bzh,Dietrich:2018uni}.

We also use the \textsc{NRHybSur3dq8} model~\cite{Varma:2018mmi} to produce proxy signals. 
This model is calibrated to NR waveforms from binaries
with mass ratios between $q=1$ and $q=8$, and spins up to $\chi = 0.8$, and allows 
us to consider signals at arbitrary points in this parameter
space. As with the BAM NR waveforms, for this study we only consider aligned-spin 
binaries, and the dominant $(2,2)$ multipole.

As an example waveform model to assess systematics we chose \d, 
for three reasons. (1) \d{} models only the dominant $(2,2)$
multipole of aligned-spin binaries, which provided a convenient reduced parameter 
space on which to test our approach; (2) \d{} is a relatively old
model with larger uncertainties than more recent models,
ensuring that our model is less accurate than our proxy signal waveforms; (3) 
\d{} was calibrated to a subset of the BAM NR waveforms that we 
use in this study listed in Table~\ref{tab:simcases}, allowing a 
consistent test of the performance of the model 
against waveforms that were treated as true signals in the model's
construction. 

In any consideration of waveform systematics it is important that the
uncertainties in the signal proxy waveforms are far smaller than the uncertainties 
in the models we are assessing, otherwise the signal uncertainties will 
contaminate our results. Error estimates for NR waveforms can be
expressed as mismatch uncertainties, which allow us to calculate their 
faithfulness indistinguishability SNR. We estimate the mismatch 
uncertainties of the BAM NR signals and the \textsc{NRHybSur3dq8} signals as 
$\sim$$10^{-4}$, i.e., $\hat{d} \sim 10^{-2}$, putting the faithfulness indistinguishability SNR
at approximately $1/\hat{d} \sim 100$. We will see in Sec.~\ref{sec:BiasSNR} 
that parameter bias SNRs can be much larger, and so we must be cautious
in interpreting these results. This is not a serious issue for this 
proof-of-principle study, where we are considering only the (2,2)-multlipole
of aligned-spin models, since these will not be used to measure properties 
of loud GW observations, but this will be a crucial point to bear in
mind when we assess systematics for state-of-the-art models in future work.

%%%%%%%%%%%%%%%%%%%%%%%%%%%%%%%%%%%%%%%%%%%%%
\subsection{Optimal model parameters} 
\label{sec:optimal-parameters}

The distance measure introduced in Sec.~\ref{sec:mismatchdistance} requires
optimization over a set of parameters \(\Theta_\text{opt}\). When computing the faithfulness
between two spin-aligned
quadrupolar signals, \(\Theta_\text{opt}\) only includes the coalescence time, \(t_c\), and 
coalescence phase, \(\varphi_c\), and in this case we express the time-
and phase-shift optimization of the match in a computationally efficient manner
using an inverse Fast Fourier Transform (iFFT)~\cite{Balasubramanian:1995bm,Ohme:2012cba},
\begin{align}
 \nonumber M_{t\varphi}(\model_1,\model_2)&=\max_{\{t_c,\varphi_c\}}\,\frac{\Braket{\model_1|\model_2}}{\left|\model_1\right|\left|\model_2\right|}\\
 &=\frac{4}{\left|\model_1\right|\left|\model_2\right|}\max_{t_c}\left|\,\text{iFFT}
\left[\frac{\tilde{h}_1 \left(f\right) \tilde{h}^{*}_2 \left(f\right)}{\tilde{S}_n\left(f\right)}\right](t_c)\right|.
\label{eq:mismatchtpopt}
\end{align}
Maximization of \(\varphi_c\) is achieved by taking the norm in 
Eq.~\eqref{eq:mismatchtpopt} and  maximization over \(t_c\) is 
done by taking the maximum component of the 
output iFFT array. We can increase the resolution of the
discrete timestep used for the timeshift optimization 
by padding the frequency-domain data before
taking the iFFT, which is especially important for signals with
only slight differences between the linear-in-frequency contributions
to their phases~\cite{Ajith:2012az, Ohme:2012cba}.

To compute the bias distances we need the appropriate
best-fit parameters between the model and signal. We write the best-fit parameters \(\thetabf\)
as the union between a set of optimized parameters \(\xi_\text{bf}\in\Theta_\text{opt}\) 
and a set of parameters held fixed, \(\bar{\theta}\in\Theta\setminus\Theta_\text{opt}\), 
such that \(\thetabf=\xi_\text{bf}\cup\bar\theta\) is
found through the \textit{minimization} of the mismatch,
\begin{equation}
\xi_\text{bf}\left(\signal;\bar\theta\right)=\argmin_{\xi\,\in\,\Theta_\text{opt}}
\mathcal{M}\left(\signal,\model(\xi; \bar\theta)\right).
\label{eq:optimal-params}
\end{equation}
For the case of computing the $N$-D~bias SNR in Eq.~\eqref{eq:ind2}, the best-fit
parameters are found by optimizing over all signal parameters, thusly \(\Theta_\text{opt}=\Theta\).
To compute the distances for individual parameter biases, for example the bias
SNR estimate for the primary mass \(m_1\), then \(\bar{\theta}=\{m_1\}\) and we minimize 
Eq.~\eqref{eq:optimal-params} over all remaining parameters, in this case 
\(\xi=\{m_2,\chi_{1z},\chi_{2z},t_c,\varphi_c\}\) for our quadrupolar, spin-aligned
model. This minimization is in practice reduced to a three- or four-dimensional 
numerical optimization over at most \(\{m_1, m_2, \chi_{1z}, \chi_{2z}\}\) 
using Eq.~\eqref{eq:mismatchtpopt} to compute the time- and phase-optimized mismatch.
We then recover \(\theta_{\text{bf}\,|\,m_1}=\xi_\text{bf}\cup\{m_1\}\) introduced in Sec.~\ref{sec:SNRissues}.

We choose to use the Nelder-Mead~\cite{10.1093/comjnl/7.4.308} algorithm implemented in the Python
library \textsc{SciPy}~\cite{2020SciPy-NMeth} to perform the numerical minimizations.
Nelder-Mead does not rely on numerical derivatives of the objective function
and generally requires only a small number of function 
evaluations to converge sufficiently to a minimum. 
We run the minimization over a spread of initial values, starting at the
true parameters of the signal \(\theta_\text{s}\) and expanding away in quadratically
increasing step sizes in each parameter to ensure at least minimal 
coverage of parameter space regions
far from the true parameters. The initial parameter guesses for the 
Nelder-Mead minimization
are precomputed and then the minimization is performed in parallel, taking the
global minimum found across all resulting values. Finally, we ensure a fine resolution
for the timestep optimization by padding the frequency-domain signals with an array of
zeros to a length equal to a large power of 
2~(\(2^{22}\))~\cite{Ohme:2012cba} and run the Nelder-Mead
algorithm with an absolute error tolerance of~\(10^{-14}\). 

Regardless of the stated error tolerance, we also check the efficacy of the optimization 
by finding the 4D~best-fit parameters using two different parameterizations,
\(\{m_1,m_2,\chi_{1z},\chi_{2z}\}\) and 
\(\{M_\text{chirp},\eta,\chi_\text{eff},\chi_\text{antisym}\}\),
and compare the effectualness in both sets of parameters. This effectualness typically
disagrees with a relative error of \(10^{-5}\), and we find that using either
set of ``best-fit'' parameters impacts the SNRs computed below 
when the SNRs reach values above \(\sim 600\). We therefore strongly suggest
caution when considering any high-SNR predictions in the tabulated data below; we
leave the values in for comparison between methods.

%%%%%%%%%%%%%%%%%%%%%%%%%%%%%%%%%%%%%%%%%%%%%
\subsection{Parameter estimation methods}
\label{sec:PE}

To demonstrate that the effectual indistinguishable SNR correctly corresponds to the biases in 
our inferred estimates for the true source parameters, we perform Bayesian inference to
estimate the \emph{posterior probability density function} for a given signal $\signal$.
Bayesian inference is the process of estimating the properties of the signal for a given model
$\model$ and observed data $\data$. A posterior probability distribution for the parameters
$\theta$, can be obtained through Bayes' theorem,

\begin{equation}
    p\left(\theta\, |\, \data, \model\right) = \frac{p\left(\theta\, |\, \model\right)\, p\left(\data\, |\, \theta, \model\right)}{\mathcal{Z}},
\end{equation}
where $p(\theta\, |\, \model)$ is the prior probability of the parameters
$\theta$ given our model $\model$, otherwise known as the prior,
$p(\data\, |\, \theta, \model)$ is the likelihood of the data given the
parameters $\theta$ and model $\model$ and
$\mathcal{Z} = \int_\Theta{p(\theta\, |\, \model)\, p(\data\, |\, \theta, \model)\, \mathrm{d}\theta}$. 
Under the noise assumptions outlined in Sec.~\ref{sec:IndistBias}, the Whittle likelihood
in gravitational-wave physics is 
proportional to~\cite{Finn:1992wt}
\begin{equation}
	\label{eq:likelihood}
	p\left(\data\, |\, \theta, \noise\right)\propto\exp\left\{-\frac{1}{2}\Braket{\data-\model(\theta)|\data-\model(\theta)}\right\}.
\end{equation} 

An aligned-spin quasi-circular binary black hole signal
is fully characterised by 11 parameters: 4 intrinsic describing the component masses $m_{1}$
and $m_{2}$ and the spins aligned with the orbital angular momenta of each black hole $\chi_{1z}$ and $\chi_{2z}$, and 7 extrinsic parameters describing the source location, inclination angle, merger time \emph{etc.}. For gravitational-wave astronomy it is difficult to analytically calculate the posterior distribution
as it requires evaluating a 11 dimensional integral. 

To further reduce the dimensionality of the evidence integral, it is possible to analytically
marginalize over some parameters~\cite{Veitch:2013aaa,Farr:2014aaa,Singer:2015ema,Singer:2016eax,Thrane:2018qnx}. In this work we marginalize over the 
luminosity distance~\cite{Singer:2015ema,Singer:2016eax} and coalescence phase of the binary~\cite{Veitch:2013aaa}. We also fix the inclination angle,
polarization and
sky location of the binary to their true values. For models that only consider the dominant 
quadrupole of aligned-spin binaries, the sky location and inclination angle only 
affect the overall amplitude of the GW and are degenerate with the luminosity distance. 
In addition, the polarization angle is completely degenerate with the coalescence phase. 
As such, we only sample over the merger time along with the masses and spins of each black hole.
We note that at high SNRs ($\gtrsim 600$) we observed non-negligible differences between the posterior distributions obtained with and without distance and phase marginalization. We therefore do not show posterior distributions for SNRs $> 600$ in subsequent sections.

Given the large parameter space of the evidence integral, stochastic
sampling~\cite{metropolis1949monte,Skilling2004,Skilling:2006} is often employed to draw samples from the unknown posterior 
distribution. Numerous tools are available to perform Bayesian inference
for gravitational-wave astronomy~\cite{Veitch:2014wba,Lange:2018pyp,Ashton:2018jfp,Biwer:2018osg,Smith:2019ucc,Ashton:2021anp,Dax:2021tsq,Tiwari:2023mzf}, and many commonly employ the nested sampling algorithm,
which iteratively evolves a set of {\emph{live points}} randomly drawn from the prior to 
converge to regions of high probability~\cite{Skilling2004,Skilling:2006}.
In this work, we perform Bayesian inference using {\sc{bilby}}~\cite{Ashton:2018jfp} with the
{\sc{dynesty}}~\cite{Speagle:2020} nested sampler.

Since we are interested in confidently identifying the 90\% credible region at potentially high SNRs,
we use 3000 live points and combine the results from 6 independent chains to obtain our
final posterior distribution. This compares to 1000 live points and 4 independent chains
commonly used by the LIGO--Virgo--KAGRA collaboration in their production
analyses~\cite{LIGOScientific:2021vkt}. We employ the {\sc{bilby}}-implemented {\sc{rwalk}} sampling algorithm with
an average of 60 steps per Markov Chain Monte Carlo, and we also assume wide and agnostic 
priors for all parameters. Specifically, we employ uniform priors on the component masses with chirp mass and mass ratio constraints. Constraints are chosen to ensure regions of high probability are sufficiently sampled, while also reducing computational cost where possible. We also assume uniform priors on the aligned-spin components of the binary~\cite[Eq. (A7) in Ref.][]{Lange:2018pyp}.

Although directly translatable to any GW detector network, in this work we focus on 
next-generation GW detectors. Specifically, we assume a single detector network consisting
of the Einstein Telescope (ET)~\cite{Punturo:2010zza} and assume a prospective PSD~\cite{Hild:2010id} when 
evaluating the inner product. Since the exact configuration of ET is
still under discussion, for simplicity we assume that ET is formed of a single L-shaped interferometer~\cite{Branchesi:2023mws}.

%%%%%%%%%%%%%%%%%%%%%%%%%%%%%%%%%%%%%%%%%%%%%
\section{Results: $N$-D Bias SNR} 
\label{sec:EffSNR}

We discussed in Sec.~\ref{sec:SNRissues} how
to identify the SNR at which the true parameters 
will be observably biased from the posterior distribution
using the appropriate distance measure 
\(\hat{d}_\text{bias}\). When this distance is used in Eq.~\eqref{eq:ind2},
the number of degrees of freedom, \(N\), is not immediately specified.
In fact the value of \(N\) depends on the dimensionality of 
the (marginalized) posterior distribution of interest~\cite{Baird:2013dbm} 
and therefore relates to the number of model 
parameters held fixed to their ``true'' values during the 
mismatch optimizations performed in finding \(\thetabf\), 
i.e. the dimensionality of~\(\bar{\theta}\).
In this section we explore the validity of Eq.~\eqref{eq:ind2} through direct
comparison to parameter estimation results and the scaling of the posterior's 90\%~CI. 

We begin by using a model with two effective degrees of freedom, 
\(m_1\) and \(m_2\),  so 
that we can view samples from the entire posterior in 
a two-dimensional scatter plot. We construct this effective model 
from \d{} by fixing the
component spins to the values of the injected signal we compare
against, listed in Table~\ref{tab:simcases}, both when sampling in
parameter estimation and when computing optimal mismatch parameters.
Of interest to us is the SNR at which the intrinsic masses are biased,
and we consider the posterior distribution marginalized over
\(\{d_L,t_c,\varphi_c\}\), yielding an effective 
2D posterior distribution in \(m_1\) and \(m_2\), such as the one plotted in 
Fig.~\ref{fig:m1m2_toy_q4aM075}. The best-fit parameters
are found by optimizing the mismatch over all parameters except the spins.
Afterwards we will extend the analysis to the full four-dimensional model.

\subsection{Principal Component Posteriors and SNR Scaling}
\label{sec:PCA}

The $N$-D~bias SNR formula dictates the scaling of the
bulk \(N\)-dimensional posterior's credible region
under the assumption that the posterior is approximately a multivariate
normal distribution. 
We can approximate this assumption on the posterior samples by 
using Principal Component Analysis~(PCA).
PCA aims to find a linear transformation between the
component directions of the signal parameters with minimal covariance by 
diagonalizing the covariance matrix (i.e., maximizing the variance in each 
parameter). 

In order to perform the PCA on 
the posterior samples, we first normalize each component of the data, shifting
the posterior to have zero mean and unit variance using the 
\textsc{scikit-learn} class \texttt{StandardScalar}. 
The resulting posterior of the PCA will then be an approximate multivariate
normal distribution with zero mean, up to nonlinear 
correlations present in the data. 
In this idealized PCA representation of the data, the 
90\%~CI will approximate an \(N\)-dimensional sphere of radius 
\([{\chi^2_N(0.1)}]^{1/2}\).

As PCA is a linear transformation that projects the data onto
axes constructed from linear combinations of the input parameters,
its ability to produce multivariate normal
samples depends on the strength of nonlinear correlations 
between the input parameters. As the Jacobian between different mass 
parameterizations is nonlinear, we can hope to improve the 
effectiveness of the PCA by choosing a 
sample parameterization that reduces the nonlinearities in sample correlations. 
Ultimately the choice of
input parameters will depend on the structure of the posterior for each of our
injection cases. Given that the PCA computation is not expensive, we choose to 
compute the PCA of our posteriors using all possible combinations of input
parameterizations (for the 4D cases, all pairs of mass parameters and all pairs of 
spin parameters), testing for multivariate normality on the PCA-transformed
posteriors using the Henze-Zirkler test with 
a significance of 0.05~\cite{henze1990class}, available as the function
\texttt{multivariate\_normality} in the \textsc{Pingouin} Python 
package~\cite{Vallat2018}. 

This test sometimes fails with our chosen significance, so
in addition we also compute the Jensen-Shannon~(JS) 
divergence~\cite{61115} between each
one-dimensional marginalized posterior of the PCA data and a zero mean unit
variance normal distribution with an equal number of samples as the posterior,
taking as representative of non-gaussianity the maximum JS divergence across all 1D
marginal posterior distributions. We finally choose a 
parameterization that minimizes this maximal JS~divergence and passes the Henze-Zirkler 
test, if available, otherwise we take the parameterization that simply minimizes
the representative JS~divergence.

Once the PCA is performed, the variance in the posterior distribution 
increases inversely with the square of
the signal SNR. This scaling is robust and we can verify that it holds 
by comparing the PCA posteriors from
injections at two different SNRs. 
The results of such a comparison for
an injection of the 2D~signal \texttt{BAM-5}, first at SNR~250 
and again at SNR~20, shows that after training the PCA transformation
on the SNR~250 posterior data and applying the same transformation to the 
SNR~20 posterior, the rescaled 90\%~CI circle of the 
SNR~250 posterior captures 89.3\% of the posterior SNR~20 samples.

\subsection{SNR Comparisons}
\label{sec:comparisons}

Given the approximate scaling of the PCA 90\%~CI, we can estimate the SNR
at which the true parameters are biased through a simple rescaling. 
Define the norm of the true injection parameters, transformed 
using the same PCA transformations trained 
on the posterior data, to be \(r_\text{inj}\).
Then the SNR at which \(r_\text{inj}\) will fall outside
the 90\%~CI is computed by rescaling the injected SNR
\(\rho_\text{inj}\) by the ratio of the 90\%~CI sphere
radius and the norm of the injection parameters,
\begin{equation}
\rho_{\text{PCA, ND}}=\frac{\sqrt{\chi^2_N\left(0.1\right)}}{r_\text{inj}}\rho_\text{inj},
\end{equation}
where, for the examples we consider, \(\rho_\text{inj}=250\).

In Table~\ref{tab:eff-snrs} we present the results of the PCA rescaling alongside
the computed faithfulness SNRs and bias SNRs for both the 2D~and 4D models.
The columns of \(\rho_\text{faith,ND}\) contain the values arising from
computing the faithfulness SNR in Eq.~\eqref{eq:ind1}. The
SNRs \(\rho_\text{bias, ND}\)
use Eq.~\eqref{eq:ind2}, and \(\rho_\text{PCA, ND}\) are the SNRs computed
by rescaling the injected SNRs of the parameter estimation samples such that
the 90\%~CI of the PCA samples contain \(r_\text{inj}\). 

The faithfulness SNR is
a consistent lower bound for the bias SNRs. This should not be a surprise,
as the faithfulness SNR includes contributions in the faithfullness mismatch
coming from signal components orthogonal to the model 
manifold that do not affect the systematic bias.
For some cases, such as \texttt{BAM-4}, the difference between the faithfulness
and bias SNRs is small. This will happen when the true signal sits close to
the model manifold near the best-fit parameters compared
to the distance between the injection and best-fit parameters, 
as we see when comparing
the values of \(\hat{d}^{\,2}_\text{s}\) and \(\hat{d}^{\,2}_\text{bf}\)
in Table~\ref{tab:simcases}. 
Under the 
linear assumption used in the PCA, we see that the bias SNR is 
consistent in reproducing an estimate for the SNR at which the posterior bulk
will no longer contain the true signal parameters within its 90\%~CI.

\begin{table}
	\renewcommand{\arraystretch}{1.2}
\begin{tabular}{@{} c|c|cc|c|cc @{}}
    \toprule
    \hspace{1em}Simulation ID\hspace{1em} & $\rho_\text{faith, 4D}$ & $\rho_\text{bias, 4D}$ & \(\rho_\text{PCA, 4D}\) & $\rho_\text{faith, 2D}$ &$\rho_{\text{bias, 2D}}$ & \(\rho_\text{PCA,2D}\)  \\
    \hline
\texttt{BAM-1}	&	57	&	60 	&	 60 & 44 & 	48	&	47  	  \\
\texttt{BAM-2}	&	98	&	122 	&	 120 & 76 & 	195	&	182  	  \\
\texttt{BAM-3}	&	67	&	79 	&	 79 & 52 & 	61	&	61  	  \\
\texttt{BAM-4}	&	51	&	54 	&	 56 & 39 & 	43	&	43  	  \\
\texttt{BAM-5}	&	33	&	34 	&	 38 & 25 & 	27	&	27  	  \\\hline
\texttt{SUR-1}	&	47	&	49  	&	 53 & 36&	47	&	47  	 \\
\texttt{SUR-2}	&	221	&	538  	&	 556 & 170&	433	&	428  	 \\
\texttt{SUR-3}	&	134	&	174  	&	 186 & 103&	213	&	214  	 \\
\texttt{SUR-4}	&	40	&	40  	&	 39 & 30&	38	&	38  	 \\
    \botrule
\end{tabular}
\caption{Table presenting 2D and 4D bias SNRs for the injected cases of study listed in 
Table~\ref{tab:simcases}. The faithfulness SNR \(\rho_\text{faith, ND}\) is computed
using the faithfulness mismatch or, equivalently, the \(\hat{d}_\text{s}^{\,2}\)
values from Table~\ref{tab:simcases}, assuming 2 or 4~free degrees of freedom in 
Eq.~\eqref{eq:ind1}. The values for \(\rho_\text{bias, ND}\) 
arise from Eq.~\eqref{eq:ind2} and the values of the bias distances given in
Table~\ref{tab:simcases}, assuming 2~and 4~free degrees of freedom. 
The SNRs \(\rho_\text{PCA, 2D}\) and \(\rho_\text{PCA, 4D}\) 
are computed by rescaling the approximate 90\% PCA posterior volume of the 
recoverd parameter estimation posterior samples for the 2D and 4D injections,
respectively.}
\label{tab:eff-snrs}
\end{table}

\begin{figure*}[t!]
	\centering
	\includegraphics[width=0.85\textwidth]{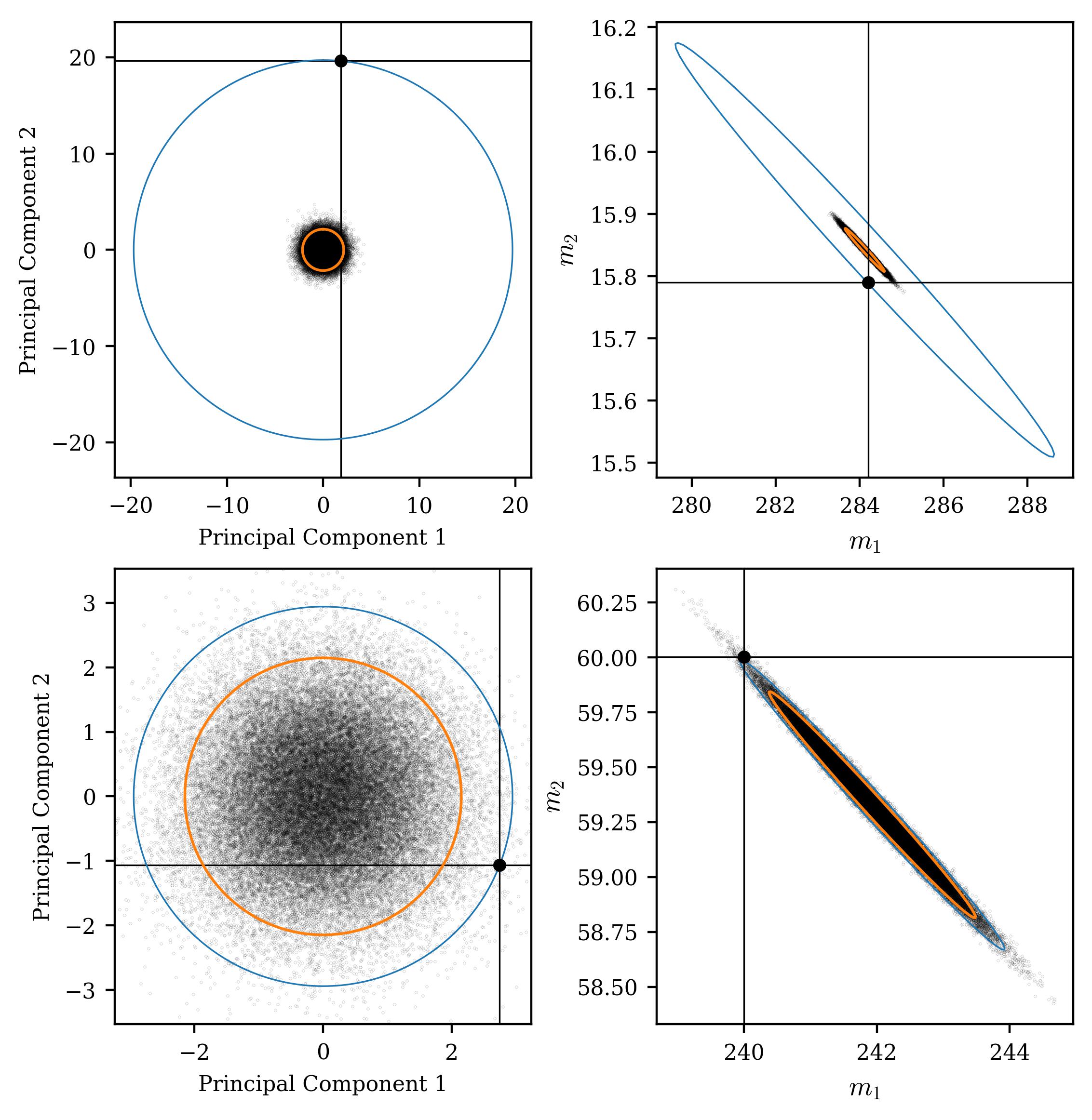}
	\caption{Posterior probability distributions for the recovery of \texttt{BAM-5} (top row) and 
	\texttt{BAM-2} (bottom row) described in Table~\ref{tab:simcases} 
	using the 2D~model restriction of \d. The left column shows the posteriors after applying
	Principal Component Analysis detailed in Sec.~\ref{sec:PCA}, with orange
	circles showing the approximate 2D 90\% credible region for the $\rho = 250$ injection. 
	The blue circles in the left column are generated using the norm of the injected signal 
	parameters (shown as the black dot)	as a radius, i.e., representing the SNR at which the
	true parameters will lie at the edge of the 90\% credible region. The right column shows the 
	samples in the physical \(m_1\)--\(m_2\) parameterization, with the blue and orange circles 
	mapped into correlation ellipses using the inverse Pricipal Component transformation.}
	\label{fig:2d-pca-cases}
\end{figure*}

For the 2D model, we can fully visualize the rescaling 
in Fig.~\ref{fig:2d-pca-cases}, where we show 
the results of performing PCA on the 2D posterior samples for SNR~250 injections 
of \texttt{BAM-5}~(top) and \texttt{BAM-2} (bottom), in the left column of the 
figure. We plot circles with radius \(r_\text{inj}\) as the blue circles. In the right
column we show the samples and rescaled contours mapped back
to the physical parameter space. For the case of \texttt{BAM-5} we see that the
norm of the injected parameters is considerably larger than the PCA 90\%~CI radius,
indicating that biased recovery occurs at much lower SNRs, roughly a factor
of 10 lower than the injected SNR according to the 
estimates in Table~\ref{tab:eff-snrs}. 

The \texttt{BAM-2}
PCA posterior shows that the biased SNR is much closer to the SNR~250 injection value,
estimated to be around SNR~180 from the posterior scaling 
and SNR~195 from Eq.~\eqref{eq:ind2}. We find that an SNR~195 injection
places the injected values on the 2D~90\%~CI boundary, as is shown below
in Fig.~\ref{fig:m1m2_toy_q4a025}. Furthermore for this case, we see that 
the true values of the simulation lie far along the semi-major axis of the 
sample correlation ellipse, meaning that the 1D~projections of the 2D~posterior
onto the \(m_1\) and \(m_2\) axes will still show bias even at the lower SNR 
required for the 2D~posterior to contain the injected values. We explore resolutions
to this below in Sec.~\ref{sec:BiasSNR}.

\begin{figure}
	\centering
	\includegraphics[width=0.48\textwidth]{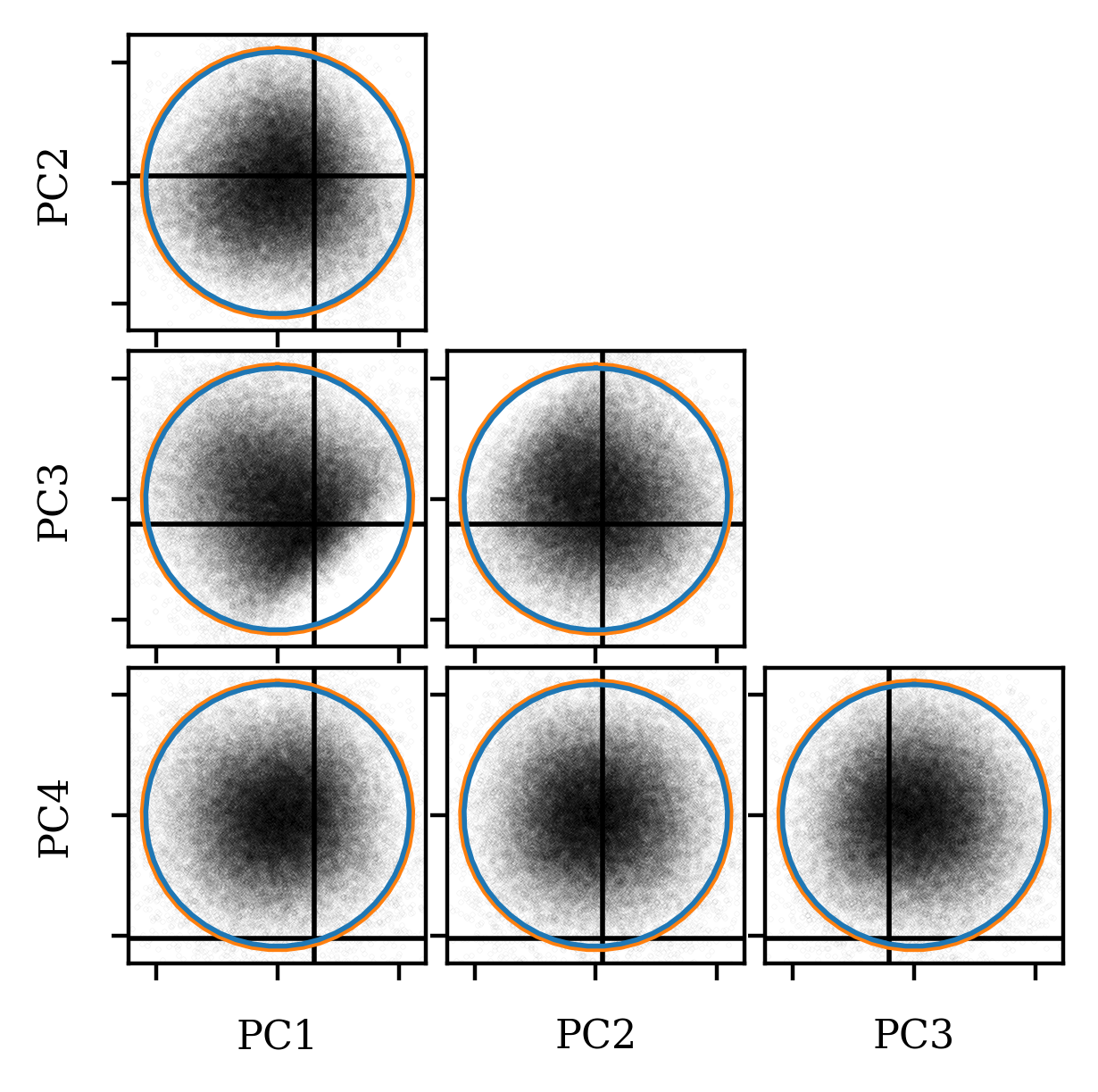}
	\caption{Two-dimensional marginal posterior projections of the four-dimensional
	posterior distribution for \texttt{BAM-3}, detailed in Table~\ref{tab:simcases},
	after Principal Component Analysis is applied to the posterior samples, with \textit{e.g.},
	PC1 denoting the first principal component. The signal
	is injected with signal-to-noise ratio (SNR) of~80, matching the predicted 4D~bias SNR
	predicted using Eq.~\eqref{eq:ind2} in Table~\ref{tab:eff-snrs}. The orange circles
	show the 2D projections of the 4D sphere with radius \([\chi^2_4(0.1)]^{1/2}\) that approximates
	the 90\%~credible region for the posterior and contains 90.7\% of the posterior samples.
	The blue circles show the 2D projections of the 4D sphere with a radius determined by the
	norm of the true signal parameters.}
	\label{fig:4d-pca-q4aM07f}
\end{figure}

Finally we compare the predictions of the PCA rescaled SNR and $N$-D bias
SNR estimates to parameter estimation results using the full 4D model.
Shown in Fig.~\ref{fig:4d-pca-q4aM07f} are
the marginalized two-dimensional projections of the full 4D~PCA posterior for
\texttt{BAM-3}, injected at the approximate
4D~bias SNR~80. At this SNR our linearized PCA approximation 
still holds (despite the noticable railing visible
in the PC1-PC3 plane), and the 4D~sphere of radius \([\chi_4^2(0.1)]^{1/2}\)
contains 90.7\% of the samples in the posterior. We also see that the norm of the 
true signal parameters in the PCA~projection matches very closely to this 
radius value. When looking across all cases of interest, we find
that the 4D sphere estimate works well at containing approximately 90\%
of the posterior samples for all cases when injected at the 4D~bias SNR values listed
in Table~\ref{tab:eff-snrs}, and this radius matches the norm of the injected values
to a relative error within 8\%~for the majority of cases, with the notable outlier
being \texttt{SUR-3} with a relative error of 15\%.

%%%%%%%%%%%%%%%%%%%%%%%%%%%%%%%%%%%%%%%%%%%%%
\section{Results: Parameter Bias SNRs} 
\label{sec:BiasSNR}

In this section we compute the one-dimensional bias SNRs for all cases of interest and present
these results in Tables~\ref{tab:2d-bias-snrs} and \ref{tab:4d-bias-snrs} 
found in Appendix~\ref{sec:1d-bias-snr-results}, 
both for the 2D~model and the full 4D~injections respectively, and comment on the results
below in Sec.~\ref{sec:eff-bias-snrs}. We also compare the results of the bias
SNR computation to Fisher analysis results in Sec.~\ref{sec:fisher-comparison}
and discuss the impacts of prior railing in Sec.~\ref{sec:prior-bounds}.

\subsection{Parameter Bias SNRs for Cases of Interest}
\label{sec:eff-bias-snrs}

The bias SNRs calculated in Sec.~\ref{sec:EffSNR} estimate the SNR required
for the $N$-dimensional injection parameter vector to lie outside of the 90\%~CI of the $N$-dimensional
posterior probability density. As discussed in Sec.~\ref{sec:SNRissues}, it
does not tell us whether any given parameter of interest is biased. We see this
fact demonstrated in Fig.~\ref{fig:m1m2_toy_q4aM075}, which displays 
the two-dimensional posterior distribution of \(m_1\) and \(m_2\) for \texttt{BAM-3}
along with the 90\%~CIs for the marginal one-dimensional posteriors of each mass
separately as arrows along the plot edge. The SNRs at which the 2D~90\%~CI just contains
the injected parameters is comparable to the value of \(\rho_\text{bias, 2D}\)
presented in Table~\ref{tab:eff-snrs}, but each individual mass parameter 
becomes biased at SNRs much greater than this value (though this may not always
be the case, as we discuss below).
To investigate the bias SNR for \(m_1\) in this example we instead 
compute the parameters \(\thetaod{m_1}\) and use them in Eq.~\eqref{eq:ind2}
with \(N=1\) in place of the full signal parameters \(\theta_\text{s}\). In this way, we 
compute the distance between the true value of \(m_1\) and the effectual
value of \(m_1\) along the one-dimensional submanifold described by choosing,
at each value of \(m_1\), the remaining signal parameters 
from \(\Theta_\text{opt}=\Theta\setminus\{m_1\}\) utilizing 
Eq.~\eqref{eq:optimal-params}. 

\begin{figure*}
	\centering
	\includegraphics[width=0.98\textwidth]{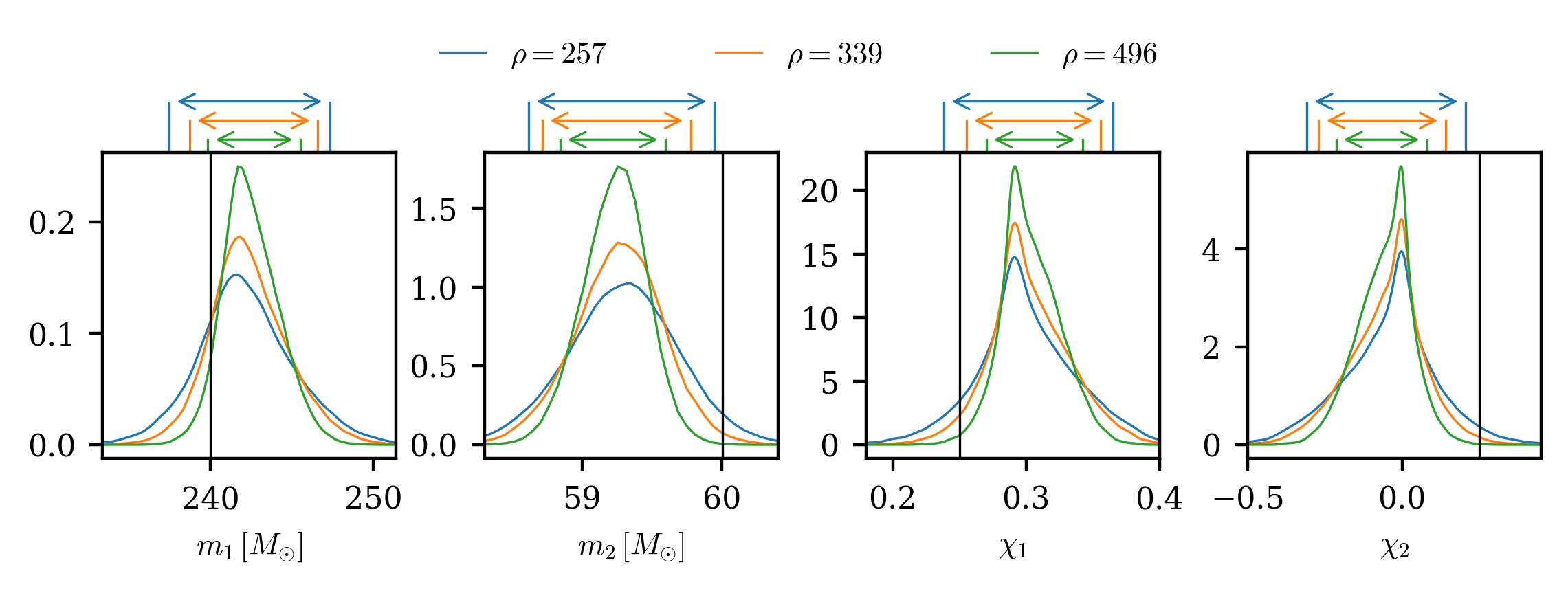}
	\includegraphics[width=0.98\textwidth]{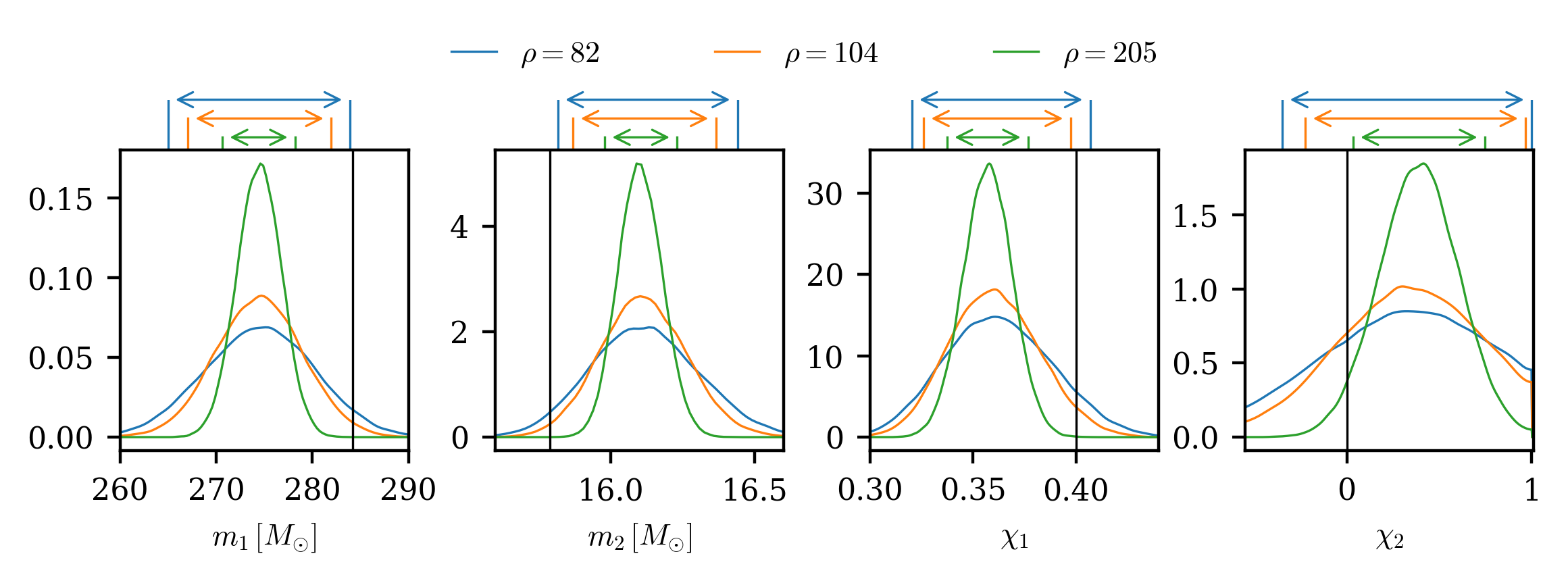}
	\caption{Measurement of the primary mass $m_{1}$, secondary mass $m_{2}$, primary spin $\chi_{1}$ and secondary spin $\chi_{2}$ for different signal-to-noise ratios $\rho$. 
	The top row shows our analysis of \texttt{BAM-2}. For this simulation, the 1D effectual SNRs for $[m_{1}, m_{2}, \chi_{1}, \chi_{2}]$ are $\rho = [496, 257, 339, 256]$ respectively. The bottom row shows our analysis of \texttt{BAM-5}. For this simulation the 1D effectual SNRs for $[m_{1}, m_{2}, \chi_{1}, \chi_{2}]$ are $\rho = [85, 82, 104, 205]$ respectively. In all cases, the black vertical line indicates the true value, the horizontal
	arrows and verticle bars display the 90\%~CIs and we sample over $m_1$, $m_2$, $\chi_1$ and $\chi_2$.}
	\label{fig:1d-posteriors}
\end{figure*}

The results for computing the one-dimensional bias SNRs for the 2D~model 
cases are presented in Table~\ref{tab:2d-bias-snrs}, and the 4D~results
are presented in Table~\ref{tab:4d-bias-snrs}. The general trend of these SNRs is that they are higher 
than the $N$-D bias SNR for each model (i.e., 2D or 4D), meaning that for many signals
of interest calculating either the faithfulness SNR or the $N$-D bias SNR will provide 
lower-bounds to the parameter bias SNRs, but these lower bounds may sometimes
be orders of magnitude too conservative. 
We plot a selection of one-dimensional marginalized posteriors for the individual
component masses and spins of the cases \texttt{BAM-2} and \texttt{BAM-5}
in Fig.~\ref{fig:1d-posteriors}, injected at varying SNRs estimated by the 
predicted parameter bias SNRs. 
The injected values of these parameters are shown as vertical
black lines and the 90\%~CI boundaries for the different SNR injections are 
shown above each figure panel. The SNRs predicted from Eq.~\eqref{eq:ind2}
provide a robust estimate for the SNR at which these individual parameters become
biased. We show the full 1D~comparison results in Figs.~\ref{fig:nr-1d-comparison} and
\ref{fig:sur-1d-comparison} in Appendix~\ref{sec:1d-bias-snr-results}.

\begin{figure}
	\centering
	\includegraphics[width=0.48\textwidth]{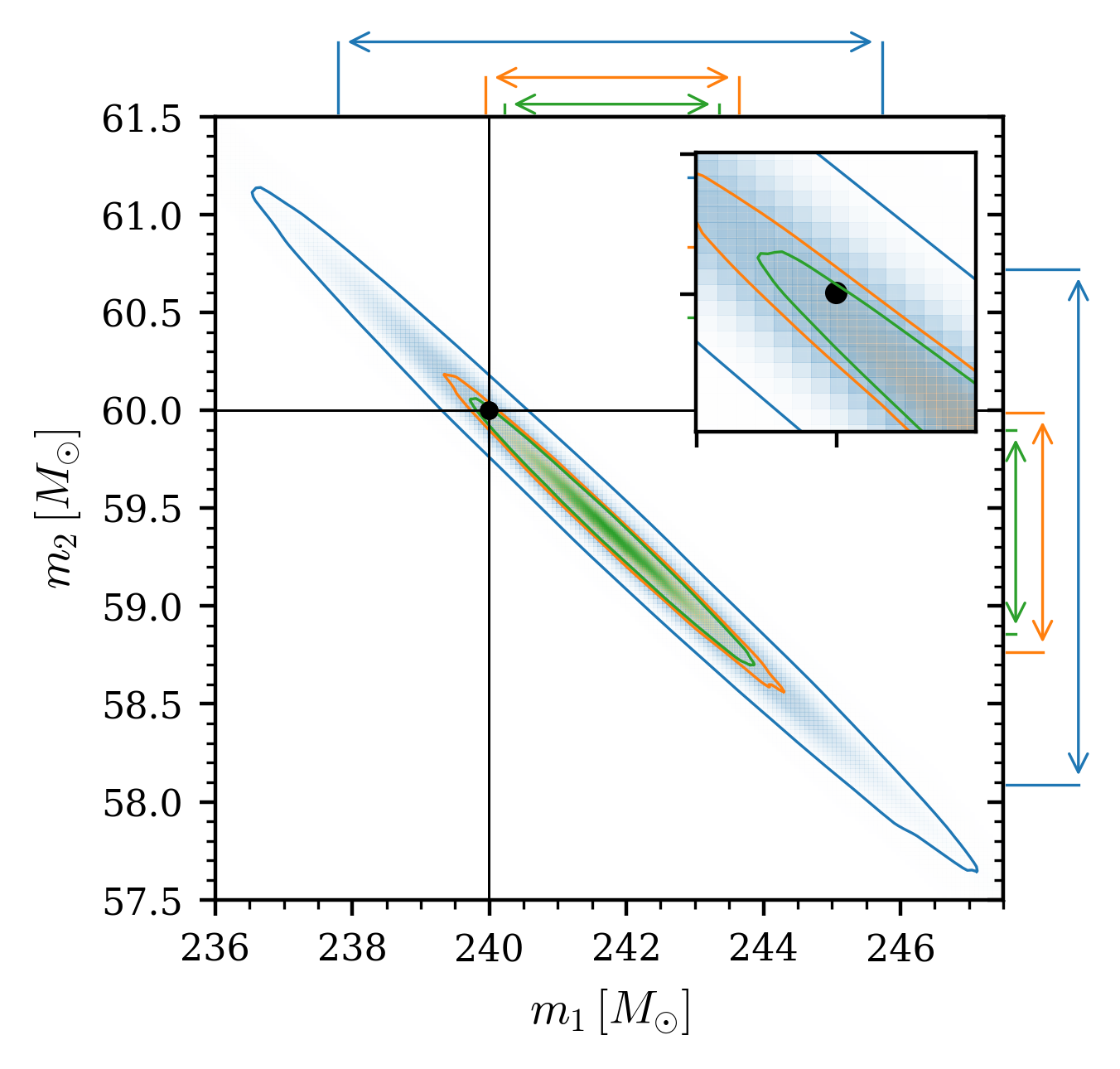}
	\caption{Measurement of the primary mass, $m_1$, and secondary mass $m_2$ for different signal-to-noise ratios $\rho$. Here we show our analysis of \texttt{BAM-2} when only sampling over $m_1$ and $m_2$ (other parameters are held fixed to their true values). The black dot indicates the true parameters, the contours
	show the 90\% credible intervals and the horizontal/vertical lines above and to the right of the Figure show the 90\% symmetric credible intervals for the 1D marginalized posteriors. The inset shows a zoomed in portion of the posterior, focusing on the correlation between the true parameters and the credible interval at which they are biased. For this simulation, the faithfulness indistinguishability SNR is 76, the 2D~bias SNR is 195, and both the primary and secondary mass are estimated to be biased at $\rho\approx 160$. We see that the effectual SNR correctly identifies the SNR at which the 2D posterior is biased (at the 90\% credible interval) and the 1D marginalized posteriors for the primary and secondary masses remain unbiased until $\rho\approx 160$.}
	\label{fig:m1m2_toy_q4a025}
\end{figure}

The variation between the parameter bias SNRs for different 
parameters has no obvious correlation
and parameter bias SNRs for a given parameter can 
vary significantly across parameter space.
We also note the occurrence of a parameter bias SNR being \textit{smaller} than the
$N$-D bias SNR, which happens for individual parameters in a few cases but is most 
prominent in the 2D~example for \texttt{BAM-3}, where 
\(\rho_\text{bias, 2D}=195\) but all of the parameter
bias SNRs for the mass parameters listed in Table~\ref{tab:2d-bias-snrs} are between
150--165. This case is shown in Fig.~\ref{fig:m1m2_toy_q4a025}
along with the posterior results for parameter estimation runs performed at SNRs
of 76~(the faithfulness SNR estimate), 160~and~195. We verify that indeed the true
parameters lie within the 90\%~CI for the two-dimensional posterior at an SNR of~195
and fall near the boundary of the one-dimensional marginalized posterior 90\%~CIs
for both individual masses at SNR~160.
In this case, then, should one wish to produce a truly conservative estimate of the
bias SNR, one should compute \(\rho_\text{bias, 1D}\) regardless of the numbers of
model degrees of freedom being measured. For the 2D~\texttt{BAM-3} case, we would
then arrive at a conservative estimate of the bias SNR to be \(\rho_\text{bias, 1D}=195\times\sqrt{4.6/2.7}=149\), which indeed is a lower bound on the parameter bias
SNRs computed for this model. This holds true for all cases examined, though again
for many of the cases listed in Tables~\ref{tab:2d-bias-snrs} and~\ref{tab:4d-bias-snrs}
these 1D~estimates are overly conservative.

\begin{figure*}
	\centering
	\includegraphics[width=\textwidth]{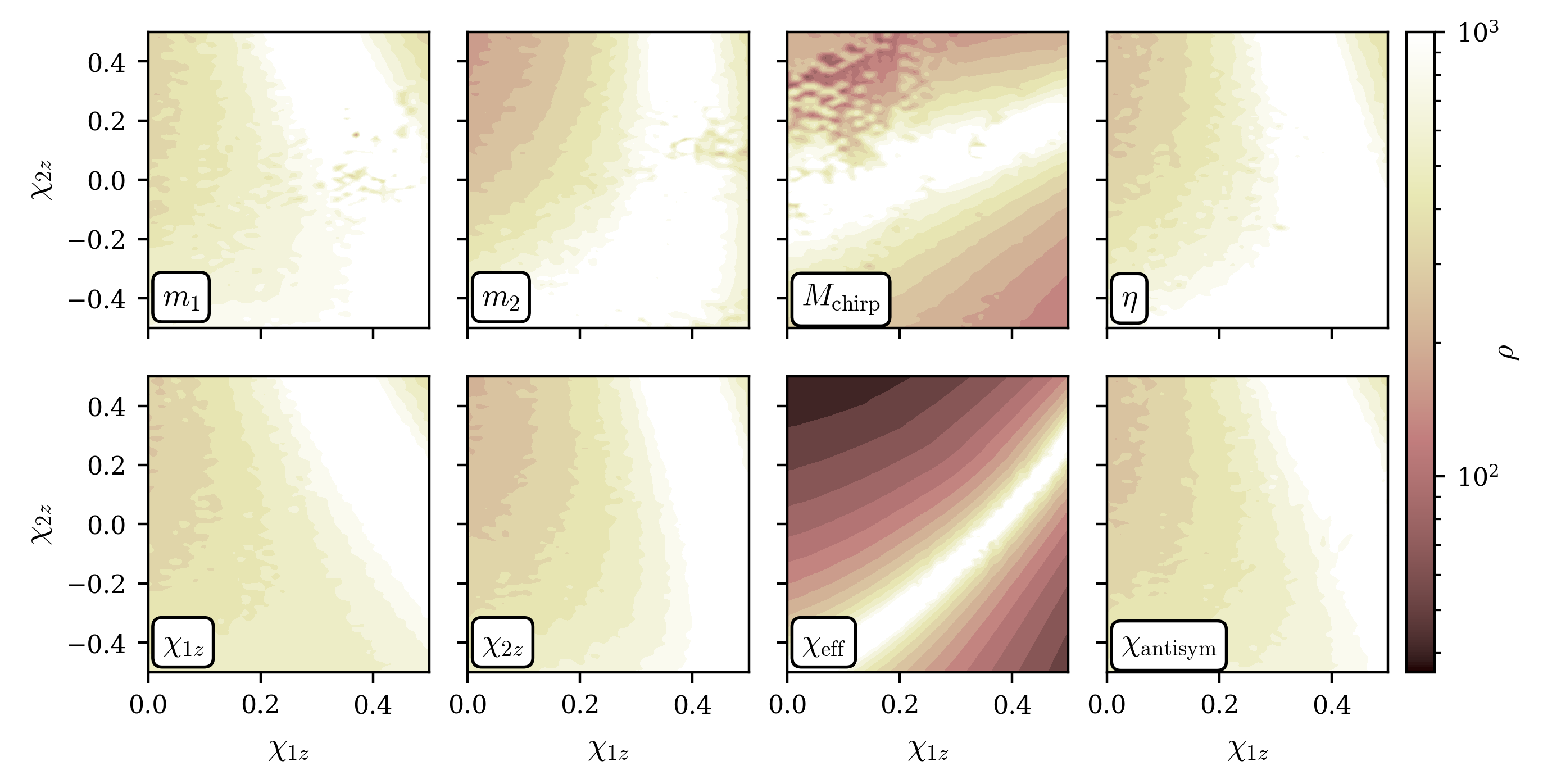}
	\caption{
	Contour plots of parameter bias signal-to-noise ratio (SNR) computed
	between the models \d{} and \textsc{NRHybSur3dq8}, plotted for a range of 
	\(\chi_{1z}\) and \(\chi_{2z}\) for fixed values of \((m_1,m_2)=(200,100)\)~\(M_\odot\).
	The top row of panels displays, from left to right, the parameter bias SNR contours 
	for \([m_1,m_2,M_\text{chirp},\eta]\), and the bottom row displays the parameter bias 
	SNR contours for the parameters \([\chi_{1z},\chi_{2z},\chi_\text{eff},
	\chi_\text{antisym}]\) from left to right. The cloudy
	structures visible in the contour plots arise both from the interpolation used
	to construct the contours and the fluctuations in the lower minimization tolerances
	used to compute the optimal parameters from Eq.~\eqref{eq:optimal-params}. We further
	cap the color range to a maximum value of \(10^3\) as SNR predictions above this
	value are not reliable from the numerical thresholds used in this work.
	}
	\label{fig:pd_sur_bias_contours}
\end{figure*}

To further investigate the variation of parameter bias SNRs of a given model 
as we move across parameter space, we 
compute the parameter bias SNRs between \d{} and \textsc{NRHybSur3dq8} for fixed values of
\((m_1,m_2)=(200,100)\)~\(M_\odot\) and ranging over equivalent spin values
shown in Fig.~\ref{fig:pd_sur_snr_contours}. We display results for six parameters
in Fig.~\ref{fig:pd_sur_bias_contours}. The top row of panels 
shows the variation of the parameter bias
SNR for \([m_1,m_2,M_\text{chirp},\eta]\)
from left to right. The bottom row of panels 
displays the bias SNR for \([\chi_{1z},\chi_{2z},\chi_\text{eff},\chi_\text{antisym}]\)
from left to right. The structures of the parameter
bias SNR contours show some similarity, especially between the component masses and
spins, \(\eta\) and \(\chi_\text{antisym}\). One notes that all parameters have relatively
high parameter bias SNRs across most of the spin parameter space 
except for \(\chi_\text{eff}\).
The structure of the parameter bias SNR for \(\chi_\text{eff}\) mimics closely the structure
of \(\hat{d}_\text{s}\) plotted in Fig.~\ref{fig:pd_sur_snr_contours}. This behavior
is also visible in the tabulated bias SNR data in Table~\ref{tab:4d-bias-snrs}
when comparing \(\rho_\text{bias, 4D}\) to the parameter bias SNR 
for \(\chi_\text{eff}\) in the 
four \texttt{SUR} cases, hinting that the \(\chi_\text{eff}\) modelling 
bias between \d{} and \textsc{NRHybSur3dq8}
is the driving systematic cause of difference between the two waveform models.

%%%%%%%%%%%%%%%%%%%%%%%%%%%%%%%%%%%%%%%%%%%%%

\subsection{Comparisons to Fisher Biases}
\label{sec:fisher-comparison}
One approach to computing bias estimates common in the GW literature is that
of Fisher analysis, which we summarize in Appendix~\ref{sec:Fisher}. The main
results of interest for this study are the estimate to the shift in measured
parameters due to systematic errors, \(\Delta\theta_\text{sys}\) in 
Eq.~\eqref{eq:sys-error}, and the definition of the Fisher matrix \(\Gamma\)
given in Eq.~\eqref{eq:fisher}. From these two quantities we can estimate
an SNR at which the Fisher bias in the parameter \(\theta^i\) 
will become larger than the Fisher estimate of the 90\%~CI by
\begin{equation}
	\rho^i_\text{Fisher}=1.645\, \rho_\text{inj}\frac{\sqrt{\Gamma_{ii}}}{\Delta\theta^i_\text{sys}}.
	\label{eq:fisher-snr}
\end{equation}
Here \(\sqrt{\Gamma_{ii}}\) (\(i\) not summed over) is used 
as the approximate standard deviation in the 
measurement of parameter \(\theta^i\)~\cite{Finn:1992wt}, the numerical factor rescales the significance 
to represent the 90\%~CI, and \(\rho_\text{inj}\) is the SNR of the injected signal
\(\signal\) used in Eq.~\eqref{eq:sys-error}. For this paper we consistently use \(\rho_\text{inj}=250\).

We present the results of \(\rho^i_\text{Fisher}\) for the cases of interest in
Table~\ref{tab:fisher-snrs} found in Appendix~\ref{sec:1d-bias-snr-results}. When
compared to the bias SNRs in Table~\ref{tab:4d-bias-snrs}, we see that the two 
methods broadly agree to within 10\% for most cases except for \texttt{BAM-3}
and \texttt{BAM-4}, which are the two cases impacted by bounded priors and discussed
in Sec.~\ref{sec:prior-bounds}, or where the bias SNR is particularly large, where
results from both analyses may become unreliable due to either 
limited numerical precision in the minimization for the bias distance calculation
or accuracy of numerical derivatives and conditioning of the Fisher matrix in
the case of the Fisher estimates. We conclude from this comparison that 
both approaches are equivalent at estimating the bias SNR for the cases we have 
considered when the Fisher analysis is done correctly (see discussions in 
Appendix~\ref{sec:Fisher} about alignment and parameter choices), 
and one should use whichever method is most convenient to calculate
when investigating for potential systematic biases.

\subsection{Impact of Bounded Priors}
\label{sec:prior-bounds}

One of the assumptions made for this work in Sec.~\ref{sec:SNRissues} is that one
may overlook the impact of priors on the posterior probability scaling when estimating
the bias SNR. We have seen that this assumption is 
upheld when comparing our bias SNR values to the posteriors resulting from
parameter estimation, even at moderately-low SNRs around 40, but have also found
two 4D~cases for which this is not true: \texttt{BAM-3}
and, to a lesser extent, \texttt{BAM-4}. These two cases are denoted with 
asterisks in Table~\ref{tab:4d-bias-snrs} and in Fig.~\ref{fig:nr-1d-comparison}.

In both of these cases the best-fitting values of the secondary spin \(\chi_{2z}\) 
for the model \d{} lie close to the physically-imposed \(\chi_{2z}=-1\) boundary,
and the railing of the posterior against this boundary produces large shifts
in the recovery of the other parameters when considering each parameter's 
one-dimensional marginalized posterior. 
The presumption that the posterior is a multivariate normal distribution
no longer holds, and instead the full posterior is a \textit{truncated} multivariate 
normal distribution~\cite{birnbaum1951effect,tallis1961moment}, in this case 
truncated in one dimension, and the severity
of the truncation will impact the recovered means and covariances of all one-dimensional
marginalized parameter posteriors. We leave to future work further handling of
bounded priors on bias SNR estimation, but make a few remarks. 

First, the parameter bias SNRs computed for these cases tend 
to \textit{overestimate} the SNR at which
parameters are correctly recovered, leading to overconfidence in model performance.
This is especially true for \texttt{BAM-3}, where the parameter estimation shows
biases at SNRs significantly lower than the estimated values from both Eq.~\eqref{eq:ind2}
and the Fisher analysis. In this instance, the best-fitting value for \(\chi_{2z}\) 
lies at the physical lower bound and the Fisher bias value for \(\Delta\chi_{2z}\)
is near the unphysical spin value of \(\chi_{2z}\approx-1.2\). Both of these facts
provide clear indicators that the prior bound is impacting our parameter bias SNR
estimates and should be watched for when applying these methods.

The second remark is that, again for these two cases, the 4D~bias SNR
estimates are \textit{not} greatly impacted by the railing posterior against
the \(\chi_{2z}\) boundary, as seen in Table~\ref{tab:eff-snrs} 
and shown in Fig.~\ref{fig:4d-pca-q4aM07f}. 
While the boundary is still clearly visible in the PCA of the
posterior samples, its impact on the total posterior scaling is seemingly minor.
Assessing how robust of an observation this is we leave to future work.

%%%%%%%%%%%%%%%%%%%%%%%%%%%%%%%%%%%%%%%%%%%%%
\section{Accuracy requirements for future detectors}
\label{sec:acc-req}

We expect to observe signals with SNRs of $O(1000)$ with next-generation detectors Einstein Telescope and Cosmic Explorer. 
This is two orders of magnitude above the $O(10)$ SNRs of LVK observations to date. How do our results translate into waveform
accuracy requirements, both for models, and for NR simulations and inspiral approximations? 
(For other studies on NR and waveform model accuracy needs for next-generation detectors, see 
Refs.~\cite{Purrer:2019jcp,Ferguson:2020xnm,Jan:2023raq}.)
The examples
in this paper are specific to a quadrupole-only aligned-spin model; we plan to extend and apply our methods
to state-of-the-art generic-binary models in future work. But a number of aspects of our study on 
indistinguishability SNRs -- from simple conservative estimates using the faithfulness through to individual-parameter bias SNRs -- allow us to 
make some general statements about accuracy requirements over the next 10-15 years. 

Our focus has been on models tuned to NR simulations, so let us first put this discussion in the context of NR simulation uncertainties 
and computational cost.
Phase errors in NR waveforms are dominated by numerical resolution; with a 4th-order accurate scheme 
(in both time and space discretisation), a factor of two improvement in resolution leads to errors reduced by a factor of 16.
The higher-resolution simulation requires eight times the memory (in a 3D code), and also double the number of time integration
steps, so the computational cost also increases by a factor of 16. This tells us that the computational cost scales roughly linearly
with the accuracy. If we require an order of magnitude improvement in accuracy, we need an order of magnitude increase in 
computational resources. For higher-order or pseudospectral codes, the scaling may be better, with a slower increase in 
computational costs, but assuming a linear scaling between accuracy and computational cost allows us to make an 
approximate translation of mismatch requirements to computational resource needs. 

In the following, therefore, we recall that the normalised difference between two waveforms relates to the mismatch as
$\hat{d} = \sqrt{\mathcal{M}}$, and errors in NR simulations (e.g., the waveform phase and amplitude) scale as $\hat{d}$. 
For the purposes of this discussion we will therefore assume that computational cost scales linearly with $1/\hat{d}$. 
It is straightforward to adjust our estimates for different computational cost scalings. 

Equation~(\ref{eq:ind1}) provides the most conservative mismatch accuracy requirement if we use $N=1$ for individual parameter measurements. 
To guarantee no parameter biases due to model inaccuracies for $\rho > 1000$, this criterion requires $\mathcal{M} \lesssim 10^{-6}$. 
Current BBH NR waveforms and waveform models quote mismatch 
uncertainties of $10^{-4}$--$10^{-2}$, for example see Refs.~\cite{Hamilton:2023qkv,Varma:2019csw,Thompson:2023ase,Ramos-Buades:2023ehm}.
This suggests a necessary improvement of between two and four orders of magnitude in mismatch uncertainty, or 1-2 orders of magnitude 
improvement in simulation accuracy and computational cost. 

As we have seen, the true bias SNRs are typically 5-10 times larger than those predicted by the most conservative estimate, due 
mostly to parameter correlations over the high-dimensional binary parameter space. The scaling will depend on both the parameter of interest
and the point in parameter space, but could be determined by studying the parameter correlations of the model, largely independently
of any accuracy analysis. 
However, we have also seen that in some cases the model error is along the principal direction of signal 
variation, and the true bias SNR can be comparable to conservative estimates, 
such as happens in the \texttt{BAM-2} case in Fig.~\ref{fig:2d-pca-cases}
and Tab.~\ref{tab:4d-bias-snrs}. 
In general, for any given model, we cannot know \textit{a priori} the distribution of bias SNRs across the parameter space; to properly 
determine the limits of a model, we must calculate the true bias SNR over a sufficiently dense sampling of the binary parameter space.
\emph{If} a model's biases are always approximately orthogonal to the principal parameter directions of the signal space, then the mismatch accuracy 
requirements will be two orders of magnitude less strict for some parameters, 
thereby only requiring 1-2 orders of magnitude improvement
in mismatch accuracy, and a factor of 3-10 increase in computational cost. 
This suggests that an important diagnostic in the 
construction of future waveform models will be the direction of parameter biases; 
it remains to be seen whether it is possible
to optimise a model's construction to ensure that parameter biases are always 
approximately orthogonal to the principal parameter directions, though 
techniques introduced to mitigate waveform modeling errors and applied
to extreme mass-ratio inspiral signal models
may well be suited to this task~\cite{Chua:2019wgs,Liu:2023oxw}.
We note that the analysis of the parameter-space variations of the parameter
bias SNRs in \d{} was possible only because we have access to a 
much more accurate model from which to construct proxy true signals, 
\textsc{NRHybSur3dq8}; cutting-edge model development will
not have that luxury.

For fiducial ``true'' signals, the
only accuracy measure available to us is the faithfulness SNR, 
and so for these signals we cannot escape the requirement of 
mismatch uncertainties of $\sim$$10^{-6}$. 
(Note, however, that NR accuracies at this level are already achievable in principle, 
as seen in the tail of the mismatch distribution in Fig. 4 of Ref.~\cite{Varma:2019csw}.) 
How smoothly the bias SNRs vary across the 
parameter space, and therefore the density of much more 
accurate NR waveforms required to fully assess a model's 
accuracy, will also depend on the details of the model. 
Another goal of modelling procedures should be to achieve bias
SNRs that vary as little as possible, and as slowly as possible, across the binary parameter space. 

Our overall conclusion would then be that current NR and 
model mismatches need to improve by up to four orders of 
magnitude for next-generation detectors, requiring roughly 
two orders of magnitude increase in computational cost. 
However, further improvements in modelling techniques, 
and a more complete understanding of the parameter correlations
for generic binaries over the full binary parameter space, 
may soften these requirements, and only modest improvements
may be necessary over the most accurate current NR simulations 
and waveform models.
We should make clear that improved accuracy is not the only factor that affects computational cost. 
We likely require much longer NR simulations (\textit{i.e.}, including many 
more inspiral orbits) than at present, and a more dense sampling of binary 
parameter space, and an extension to more extreme parts of 
parameter space (higher mass ratios, routine simulations
of near-extreme-spin black holes, and eccentric orbits). 
See, for example, Sec.~4.1.5 of Ref.~\cite{LISAConsortiumWaveformWorkingGroup:2023arg}
for a discussion of the scaling of NR computational costs.

\section{Conclusions} 
\label{sec:conclusions}

We have discussed a common estimate of the indistinguishability SNR of BBH waveforms and waveform models, based on the 
mismatch of a signal against a model evaluated at the signal source's parameters, or the mismatch uncertainty of a waveform. 
We also stress that the square root of the mismatch, $\hat{d} = \sqrt{\mathcal{M}}$, which is the normalised distance between two
waveforms, is a more intuitive measure of waveform differences.
The standard indistinguishability SNR estimate is known to be conservative, sometimes by as much as an order of magnitude. This
is because (a) measurement biases relate instead to the difference between the model at its true parameters $\thetasig$ and the model at 
the best-fit parameters $\thetabf$ that give the best agreement between the model and signal; see Fig.~\ref{fig:manifolds}, and
(b) the correct indistinguishability SNR calculated from the distance $\hat{d}_{\rm bf}$ is the SNR at which the true parameters lie
outside an $N$-D confidence surface, where $N$ is the number of fixed parameters in the mismatch calculation; it cannot be 
used to estimate the indistinguishability SNR for single parameters, except as a conservative lower bound, calculated using 
one degree of freedom in $\chi^2$ in Eq.~(\ref{eq:ind2}). The correct indistinguishability SNR for each parameter, which we call
the \emph{parameter bias SNR}, is calculated by optimising all other parameters in the mismatch calculation (keeping the parameter
we are interested in fixed), and using the distance between the model at those parameters and the true parameters to 
calculate the indistinguishability SNR in Eq.~(\ref{eq:ind2}) with one degree of freedom. 

We have illustrated that this approach provides accurate estimates of both the $N$-D and 1D parameter bias SNRs. For the $N$-D
case we used a PCA analysis to demonstrate that the $N$-D 90\% CI in a parameter-estimation analysis agrees well with that predicted
from the $N$-D bias SNR. In the case of parameter bias SNRs, we performed an extensive set of parameter estimation analyses to 
confirm that the parameter-bias SNRs calculated from the appropriate normalised distance (mismatch) correctly predicted the SNR
at which the true value of each parameter would lie on the 90\% CI in a measurement. We also compared with estimates from 
Fisher methods, and found that both methods were in good agreement for the cases we considered, with the caveat that both 
methods will fail if the best-fit parameters rail against a parameter boundary. 

Previous works have typically used $\hat{d}_{\rm s}$ (in our notation from Fig.~\ref{fig:manifolds}), and chosen the number of 
degrees of freedom in Eq.~(\ref{eq:ind1}) in either an ad-hoc manner, or based on the number of intrinsic parameters in the 
system~\cite{Baird:2013dbm,Chatziioannou:2017tdw,Purrer:2019jcp,Hannam:2021pit,Scheel:2025jct}. As we have illustrated, this \emph{does not} in general predict the correct parameter bias SNR, and, although the
answer is often lower than the true parameter bias SNRs, it is not necessarily so; unless one identifies the principal parameter
directions for the given point in parameter space, the relationship between common bias SNR estimates, e.g., $N/(2\rho^2)$,
and the true parameter SNR biases is unknown.

In this work we restricted examples to the simple test case of the (2,2)-mode from aligned-spin binaries. In future work we aim to
extend these results to state-of-the-art generic models, to provide robust statements on the reliability of these models across
the binary parameter space. For now we can nonetheless make broad statements about the required model 
accuracy, and levels of accuracy improvements, for future GW observatories. We estimate that model accuracy must improve
by up to two orders of magnitude for next-generation detectors, but, depending on the details of model construction, only 
modest improvements may be sufficient. 

We caution, however, that the methods we have discussed here, and the statements we have made about future accuracy needs,
apply only to situations where we can calculate a sufficiently accurate ``true'' signal against which to evaluate models. This is
currently limited to the last orbits and merger of binary black hole systems. We do not have a means to calculate long-duration fully
general-relativistic inspiral, and for systems with matter (binary neutron stars or black-hole--neutron-star binaries) we have
neither a full understanding of all physical processes involved, nor as yet sufficiently accurate numerical-relativity codes to 
calculate true waveforms. Quantifying the necessary level of modelling accuracy and physical completeness for future
science goals is an important open question in source modelling, and requires a great deal of further work over the next decade.

%%%%%%%%%%%%%%%%%%%%%%%%%%%%%%%%%%%%%%%%%%%%%
\section{Acknowledgements}

The authors would like to thank Alvin Chua, Stephen Fairhurst and Frank Ohme for enlightening discussions
on waveform systematics and bias estimation. We thank Jannik Mielke for comments
during the LIGO-Virgo-KAGRA internal review.

J.T. acknowledges support from the NASA LISA Preparatory Science grant 20-LPS20-0005. C.H. thanks the UKRI Future Leaders Fellowship for support through the grant MR/T01881X/1. E.F-J and M.H. were supported in part by Science and Technology Facilities Council (STFC) grant ST/V00154X/1.

This research used the supercomputing facilities at Cardiff University operated by Advanced Research Computing at Cardiff 
(ARCCA) on behalf of the Cardiff Supercomputing Facility and the HPC Wales and Supercomputing Wales (SCW) projects. 
We acknowledge the support of the latter, which is part-funded by the European Regional Development Fund (ERDF) via the 
Welsh Government. In part the computational 
resources at Cardiff University were also supported by STFC grant ST/I006285/1. We are also
grateful for the Sciama High Performance Compute (HPC) cluster, which is supported by the ICG, SEPNet and the University of Portsmouth. The authors are grateful for computational resources provided by the LIGO Laboratory and supported by National Science Foundation Grants PHY-0757058 and PHY-0823459. The authors acknowledge the use of the IRIDIS High Performance Computing Facility, and associated support services at the University of Southampton, in the completion of this work.

Various plots and analyses in this paper were made using Python software packages \texttt{LALSuite}~\cite{lalsuite}, \texttt{PyCBC}~\cite{alex_nitz_2024_10473621}, PESummary~\cite{Hoy:2020vys}, \texttt{Matplotlib}~\cite{Hunter:2007}, \texttt{Numpy}~\cite{harris2020array}, and \texttt{Scipy}~\cite{2020SciPy-NMeth}.

%%%%%%%%%%%%%%%%%%%%%%%%%%%%%%%%%%%%%%%%%%%%%
\appendix

\section{Fisher Uncertainty and Bias}
\label{sec:Fisher}

We briefly review the formalism for approximating modeling bias in the context
of GW presented in Refs.~\cite{Flanagan:1997kp,Cutler:2007mi}.
When considering the errors introduced in GW parameter inference, it is convenient
to expand the GW signal about the parameters \(\thetabf\) that maximize 
the likelihood in Eq.~\eqref{eq:likelihood},
\begin{equation}
	\label{eq:max-like-condition}
\Braket{\partial_i \model(\thetabf) | \data - \model(\thetabf)} = 0,
\end{equation}
where \(\partial_i \model(\theta) \equiv \partial \model(\theta) / \partial \theta^i\). Defining
\(\Delta\theta^i\equiv(\thetabf-\theta)^i\), the theoretical signal 
model \(\model\) may be expanded about \(\thetabf\) as
\begin{equation}
\model(\theta) = \model(\thetabf)+\partial_i\model(\thetabf)\Delta\theta^i
+\frac12\partial_i\partial_j\model(\thetabf)\Delta\theta^i\Delta\theta^j+\cdots.
\end{equation}
If we assert that the difference between the true and best-fit parameters
is small, we are enforcing the \textit{linear signal approximation} by truncating
the expansion at \(\mathcal{O}(\Delta\theta^2)\),
\begin{equation}
\model(\thetasig) \approx \model(\thetabf)+
\partial_i\model(\thetabf)\Delta\theta^i+\mathcal{O}(\Delta\theta^2).
\label{eq:linearsigappx}
\end{equation}
It follows then directly from the above approximation that the difference between the data and
the model evaluated at the maximum likelihood parameters leads to 
two distinct biases arising from statistical and systematic errors in \(\theta\)
as, respectively,
\begin{align}
\label{eq:stat-error}\Delta \theta^i_\text{stat}&=\left(\Gamma^{-1}(\thetabf)\right)^{ij}\Braket{\partial_j \model(\thetabf) | \noise},\\
\label{eq:sys-error}\Delta \theta^i_\text{sys}&=\left(\Gamma^{-1}(\thetabf)\right)^{ij}\Braket{\partial_j \model(\thetabf) | \signal - \model(\thetasig)},
\end{align}
where \(\Gamma\) is the \textit{Fisher information matrix}
\begin{equation}
\label{eq:fisher}
\Gamma_{ij}(\theta)=\Braket{\partial_i \model(\theta) | \partial_j \model(\theta)}.
\end{equation}

When working with loud signals,
the error from Eq.~\eqref{eq:stat-error} becomes subdominant to Eq.~\eqref{eq:sys-error} 
and we may write the bias-to-variance ratio condition as
\(\Delta\theta^i_\text{sys}/\sigma^i \leq 1,\) having made the 
usual approximation that the variance of any measured parameter \(\theta^i\) is
\(\sigma^i \approx \sqrt{(\Gamma^{-1})^{ii}}\)~\cite{Finn:1992wt}.

The authors of Refs.~\cite{Flanagan:1997kp,Cutler:2007mi} use the fact
that \(h(\theta_\text{true}) - h(\thetabf)\) can be well-approximated by its leading
term in the Taylor expansion above to recover the Fisher matrix in Eq.~\eqref{eq:sys-error}, but
one can go in the opposite direction, directly substituting into Eq.~\eqref{eq:max-like-condition}
that the directional derivative of \(h\) along \(\Delta\theta^i\) is approximated 
by the difference of the signals, in which case one arrives at
\begin{equation}
	|h(\thetabf)|^2 - \Braket{h(\theta_\text{s}) | h(\thetabf)} = 
	\Braket{h(\thetabf) | s} - \Braket{h(\theta_\text{s}) | s},
\end{equation}
recovering the result used to derive Eq.~\eqref{eq:mismatch-diff-snr}
when the SNRs of the signals are all comparable and thus contribute to an overall
scaling of both sides, which can be removed. This expression also assumes (as was
done for Eqs.~\eqref{eq:stat-error} and~\eqref{eq:sys-error}) that higher-order terms (e.g. \(\Delta\theta^i\Delta\theta^j\partial_i\partial_j h\))
can be sufficiently ignored, which is a statement about the curvature effects in the model manifold.

Under the linear signal approximation between \(\hat{\model}_1\) and \(\hat{\model}_2\)
from Eq.~\eqref{eq:linearsigappx} we find that the distance formula in Eq.~\eqref{eq:diffToDistance}
simplifies to
\begin{equation}
\hat{d}^{\,2}\left(\hat{\model}_1,\hat{\model}_2\right) \approx \frac12\widehat{\Gamma}_{ij}\left(\theta_1\right)\Delta\theta^i\Delta\theta^j,
\end{equation}
which is the (local) half squared geodetic distance between the two signals on the signal
manifold, known as Synge's world function~\cite{Synge:1960ueh,Owen:1995tm},
thereby further justifying our interpretation of \(\hat{d}\) as a distance. This
distance is also related to the Mahalanobis distance~\cite{Mahalanobis} away from
\(\theta_\text{bf}\).

The estimate for \(\Delta\theta^i_\text{sys}\) in Eq.~\eqref{eq:sys-error} 
is commonly referred to as the Cutler-Vallisneri~(CV)
criterion for the systematic bias, based on the authors of Ref.~\cite{Cutler:2007mi},
and in that work it is discussed how the validity of Eq.~\eqref{eq:sys-error}
depends heavily on the phase difference between \(\signal\) and \(h(\theta_\text{s})\).
Recent work~\cite{Dhani:2024jja} has shown that the CV~criterion can be improved
through the use of an alignment procedure that performs a time and phase shift
separately between \(\signal\) and both \(h(\theta_\text{bf})\) and 
\(h(\theta_\text{s})\) in Eq.~\eqref{eq:sys-error}, thereby
helping to ensure that any potential phase differences between the signal and
model evaluations is minimized. 

\begin{figure*}
	\centering
	\includegraphics[width=\textwidth]{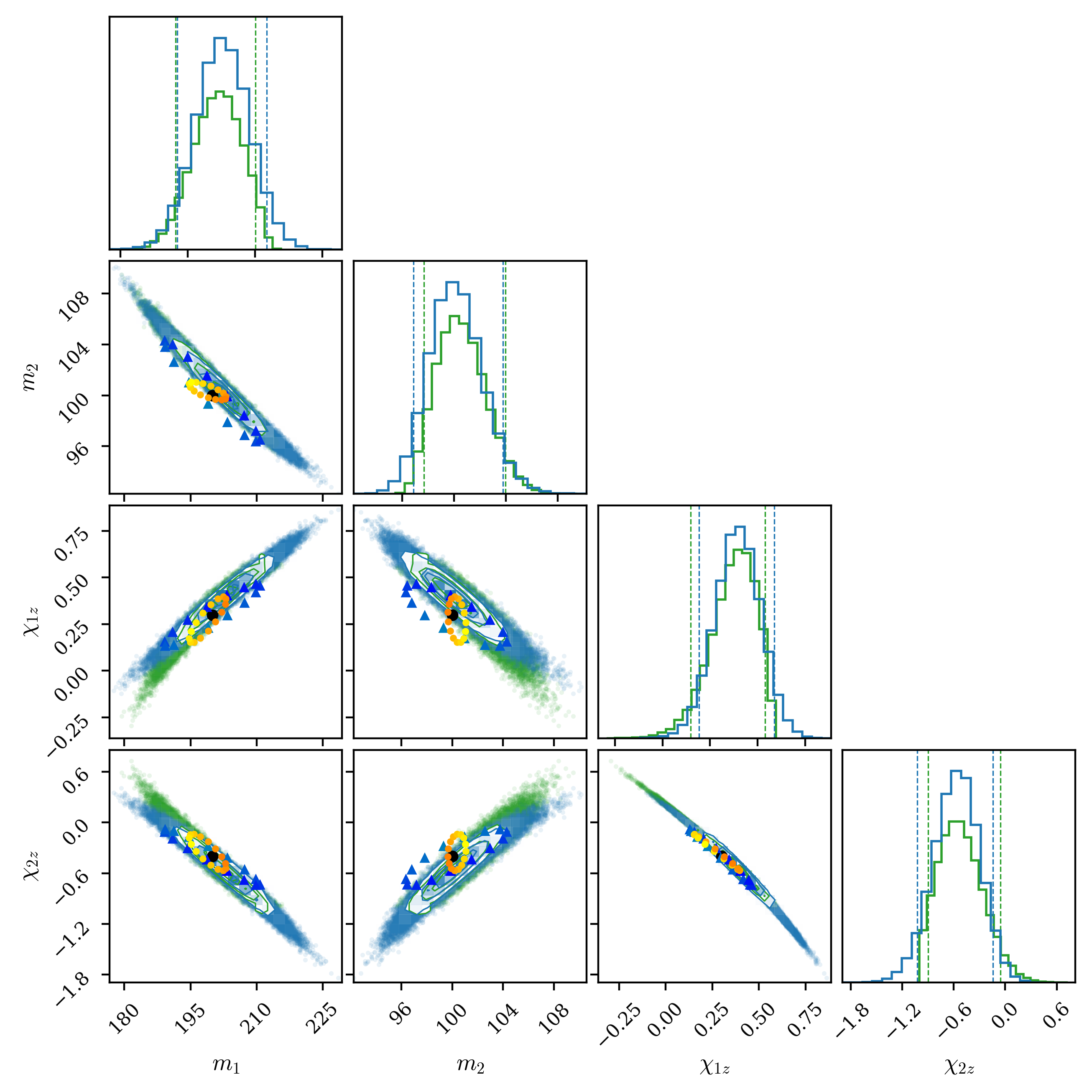}
	\caption{
	The posterior distribution for \texttt{SUR-3} shown in green alongside
	a multivariate normal distribution with mean and covariance computed using
	Fisher analysis as outlined in Appendix~\ref{sec:Fisher} having applied 
	a time and phase shift alignment. The blue-green triangles show the impact
	of a phase shift ranging between \([0,2\pi)\) applied to \(\model(\theta_\text{bf})\)
	in Eq.~\eqref{eq:sys-error}, while the orange-yellow dots show the impact of the
	same phase shift applied to \(\model(\theta_\text{s})\) in the same equation.
	}
	\label{fig:fisher-phase-shift}
\end{figure*}

The impact of this alignment is to bring the Fisher bias estimate close to the 
true bias values we might see in parameter estimation. The time shift and phase shift
both rotate the Fisher bias about the true
values of the signal by shifting the mean, as visualized in Fig.~\ref{fig:fisher-phase-shift}. 
Here the posterior probability distribution for 
\texttt{SUR-3}, injected at an SNR of 250, is plotted in green
alongside a multivariate normal distribution in blue produced with mean 
\(\theta_\text{s} + \Delta\theta_\text{sys}\) and covariance \(\Gamma^{-1}\)
computed using Eqs.~\eqref{eq:sys-error} and \eqref{eq:fisher}, after employing
the alignment procedure. We see that the Fisher approximation works well in 
reproducing the posterior except near some of the posterior tails,
where nonlinear correlations begin to appear. 
The black dots denote the location of the true parameters. 

The variation in the mean of the Fisher samples under different phase shifts is shown in 
Fig.~\ref{fig:fisher-phase-shift}. After aligning the signals in time, we 
apply an arbitrary phase shift ranging
between \([0,2\pi)\), shown as dots when applied to \(h(\theta_\text{s})\) and
as triangles when applied to \(h(\theta_\text{bf})\). We can see that the phase shift
rotates the Fisher bias about the true parameters. Finally, the importance
of this alignement procedure is made clear in Table~\ref{tab:fisher-snrs-noalign},
where we have computed the same estimates of \(\rho_\text{Fisher}\) as in Table~\ref{tab:fisher-snrs}
expect without using the alignment procedure. One notices that the bias SNRs in
this case are dramatically lower, implying that the larger phase difference in
\(\signal-\model(\theta_\text{s})\) dramatically decreases our estimation of model accuracy.

We also remark on the choice of parameters used in Eq.~\eqref{eq:sys-error}. As
derived, this equation requires us to evaluate the model \(\model\) at both the
best fitting point in parameter space, \(\theta_\text{bf}\), when computing
the waveform derivatives and Fisher matrix, and at the true
injection values \(\theta_\text{s}\) when computing the signal difference 
\(\signal - \model(\theta_\text{s})\). Often times in the literature one sees this fact
overlooked or, at least, not clearly distinguished, and the impact of using one
set of parameters rather than both is something we wish to clarify. We compute
the Fisher SNRs in Eq.~\eqref{eq:fisher-snr} using either only the parameters 
\(\theta_\text{s}\) or \(\theta_\text{bf}\) and the alignment procedure, with
results presented in Table~\ref{tab:fisher-snrs-only-inj} for \(\theta_\text{s}\)
and in Table~\ref{tab:fisher-snrs-only-maxl} for \(\theta_\text{bf}\).

When only using \(\theta_\text{s}\) in Eq.~\eqref{eq:sys-error}, the bias estimates
produced increase, thereby moderately lowering the estimated SNR at which the 
model will show bias. The results compared to those in Table~\ref{tab:fisher-snrs}
show larger differences between the two Fisher calculations than between the 
Fisher analysis and the bias distance estimates discussed in 
Sec.~\ref{sec:fisher-comparison}, with an average relative difference of 25\%.
The results of using only \(\theta_\text{bf}\) are expectedly worse, where the 
improved difference between signals \(\signal - \model(\theta_\text{bf})\) greatly
underestimates the bias, causing the bias SNR to greatly overestimate the accuracy
of the model. The only case for which this doesn't hold is \texttt{BAM-3}, which 
is impacted severly by the prior bound on \(\chi_{2z}\) and discussed in Sec.~\ref{sec:prior-bounds}.
The condition numbers of the Fisher matrices computed in this analysis are large,
but we have verified the robustness of the results in the tables provided in 
Appendix~\ref{sec:1d-bias-snr-results} by comparing the results of the bias SNR estimates 
after adding uniform random noise several orders
of magnitude larger than the inverse condition number
to each Fisher matrix before inversion~\cite{Vallisneri:2007ev}. Signal-to-noise ratio
estimates above 1000 are more sensitive to this added noise, but the leading-order results
hold in these cases.

Our overall conclusion from this exercise is that care should be taken when computing
bias estimates with Fisher analyses to apply the alignment procedure and use both
appropriate sets of parameters in Eq.~\eqref{eq:sys-error}.

\section{Waveform Model Derivatives}
\label{sec:model-diffs}

The Fisher analysis outlined in Appendix~\ref{sec:Fisher} requires 
differentiating the waveform model \(h\), and we discuss
our approach to waveform differentiation in this section.
The waveform model we use in this analysis, \d, is written in C-code inside 
of the \textsc{LALSuite} software library~\cite{lalsuite} and is not readily
ammenable to modern approaches to function differentiation like 
autodifferentiation~\cite{doi:10.1137/080743627}. While Python libraries exist
to compute derivatives of \d{}, such as \textsc{Ripple}~\cite{Edwards:2023sak},
we decided to implement a simpler framework for waveform derivatives.

Certain parameters in \(\theta\) are straightforward to differentiate with respect
to in \d{} due to the simple functional dependence of \(h\) on these parameters. 
This fact is (implicitly) outlined in Appendendix~A of Ref.~\cite{Dhani:2024jja}.
For the luminosity distance \(d_L\), coalescence time \(t_c\) and coalescence phase
\(\varphi_c\), the partial derivative of \(h\) can be analytically written as
\begin{align}
	\frac{\partial h}{\partial d_L} & = -\frac{h}{d_L},\\
	\frac{\partial h}{\partial t_c} & = -2\pi i f h,\\
	\frac{\partial h}{\partial \varphi_c} & = -i h.
\end{align}

For all other parameters no simple functional dependence exists, so we compute
these derivatives numerically using fourth-order finite difference
stencils. For a function \(f(x)\) and some finite step size \(\Delta x\), the centered 
fourth-order finite difference stencil is
\begin{equation}
\frac{\mathrm{d}f}{\mathrm{d}x} \approx \frac{f(x-2\Delta x) - 8 f(x-\Delta x) + 8 f(x+\Delta x) - f(x + 2\Delta x)}{12\Delta x}.
\end{equation}
On rare occasions, in particular near the boundaries of parameter priors, we 
may need to use a forward or backward directed stencil instead of the centered 
stencil. These expressions are given by,
\begin{widetext}
\begin{align}
\left.\frac{\mathrm{d}f}{\mathrm{d}x}\right|_\text{forward} & \approx -\frac{3f(x+4\Delta x)-16f(x+3\Delta x)+36f(x+2\Delta x)-48f(x+\Delta x)+25f(x)}{12\Delta x},\\
\left.\frac{\mathrm{d}f}{\mathrm{d}x}\right|_\text{backward} & \approx \frac{3f(x-4\Delta x)-16f(x-3\Delta x)+36f(x-2\Delta x)-48f(x-\Delta x)+25f(x)}{12\Delta x}.
\end{align}
\end{widetext}

The stencils all require specification of a step size \(\Delta x\). For the 
differentiation of \d{} with respect to various parameters, we don't know \textit{a prior}
what appropriate step size to choose at any given point in parameter space. Instead
we guess an initial step size, \(\Delta \theta^i = 2^{-11}\), for the chosen 
parameter \(\theta^i\) and compute derivatives
at this chosen \(\Delta \theta^i\) as well as at a coarser resolution \(2\Delta \theta^i\) 
and a finer resolution \(\Delta \theta^i/2\).
After computing the numerical derivative at these three initial step sizes, we
inspect the relative difference of the overlap Eq.~\eqref{eq:inner} between each
increasingly finer resolution. If the relative difference in the overlap is below \(10^{-8}\),
we choose the middle step size resolution \(\Delta\theta^i\). If not, we decrease
all step sizes by 2 and repeat until convergence or after six iterations.

\section{Bias SNR Results}
\label{sec:1d-bias-snr-results}

We present the tabulated parameter bias SNRs which are discussed
in Sec.~\ref{sec:BiasSNR} and Appendix~\ref{sec:Fisher}, along with complete 1D
posterior plots for the parameter bias estimates given in Sec.~\ref{sec:BiasSNR}.

\begin{figure*}[t!]
    \includegraphics[height=0.48\textheight]{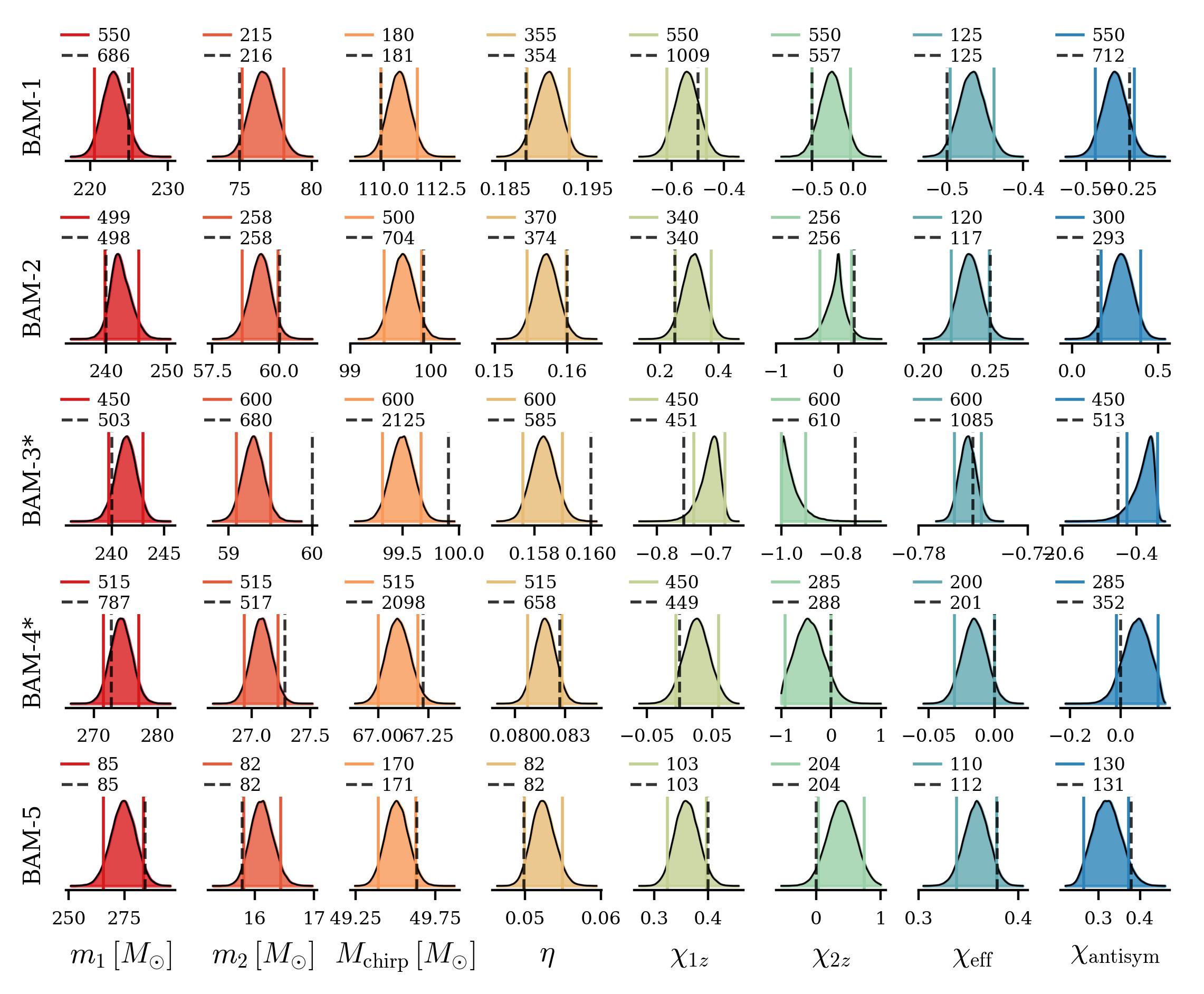}
    \caption{One-dimensional marginalized posteriors for parameter estimation of the five
	\texttt{BAM} cases, one in each row, listed in Table~\ref{tab:simcases}. Each 
	column corresponds to a separate parameter listed at the bottom of the figure. 
	We plot the true injected value
	as a dashed black line and the 90\%~CI as solid verticle lines. The legend
	for each plot shows the injected SNR next to the sold line and the predicted
	parameter bias SNR next to the dashed line, as discussed in Sec.~\ref{sec:eff-bias-snrs}. The asterisks denote cases where 
	railing in \(\chi_{2z}\) influences the SNR prediction, as discussed in
	Sec.~\ref{sec:prior-bounds}.}
	\label{fig:nr-1d-comparison}
\end{figure*}

\begin{figure*}[t!]
   \includegraphics[height=0.37\textheight]{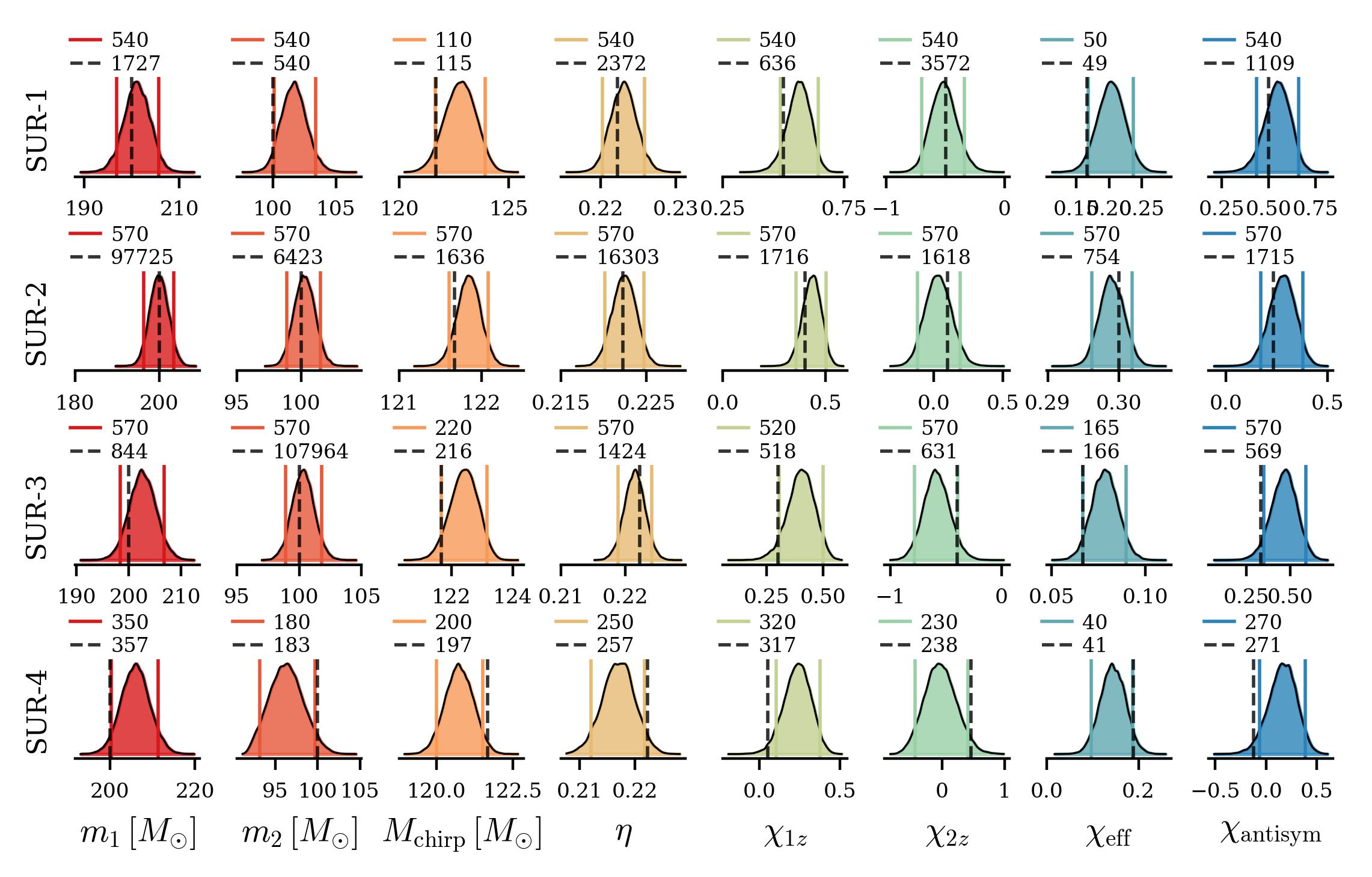}
   \caption{One-dimensional marginalized posteriors for parameter estimation of the four
	\texttt{SUR} cases, one in each row, listed in Table~\ref{tab:simcases}. Each 
	column corresponds to a separate parameter listed at the bottom of the figure. 
	We plot the true injected value
	as a dashed black line and the 90\%~CI as solid verticle lines. The legend
	for each plot shows the injected SNR next to the sold line and the predicted
	parameter bias SNR next to the dashed line, as discussed in Sec.~\ref{sec:eff-bias-snrs}.}
	\label{fig:sur-1d-comparison}
\end{figure*}

\begin{table}
	\renewcommand{\arraystretch}{1.2}
\begin{tabular}{@{} c|c|cccc @{}}
    \toprule
    \hspace{1em}Simulation ID\hspace{1em} & $\rho_{\text{bias, 2D}}$ & $m_1$ & $m_2$ & $M_\text{chirp}$ & $\eta$  \\
    \hline
\texttt{BAM-1}	& 	48	&	101	&	173	&	371	&	138	  \\
\texttt{BAM-2}	& 	195	&	164	&	157	&	151	&	159	  \\
\texttt{BAM-3}	& 	61	&	285	&	2449	&	523	&	1079	  \\
\texttt{BAM-4}	& 	43	&	75	&	112	&	165	&	96	  \\
\texttt{BAM-5}	& 	27	&	730	&	127	&	66	&	204	  \\ \hline
\texttt{SUR-1}	& 	47	&	271	&	148	&	45	&	197	  \\
\texttt{SUR-2}	& 	433	&	345	&	333	&	360	&	337	  \\
\texttt{SUR-3}	& 	213	&	382	&	492	&	323	&	431	  \\
\texttt{SUR-4}	& 	38	&	81	&	120	&	612	&	99	  \\
    \botrule
\end{tabular}
\caption{Values of the parameter bias SNRs computed for the 2D~restriction of
\d{} for all four mass parameters considered in this work. We replicate the 
values of the 2D~bias SNR from Table~\ref{tab:eff-snrs} for comparison.
We leave all values as computed for comparison but caution that SNR values above 500 
may not be reliable given the numerical accuracy thresholds used in this work.}
\label{tab:2d-bias-snrs}
\end{table}

\begin{table*}
	\renewcommand{\arraystretch}{1.2}
\begin{tabular}{@{} c|c|cccccccc @{}}
    \toprule
    \hspace{1em}Simulation ID\hspace{1em} & $\rho_\text{bias, 4D}$ & $m_1$ & $m_2$ & $M_\text{chirp}$ & $\eta$ & $\chi_{1z}$ & $\chi_{2z}$ & \(\chi_\text{eff}\) & \(\chi_\text{antisym}\) \\
    \hline
\texttt{BAM-1}	&	60 	&	686	&	216	&	181	&	354	&	1009	&	557	&	125	&	712	  \\
\texttt{BAM-2}	&	122 	&	498	&	258	&	704	&	374	&	340	&	256	&	117	&	293	  \\
\texttt{BAM-3}*	&	79 	&	503	&	680	&	2125	&	585	&	451	&	610	&	1085	&	513	  \\
\texttt{BAM-4}*	&	54 	&	787	&	517	&	2098	&	658	&	449	&	288	&	201	&	352	  \\
\texttt{BAM-5}	&	34 	&	85	&	82	&	171	&	82	&	103	&	204	&	112	&	131	  \\ \hline
\texttt{SUR-1}	&	49 	&	1727	&	540	&	115	&	2372	&	636	&	3572	&	49	&	1109	  \\
\texttt{SUR-2}	&	538 	&	97725	&	6423	&	1636	&	16303	&	1716	&	1618	&	754	&	1715	  \\
\texttt{SUR-3}	&	174 	&	844	&	107964	&	216	&	1424	&	518	&	631	&	166	&	569	  \\
\texttt{SUR-4}	&	40 	&	357	&	183	&	197	&	257	&	317	&	238	&	41	&	271	  \\
    \botrule
\end{tabular}
\caption{Values of the parameter bias SNRs computed for \d{} for 
all mass and spin parameters considered in this work. We replicate the 
values of the 4D~bias SNR from Table~\ref{tab:eff-snrs} for comparison.
The asterisks denote signals for which the parameter estimation is
heavily impacted by the \(\chi_{2z}\) prior bound, discussed in Sec.~\ref{sec:prior-bounds}.
We leave all values as computed for comparison but caution that SNR values above 500 
may not be reliable given the numerical accuracy thresholds used in this work.}
\label{tab:4d-bias-snrs}
\end{table*}

\begin{table*}
	\renewcommand{\arraystretch}{1.2}
\begin{tabular}{@{} c|cccccccc @{}}
    \toprule
    \hspace{1em}Simulation ID\hspace{1em} & $m_1$ & $m_2$ & $M_\text{chirp}$ & $\eta$ & $\chi_{1z}$ & $\chi_{2z}$ & \(\chi_\text{eff}\) & \(\chi_\text{antisym}\) \\
    \hline
\texttt{BAM-1}	&	623	&	212	&	198	&	336	&	969	&	567	&	134	&	732	  \\
\texttt{BAM-2}	&	475	&	250	&	747	&	360	&	330	&	249	&	114	&	284	  \\
\texttt{BAM-3}*	&	310	&	407	&	25184	&	326	&	251	&	282	&	1221	&	265	  \\
\texttt{BAM-4}*	&	1293	&	733	&	1570	&	1011	&	532	&	303	&	180	&	383	  \\
\texttt{BAM-5}	&	80	&	75	&	174	&	78	&	100	&	221	&	101	&	128	  \\ \hline
\texttt{SUR-1}	&	24164	&	422	&	115	&	974	&	874	&	3958	&	51	&	2569	  \\
\texttt{SUR-2}	&	23602	&	8316	&	1578	&	31018	&	1589	&	1526	&	764	&	1620	  \\
\texttt{SUR-3}	&	900	&	11074	&	218	&	1609	&	517	&	675	&	168	&	585	  \\
\texttt{SUR-4}	&	455	&	208	&	186	&	307	&	377	&	255	&	40	&	303	  \\
    \botrule
\end{tabular}
\caption{Fisher bias SNRs computed from Eq.~\eqref{eq:fisher-snr} 
and applying the time and phase shift alignment procedure outlined in 
Appendix~\ref{sec:Fisher}.
The asterisks denote signals for which the parameter estimation is
heavily impacted by the \(\chi_{2z}\) prior bound, discussed in Sec.~\ref{sec:prior-bounds}.
We leave all values as computed for comparison but caution that SNR values above 500 
may not be reliable given the numerical accuracy thresholds used in this work.}
\label{tab:fisher-snrs}
\end{table*}

\begin{table*}
	\renewcommand{\arraystretch}{1.2}
\begin{tabular}{@{} c|cccccccc @{}}
    \toprule
    \hspace{1em}Simulation ID\hspace{1em} & $m_1$ & $m_2$ & $M_\text{chirp}$ & $\eta$ & $\chi_{1z}$ & $\chi_{2z}$ & \(\chi_\text{eff}\) & \(\chi_\text{antisym}\) \\
    \hline
\texttt{BAM-1}	&	302	&	19	&	11	&	40	&	180	&	62	&	10	&	94	  \\
\texttt{BAM-2}	&	7	&	8	&	6	&	7	&	6	&	6	&	17	&	6	  \\
\texttt{BAM-3}*	&	27	&	19	&	30	&	21	&	149	&	271	&	97	&	322	  \\
\texttt{BAM-4}*	&	32	&	21	&	99	&	27	&	14	&	9	&	7	&	11	  \\
\texttt{BAM-5}	&	7	&	7	&	11	&	7	&	11	&	43	&	9	&	16	  \\ \hline
\texttt{SUR-1}	&	20	&	26	&	12	&	22	&	19	&	20	&	20	&	20	  \\
\texttt{SUR-2}	&	152	&	103	&	15	&	1321	&	98	&	422	&	7	&	182	  \\
\texttt{SUR-3}	&	6	&	6	&	5	&	6	&	5	&	5	&	6	&	5	  \\
\texttt{SUR-4}	&	3	&	2	&	4	&	3	&	3	&	3	&	475	&	3	  \\
    \botrule
\end{tabular}
\caption{Fisher bias SNRs computed from Eq.~\eqref{eq:fisher-snr} but without
applying the time and phase shift alignment procedure outlined in Appendix~\ref{sec:Fisher}.
The asterisks denote signals for which the parameter estimation is
heavily impacted by the \(\chi_{2z}\) prior bound, discussed in Sec.~\ref{sec:prior-bounds}.
We leave all values as computed for comparison but caution that SNR values above 500 
may not be reliable given the numerical accuracy thresholds used in this work.}
\label{tab:fisher-snrs-noalign}
\end{table*}

\begin{table*}
	\renewcommand{\arraystretch}{1.2}
\begin{tabular}{@{} c|cccccccc @{}}
    \toprule
    \hspace{1em}Simulation ID\hspace{1em} & $m_1$ & $m_2$ & $M_\text{chirp}$ & $\eta$ & $\chi_{1z}$ & $\chi_{2z}$ & \(\chi_\text{eff}\) & \(\chi_\text{antisym}\) \\
    \hline
\texttt{BAM-1}	&	601	&	197	&	180	&	316	&	798	&	475	&	121	&	607	  \\
\texttt{BAM-2}	&	420	&	236	&	901	&	328	&	298	&	231	&	97	&	262	  \\
\texttt{BAM-3}*	&	336	&	573	&	2589	&	398	&	272	&	323	&	2619	&	295	  \\
\texttt{BAM-4}*	&	806	&	519	&	1485	&	667	&	472	&	295	&	195	&	361	  \\
\texttt{BAM-5}	&	79	&	80	&	132	&	78	&	87	&	146	&	115	&	105	  \\ \hline
\texttt{SUR-1}	&	751	&	999	&	107	&	2788	&	479	&	1214	&	49	&	705	  \\
\texttt{SUR-2}	&	36274	&	7355	&	1611	&	24298	&	1655	&	1561	&	737	&	1660	  \\
\texttt{SUR-3}	&	747	&	7216	&	212	&	1176	&	485	&	585	&	165	&	530	  \\
\texttt{SUR-4}	&	220	&	139	&	266	&	177	&	210	&	172	&	39	&	189	  \\
    \botrule
\end{tabular}
\caption{Fisher bias SNRs computed from Eq.~\eqref{eq:fisher-snr} using only the
parameters \(\theta_\text{s}\) and applying the time and phase shift alignment procedure outlined in 
Appendix~\ref{sec:Fisher}.
The asterisks denote signals for which the parameter estimation is
heavily impacted by the \(\chi_{2z}\) prior bound, discussed in Sec.~\ref{sec:prior-bounds}.
We leave all values as computed for comparison but caution that SNR values above 500 
may not be reliable given the numerical accuracy thresholds used in this work.}
\label{tab:fisher-snrs-only-inj}
\end{table*}

\begin{table*}
	\renewcommand{\arraystretch}{1.2}
\begin{tabular}{@{} c|cccccccc @{}}
    \toprule
    \hspace{1em}Simulation ID\hspace{1em} & $m_1$ & $m_2$ & $M_\text{chirp}$ & $\eta$ & $\chi_{1z}$ & $\chi_{2z}$ & \(\chi_\text{eff}\) & \(\chi_\text{antisym}\) \\
    \hline
\texttt{BAM-1}	&	22504	&	17353	&	5114	&	417209	&	6053	&	5335	&	3596	&	5537	  \\
\texttt{BAM-2}	&	10129	&	7786	&	30028	&	9102	&	14947	&	15737	&	11788	&	15214	  \\
\texttt{BAM-3}*	&	843	&	466	&	568	&	551	&	660	&	559	&	490	&	603	  \\
\texttt{BAM-4}*	&	64329	&	9900	&	6250	&	21901	&	230639	&	174300	&	27077	&	9696014	  \\
\texttt{BAM-5}	&	14638	&	17164	&	15499	&	15271	&	28389	&	76884	&	11574	&	77380	  \\ \hline
\texttt{SUR-1}	&	9066	&	8823	&	13757	&	8935	&	9632	&	9482	&	42356	&	9542	  \\
\texttt{SUR-2}	&	14469	&	12252	&	86069	&	12387	&	9163	&	8487	&	4433	&	8432	  \\
\texttt{SUR-3}	&	2162	&	2602	&	1641	&	2314	&	1985	&	2077	&	2186	&	2029	  \\
\texttt{SUR-4}	&	3375	&	3683	&	3286	&	3476	&	3590	&	3700	&	4524	&	3641	  \\
    \botrule
\end{tabular}
\caption{Fisher bias SNRs computed from Eq.~\eqref{eq:fisher-snr} using only the
parameters \(\theta_\text{bf}\) and applying the time and phase shift alignment procedure outlined in 
Appendix~\ref{sec:Fisher}.
The asterisks denote signals for which the parameter estimation is
heavily impacted by the \(\chi_{2z}\) prior bound, discussed in Sec.~\ref{sec:prior-bounds}.
We leave all values as computed for comparison but caution that SNR values above 500 
may not be reliable given the numerical accuracy thresholds used in this work.}
\label{tab:fisher-snrs-only-maxl}
\end{table*}

\bibliography{references.bib}

\begin{thebibliography}{88}%
\makeatletter
\providecommand \@ifxundefined [1]{%
 \@ifx{#1\undefined}
}%
\providecommand \@ifnum [1]{%
 \ifnum #1\expandafter \@firstoftwo
 \else \expandafter \@secondoftwo
 \fi
}%
\providecommand \@ifx [1]{%
 \ifx #1\expandafter \@firstoftwo
 \else \expandafter \@secondoftwo
 \fi
}%
\providecommand \natexlab [1]{#1}%
\providecommand \enquote  [1]{``#1''}%
\providecommand \bibnamefont  [1]{#1}%
\providecommand \bibfnamefont [1]{#1}%
\providecommand \citenamefont [1]{#1}%
\providecommand \href@noop [0]{\@secondoftwo}%
\providecommand \href [0]{\begingroup \@sanitize@url \@href}%
\providecommand \@href[1]{\@@startlink{#1}\@@href}%
\providecommand \@@href[1]{\endgroup#1\@@endlink}%
\providecommand \@sanitize@url [0]{\catcode `\\12\catcode `\$12\catcode
  `\&12\catcode `\#12\catcode `\^12\catcode `\_12\catcode `\%12\relax}%
\providecommand \@@startlink[1]{}%
\providecommand \@@endlink[0]{}%
\providecommand \url  [0]{\begingroup\@sanitize@url \@url }%
\providecommand \@url [1]{\endgroup\@href {#1}{\urlprefix }}%
\providecommand \urlprefix  [0]{URL }%
\providecommand \Eprint [0]{\href }%
\providecommand \doibase [0]{http://dx.doi.org/}%
\providecommand \selectlanguage [0]{\@gobble}%
\providecommand \bibinfo  [0]{\@secondoftwo}%
\providecommand \bibfield  [0]{\@secondoftwo}%
\providecommand \translation [1]{[#1]}%
\providecommand \BibitemOpen [0]{}%
\providecommand \bibitemStop [0]{}%
\providecommand \bibitemNoStop [0]{.\EOS\space}%
\providecommand \EOS [0]{\spacefactor3000\relax}%
\providecommand \BibitemShut  [1]{\csname bibitem#1\endcsname}%
\let\auto@bib@innerbib\@empty
%</preamble>
\bibitem [{\citenamefont {Abbott}\ \emph
  {et~al.}(2016{\natexlab{a}})\citenamefont {Abbott} \emph
  {et~al.}}]{LIGOScientific:2016aoc}%
  \BibitemOpen
  \bibfield  {author} {\bibinfo {author} {\bibfnamefont {B.~P.}\ \bibnamefont
  {Abbott}} \emph {et~al.} (\bibinfo {collaboration} {LIGO Scientific,
  Virgo}),\ }\href {\doibase 10.1103/PhysRevLett.116.061102} {\bibfield
  {journal} {\bibinfo  {journal} {Phys. Rev. Lett.}\ }\textbf {\bibinfo
  {volume} {116}},\ \bibinfo {pages} {061102} (\bibinfo {year}
  {2016}{\natexlab{a}})},\ \Eprint {http://arxiv.org/abs/1602.03837}
  {arXiv:1602.03837 [gr-qc]} \BibitemShut {NoStop}%
\bibitem [{\citenamefont {Abbott}\ \emph
  {et~al.}(2016{\natexlab{b}})\citenamefont {Abbott} \emph
  {et~al.}}]{LIGOScientific:2016dsl}%
  \BibitemOpen
  \bibfield  {author} {\bibinfo {author} {\bibfnamefont {B.~P.}\ \bibnamefont
  {Abbott}} \emph {et~al.} (\bibinfo {collaboration} {LIGO Scientific,
  Virgo}),\ }\href {\doibase 10.1103/PhysRevX.6.041015} {\bibfield  {journal}
  {\bibinfo  {journal} {Phys. Rev. X}\ }\textbf {\bibinfo {volume} {6}},\
  \bibinfo {pages} {041015} (\bibinfo {year} {2016}{\natexlab{b}})},\ \bibinfo
  {note} {[Erratum: Phys.Rev.X 8, 039903 (2018)]},\ \Eprint
  {http://arxiv.org/abs/1606.04856} {arXiv:1606.04856 [gr-qc]} \BibitemShut
  {NoStop}%
\bibitem [{\citenamefont {Abbott}\ \emph {et~al.}(2021)\citenamefont {Abbott}
  \emph {et~al.}}]{LIGOScientific:2020ibl}%
  \BibitemOpen
  \bibfield  {author} {\bibinfo {author} {\bibfnamefont {R.}~\bibnamefont
  {Abbott}} \emph {et~al.} (\bibinfo {collaboration} {LIGO Scientific,
  Virgo}),\ }\href {\doibase 10.1103/PhysRevX.11.021053} {\bibfield  {journal}
  {\bibinfo  {journal} {Phys. Rev. X}\ }\textbf {\bibinfo {volume} {11}},\
  \bibinfo {pages} {021053} (\bibinfo {year} {2021})},\ \Eprint
  {http://arxiv.org/abs/2010.14527} {arXiv:2010.14527 [gr-qc]} \BibitemShut
  {NoStop}%
\bibitem [{\citenamefont {Abbott}\ \emph {et~al.}(2024)\citenamefont {Abbott}
  \emph {et~al.}}]{LIGOScientific:2021usb}%
  \BibitemOpen
  \bibfield  {author} {\bibinfo {author} {\bibfnamefont {R.}~\bibnamefont
  {Abbott}} \emph {et~al.} (\bibinfo {collaboration} {LIGO Scientific,
  VIRGO}),\ }\href {\doibase 10.1103/PhysRevD.109.022001} {\bibfield  {journal}
  {\bibinfo  {journal} {Phys. Rev. D}\ }\textbf {\bibinfo {volume} {109}},\
  \bibinfo {pages} {022001} (\bibinfo {year} {2024})},\ \Eprint
  {http://arxiv.org/abs/2108.01045} {arXiv:2108.01045 [gr-qc]} \BibitemShut
  {NoStop}%
\bibitem [{\citenamefont {Abbott}\ \emph {et~al.}(2023)\citenamefont {Abbott}
  \emph {et~al.}}]{LIGOScientific:2021vkt}%
  \BibitemOpen
  \bibfield  {author} {\bibinfo {author} {\bibfnamefont {R.}~\bibnamefont
  {Abbott}} \emph {et~al.} (\bibinfo {collaboration} {LIGO Scientific, Virgo,
  KAGRA}),\ }\href {\doibase 10.1103/PhysRevX.13.041039} {\bibfield  {journal}
  {\bibinfo  {journal} {Phys. Rev. X}\ }\textbf {\bibinfo {volume} {13}},\
  \bibinfo {pages} {041039} (\bibinfo {year} {2023})},\ \Eprint
  {http://arxiv.org/abs/2111.03606} {arXiv:2111.03606 [gr-qc]} \BibitemShut
  {NoStop}%
\bibitem [{\citenamefont {Abbott}\ \emph {et~al.}(2020)\citenamefont {Abbott},
  \citenamefont {Abbott}, \citenamefont {Ackley}, \citenamefont {Adams},
  \citenamefont {Adya}, \citenamefont {Affeldt}, \citenamefont {Agathos} \emph
  {et~al.}}]{abbott2020prospects}%
  \BibitemOpen
  \bibfield  {author} {\bibinfo {author} {\bibfnamefont {R.}~\bibnamefont
  {Abbott}}, \bibinfo {author} {\bibfnamefont {T.}~\bibnamefont {Abbott}},
  \bibinfo {author} {\bibfnamefont {K.}~\bibnamefont {Ackley}}, \bibinfo
  {author} {\bibfnamefont {C.}~\bibnamefont {Adams}}, \bibinfo {author}
  {\bibfnamefont {V.}~\bibnamefont {Adya}}, \bibinfo {author} {\bibfnamefont
  {C.}~\bibnamefont {Affeldt}}, \bibinfo {author} {\bibfnamefont
  {M.}~\bibnamefont {Agathos}},  \emph {et~al.},\ }\href@noop {} {\bibfield
  {journal} {\bibinfo  {journal} {Living reviews in relativity}\ }\textbf
  {\bibinfo {volume} {23}},\ \bibinfo {pages} {1} (\bibinfo {year}
  {2020})}\BibitemShut {NoStop}%
\bibitem [{\citenamefont {Reitze}\ \emph {et~al.}(2019)\citenamefont {Reitze}
  \emph {et~al.}}]{Reitze:2019iox}%
  \BibitemOpen
  \bibfield  {author} {\bibinfo {author} {\bibfnamefont {D.}~\bibnamefont
  {Reitze}} \emph {et~al.},\ }\href@noop {} {\bibfield  {journal} {\bibinfo
  {journal} {Bull. Am. Astron. Soc.}\ }\textbf {\bibinfo {volume} {51}},\
  \bibinfo {pages} {035} (\bibinfo {year} {2019})},\ \Eprint
  {http://arxiv.org/abs/1907.04833} {arXiv:1907.04833 [astro-ph.IM]}
  \BibitemShut {NoStop}%
\bibitem [{\citenamefont {Evans}\ \emph {et~al.}(2023)\citenamefont {Evans}
  \emph {et~al.}}]{Evans:2023euw}%
  \BibitemOpen
  \bibfield  {author} {\bibinfo {author} {\bibfnamefont {M.}~\bibnamefont
  {Evans}} \emph {et~al.},\ }\href@noop {} {\  (\bibinfo {year} {2023})},\
  \Eprint {http://arxiv.org/abs/2306.13745} {arXiv:2306.13745 [astro-ph.IM]}
  \BibitemShut {NoStop}%
\bibitem [{\citenamefont {Punturo}\ \emph {et~al.}(2010)\citenamefont {Punturo}
  \emph {et~al.}}]{Punturo:2010zza}%
  \BibitemOpen
  \bibfield  {author} {\bibinfo {author} {\bibfnamefont {M.}~\bibnamefont
  {Punturo}} \emph {et~al.},\ }\href {\doibase 10.1088/0264-9381/27/8/084007}
  {\bibfield  {journal} {\bibinfo  {journal} {Class. Quant. Grav.}\ }\textbf
  {\bibinfo {volume} {27}},\ \bibinfo {pages} {084007} (\bibinfo {year}
  {2010})}\BibitemShut {NoStop}%
\bibitem [{\citenamefont {Hild}\ \emph {et~al.}(2011)\citenamefont {Hild} \emph
  {et~al.}}]{Hild:2010id}%
  \BibitemOpen
  \bibfield  {author} {\bibinfo {author} {\bibfnamefont {S.}~\bibnamefont
  {Hild}} \emph {et~al.},\ }\href {\doibase 10.1088/0264-9381/28/9/094013}
  {\bibfield  {journal} {\bibinfo  {journal} {Class. Quant. Grav.}\ }\textbf
  {\bibinfo {volume} {28}},\ \bibinfo {pages} {094013} (\bibinfo {year}
  {2011})},\ \Eprint {http://arxiv.org/abs/1012.0908} {arXiv:1012.0908 [gr-qc]}
  \BibitemShut {NoStop}%
\bibitem [{\citenamefont {Maggiore}\ \emph {et~al.}(2020)\citenamefont
  {Maggiore} \emph {et~al.}}]{ET:2019dnz}%
  \BibitemOpen
  \bibfield  {author} {\bibinfo {author} {\bibfnamefont {M.}~\bibnamefont
  {Maggiore}} \emph {et~al.} (\bibinfo {collaboration} {ET}),\ }\href {\doibase
  10.1088/1475-7516/2020/03/050} {\bibfield  {journal} {\bibinfo  {journal}
  {JCAP}\ }\textbf {\bibinfo {volume} {03}},\ \bibinfo {pages} {050} (\bibinfo
  {year} {2020})},\ \Eprint {http://arxiv.org/abs/1912.02622} {arXiv:1912.02622
  [astro-ph.CO]} \BibitemShut {NoStop}%
\bibitem [{\citenamefont {Abac}\ \emph {et~al.}(2025)\citenamefont {Abac} \emph
  {et~al.}}]{Abac:2025saz}%
  \BibitemOpen
  \bibfield  {author} {\bibinfo {author} {\bibfnamefont {A.}~\bibnamefont
  {Abac}} \emph {et~al.},\ }\href@noop {} {\  (\bibinfo {year} {2025})},\
  \Eprint {http://arxiv.org/abs/2503.12263} {arXiv:2503.12263 [gr-qc]}
  \BibitemShut {NoStop}%
\bibitem [{\citenamefont {Amaro-Seoane}\ \emph {et~al.}(2017)\citenamefont
  {Amaro-Seoane} \emph {et~al.}}]{LISA:2017pwj}%
  \BibitemOpen
  \bibfield  {author} {\bibinfo {author} {\bibfnamefont {P.}~\bibnamefont
  {Amaro-Seoane}} \emph {et~al.} (\bibinfo {collaboration} {LISA}),\
  }\href@noop {} {\  (\bibinfo {year} {2017})},\ \Eprint
  {http://arxiv.org/abs/1702.00786} {arXiv:1702.00786 [astro-ph.IM]}
  \BibitemShut {NoStop}%
\bibitem [{\citenamefont {Babak}\ \emph {et~al.}(2021)\citenamefont {Babak},
  \citenamefont {Petiteau},\ and\ \citenamefont {Hewitson}}]{Babak:2021mhe}%
  \BibitemOpen
  \bibfield  {author} {\bibinfo {author} {\bibfnamefont {S.}~\bibnamefont
  {Babak}}, \bibinfo {author} {\bibfnamefont {A.}~\bibnamefont {Petiteau}}, \
  and\ \bibinfo {author} {\bibfnamefont {M.}~\bibnamefont {Hewitson}},\
  }\href@noop {} {\  (\bibinfo {year} {2021})},\ \Eprint
  {http://arxiv.org/abs/2108.01167} {arXiv:2108.01167 [astro-ph.IM]}
  \BibitemShut {NoStop}%
\bibitem [{\citenamefont {Colpi}\ \emph {et~al.}(2024)\citenamefont {Colpi}
  \emph {et~al.}}]{LISA:2024hlh}%
  \BibitemOpen
  \bibfield  {author} {\bibinfo {author} {\bibfnamefont {M.}~\bibnamefont
  {Colpi}} \emph {et~al.} (\bibinfo {collaboration} {LISA}),\ }\href@noop {} {\
   (\bibinfo {year} {2024})},\ \Eprint {http://arxiv.org/abs/2402.07571}
  {arXiv:2402.07571 [astro-ph.CO]} \BibitemShut {NoStop}%
\bibitem [{\citenamefont {Finn}(1992)}]{Finn:1992wt}%
  \BibitemOpen
  \bibfield  {author} {\bibinfo {author} {\bibfnamefont {L.~S.}\ \bibnamefont
  {Finn}},\ }\href {\doibase 10.1103/PhysRevD.46.5236} {\bibfield  {journal}
  {\bibinfo  {journal} {Phys. Rev. D}\ }\textbf {\bibinfo {volume} {46}},\
  \bibinfo {pages} {5236} (\bibinfo {year} {1992})},\ \Eprint
  {http://arxiv.org/abs/gr-qc/9209010} {arXiv:gr-qc/9209010} \BibitemShut
  {NoStop}%
\bibitem [{\citenamefont {Lindblom}\ \emph {et~al.}(2008)\citenamefont
  {Lindblom}, \citenamefont {Owen},\ and\ \citenamefont
  {Brown}}]{lindblom:2008cm}%
  \BibitemOpen
  \bibfield  {author} {\bibinfo {author} {\bibfnamefont {L.}~\bibnamefont
  {Lindblom}}, \bibinfo {author} {\bibfnamefont {B.~J.}\ \bibnamefont {Owen}},
  \ and\ \bibinfo {author} {\bibfnamefont {D.~A.}\ \bibnamefont {Brown}},\
  }\href {\doibase 10.1103/PhysRevD.78.124020} {\bibfield  {journal} {\bibinfo
  {journal} {Phys.Rev.}\ }\textbf {\bibinfo {volume} {D78}},\ \bibinfo {pages}
  {124020} (\bibinfo {year} {2008})},\ \Eprint {http://arxiv.org/abs/0809.3844}
  {arXiv:0809.3844 [gr-qc]} \BibitemShut {NoStop}%
\bibitem [{\citenamefont {McWilliams}\ \emph {et~al.}(2010)\citenamefont
  {McWilliams}, \citenamefont {Kelly},\ and\ \citenamefont
  {Baker}}]{McWilliams:2010eq}%
  \BibitemOpen
  \bibfield  {author} {\bibinfo {author} {\bibfnamefont {S.~T.}\ \bibnamefont
  {McWilliams}}, \bibinfo {author} {\bibfnamefont {B.~J.}\ \bibnamefont
  {Kelly}}, \ and\ \bibinfo {author} {\bibfnamefont {J.~G.}\ \bibnamefont
  {Baker}},\ }\href {\doibase 10.1103/PhysRevD.82.024014} {\bibfield  {journal}
  {\bibinfo  {journal} {Phys. Rev. D}\ }\textbf {\bibinfo {volume} {82}},\
  \bibinfo {pages} {024014} (\bibinfo {year} {2010})},\ \Eprint
  {http://arxiv.org/abs/1004.0961} {arXiv:1004.0961 [gr-qc]} \BibitemShut
  {NoStop}%
\bibitem [{\citenamefont {Hannam}\ \emph {et~al.}(2010)\citenamefont {Hannam},
  \citenamefont {Husa}, \citenamefont {Ohme},\ and\ \citenamefont
  {Ajith}}]{Hannam:2010ky}%
  \BibitemOpen
  \bibfield  {author} {\bibinfo {author} {\bibfnamefont {M.}~\bibnamefont
  {Hannam}}, \bibinfo {author} {\bibfnamefont {S.}~\bibnamefont {Husa}},
  \bibinfo {author} {\bibfnamefont {F.}~\bibnamefont {Ohme}}, \ and\ \bibinfo
  {author} {\bibfnamefont {P.}~\bibnamefont {Ajith}},\ }\href {\doibase
  10.1103/PhysRevD.82.124052} {\bibfield  {journal} {\bibinfo  {journal} {Phys.
  Rev. D}\ }\textbf {\bibinfo {volume} {82}},\ \bibinfo {pages} {124052}
  (\bibinfo {year} {2010})},\ \Eprint {http://arxiv.org/abs/1008.2961}
  {arXiv:1008.2961 [gr-qc]} \BibitemShut {NoStop}%
\bibitem [{\citenamefont {{Baird}}\ \emph {et~al.}(2013)\citenamefont
  {{Baird}}, \citenamefont {{Fairhurst}}, \citenamefont {{Hannam}},\ and\
  \citenamefont {{Murphy}}}]{Baird:2013dbm}%
  \BibitemOpen
  \bibfield  {author} {\bibinfo {author} {\bibfnamefont {E.}~\bibnamefont
  {{Baird}}}, \bibinfo {author} {\bibfnamefont {S.}~\bibnamefont
  {{Fairhurst}}}, \bibinfo {author} {\bibfnamefont {M.}~\bibnamefont
  {{Hannam}}}, \ and\ \bibinfo {author} {\bibfnamefont {P.}~\bibnamefont
  {{Murphy}}},\ }\href {\doibase 10.1103/PhysRevD.87.024035} {\bibfield
  {journal} {\bibinfo  {journal} {\prd}\ }\textbf {\bibinfo {volume} {87}},\
  \bibinfo {eid} {024035} (\bibinfo {year} {2013})},\ \Eprint
  {http://arxiv.org/abs/1211.0546} {arXiv:1211.0546 [gr-qc]} \BibitemShut
  {NoStop}%
\bibitem [{\citenamefont {Chatziioannou}\ \emph {et~al.}(2017)\citenamefont
  {Chatziioannou}, \citenamefont {Klein}, \citenamefont {Yunes},\ and\
  \citenamefont {Cornish}}]{Chatziioannou:2017tdw}%
  \BibitemOpen
  \bibfield  {author} {\bibinfo {author} {\bibfnamefont {K.}~\bibnamefont
  {Chatziioannou}}, \bibinfo {author} {\bibfnamefont {A.}~\bibnamefont
  {Klein}}, \bibinfo {author} {\bibfnamefont {N.}~\bibnamefont {Yunes}}, \ and\
  \bibinfo {author} {\bibfnamefont {N.}~\bibnamefont {Cornish}},\ }\href
  {\doibase 10.1103/PhysRevD.95.104004} {\bibfield  {journal} {\bibinfo
  {journal} {Phys. Rev. D}\ }\textbf {\bibinfo {volume} {95}},\ \bibinfo
  {pages} {104004} (\bibinfo {year} {2017})},\ \Eprint
  {http://arxiv.org/abs/1703.03967} {arXiv:1703.03967 [gr-qc]} \BibitemShut
  {NoStop}%
\bibitem [{\citenamefont {Toubiana}\ and\ \citenamefont
  {Gair}(2024)}]{Toubiana:2024car}%
  \BibitemOpen
  \bibfield  {author} {\bibinfo {author} {\bibfnamefont {A.}~\bibnamefont
  {Toubiana}}\ and\ \bibinfo {author} {\bibfnamefont {J.~R.}\ \bibnamefont
  {Gair}},\ }\href@noop {} {\  (\bibinfo {year} {2024})},\ \Eprint
  {http://arxiv.org/abs/2401.06845} {arXiv:2401.06845 [gr-qc]} \BibitemShut
  {NoStop}%
\bibitem [{\citenamefont {Abbott}\ \emph {et~al.}(2017)\citenamefont {Abbott}
  \emph {et~al.}}]{LIGOScientific:2016ebw}%
  \BibitemOpen
  \bibfield  {author} {\bibinfo {author} {\bibfnamefont {B.~P.}\ \bibnamefont
  {Abbott}} \emph {et~al.} (\bibinfo {collaboration} {LIGO Scientific,
  Virgo}),\ }\href {\doibase 10.1088/1361-6382/aa6854} {\bibfield  {journal}
  {\bibinfo  {journal} {Class. Quant. Grav.}\ }\textbf {\bibinfo {volume}
  {34}},\ \bibinfo {pages} {104002} (\bibinfo {year} {2017})},\ \Eprint
  {http://arxiv.org/abs/1611.07531} {arXiv:1611.07531 [gr-qc]} \BibitemShut
  {NoStop}%
\bibitem [{\citenamefont {P\"urrer}\ and\ \citenamefont
  {Haster}(2020)}]{Purrer:2019jcp}%
  \BibitemOpen
  \bibfield  {author} {\bibinfo {author} {\bibfnamefont {M.}~\bibnamefont
  {P\"urrer}}\ and\ \bibinfo {author} {\bibfnamefont {C.-J.}\ \bibnamefont
  {Haster}},\ }\href {\doibase 10.1103/PhysRevResearch.2.023151} {\bibfield
  {journal} {\bibinfo  {journal} {Phys. Rev. Res.}\ }\textbf {\bibinfo {volume}
  {2}},\ \bibinfo {pages} {023151} (\bibinfo {year} {2020})},\ \Eprint
  {http://arxiv.org/abs/1912.10055} {arXiv:1912.10055 [gr-qc]} \BibitemShut
  {NoStop}%
\bibitem [{\citenamefont {Varma}\ \emph
  {et~al.}(2019{\natexlab{a}})\citenamefont {Varma}, \citenamefont {Field},
  \citenamefont {Scheel}, \citenamefont {Blackman}, \citenamefont {Gerosa},
  \citenamefont {Stein}, \citenamefont {Kidder},\ and\ \citenamefont
  {Pfeiffer}}]{Varma:2019csw}%
  \BibitemOpen
  \bibfield  {author} {\bibinfo {author} {\bibfnamefont {V.}~\bibnamefont
  {Varma}}, \bibinfo {author} {\bibfnamefont {S.~E.}\ \bibnamefont {Field}},
  \bibinfo {author} {\bibfnamefont {M.~A.}\ \bibnamefont {Scheel}}, \bibinfo
  {author} {\bibfnamefont {J.}~\bibnamefont {Blackman}}, \bibinfo {author}
  {\bibfnamefont {D.}~\bibnamefont {Gerosa}}, \bibinfo {author} {\bibfnamefont
  {L.~C.}\ \bibnamefont {Stein}}, \bibinfo {author} {\bibfnamefont {L.~E.}\
  \bibnamefont {Kidder}}, \ and\ \bibinfo {author} {\bibfnamefont {H.~P.}\
  \bibnamefont {Pfeiffer}},\ }\href {\doibase 10.1103/PhysRevResearch.1.033015}
  {\bibfield  {journal} {\bibinfo  {journal} {Phys. Rev. Research.}\ }\textbf
  {\bibinfo {volume} {1}},\ \bibinfo {pages} {033015} (\bibinfo {year}
  {2019}{\natexlab{a}})},\ \Eprint {http://arxiv.org/abs/1905.09300}
  {arXiv:1905.09300 [gr-qc]} \BibitemShut {NoStop}%
\bibitem [{\citenamefont {Cutler}\ and\ \citenamefont
  {Vallisneri}(2007)}]{Cutler:2007mi}%
  \BibitemOpen
  \bibfield  {author} {\bibinfo {author} {\bibfnamefont {C.}~\bibnamefont
  {Cutler}}\ and\ \bibinfo {author} {\bibfnamefont {M.}~\bibnamefont
  {Vallisneri}},\ }\href {\doibase 10.1103/PhysRevD.76.104018} {\bibfield
  {journal} {\bibinfo  {journal} {Phys. Rev. D}\ }\textbf {\bibinfo {volume}
  {76}},\ \bibinfo {pages} {104018} (\bibinfo {year} {2007})},\ \Eprint
  {http://arxiv.org/abs/0707.2982} {arXiv:0707.2982 [gr-qc]} \BibitemShut
  {NoStop}%
\bibitem [{\citenamefont {Hu}\ and\ \citenamefont {Veitch}(2022)}]{Hu:2022rjq}%
  \BibitemOpen
  \bibfield  {author} {\bibinfo {author} {\bibfnamefont {Q.}~\bibnamefont
  {Hu}}\ and\ \bibinfo {author} {\bibfnamefont {J.}~\bibnamefont {Veitch}},\
  }\href {\doibase 10.1103/PhysRevD.106.044042} {\bibfield  {journal} {\bibinfo
   {journal} {Phys. Rev. D}\ }\textbf {\bibinfo {volume} {106}},\ \bibinfo
  {pages} {044042} (\bibinfo {year} {2022})},\ \Eprint
  {http://arxiv.org/abs/2205.08448} {arXiv:2205.08448 [gr-qc]} \BibitemShut
  {NoStop}%
\bibitem [{\citenamefont {Markovic}(1993)}]{Markovic:1993cr}%
  \BibitemOpen
  \bibfield  {author} {\bibinfo {author} {\bibfnamefont {D.}~\bibnamefont
  {Markovic}},\ }\href {\doibase 10.1103/PhysRevD.48.4738} {\bibfield
  {journal} {\bibinfo  {journal} {Phys. Rev. D}\ }\textbf {\bibinfo {volume}
  {48}},\ \bibinfo {pages} {4738} (\bibinfo {year} {1993})}\BibitemShut
  {NoStop}%
\bibitem [{\citenamefont {Cutler}\ and\ \citenamefont
  {Flanagan}(1994)}]{Cutler:1994ys}%
  \BibitemOpen
  \bibfield  {author} {\bibinfo {author} {\bibfnamefont {C.}~\bibnamefont
  {Cutler}}\ and\ \bibinfo {author} {\bibfnamefont {E.~E.}\ \bibnamefont
  {Flanagan}},\ }\href {\doibase 10.1103/PhysRevD.49.2658} {\bibfield
  {journal} {\bibinfo  {journal} {Phys. Rev. D}\ }\textbf {\bibinfo {volume}
  {49}},\ \bibinfo {pages} {2658} (\bibinfo {year} {1994})},\ \Eprint
  {http://arxiv.org/abs/gr-qc/9402014} {arXiv:gr-qc/9402014} \BibitemShut
  {NoStop}%
\bibitem [{\citenamefont {Owen}(1996)}]{Owen:1995tm}%
  \BibitemOpen
  \bibfield  {author} {\bibinfo {author} {\bibfnamefont {B.~J.}\ \bibnamefont
  {Owen}},\ }\href {\doibase 10.1103/PhysRevD.53.6749} {\bibfield  {journal}
  {\bibinfo  {journal} {Phys. Rev. D}\ }\textbf {\bibinfo {volume} {53}},\
  \bibinfo {pages} {6749} (\bibinfo {year} {1996})},\ \Eprint
  {http://arxiv.org/abs/gr-qc/9511032} {arXiv:gr-qc/9511032} \BibitemShut
  {NoStop}%
\bibitem [{\citenamefont {Hamilton}\ \emph {et~al.}(2024)\citenamefont
  {Hamilton} \emph {et~al.}}]{Hamilton:2023qkv}%
  \BibitemOpen
  \bibfield  {author} {\bibinfo {author} {\bibfnamefont {E.}~\bibnamefont
  {Hamilton}} \emph {et~al.},\ }\href {\doibase 10.1103/PhysRevD.109.044032}
  {\bibfield  {journal} {\bibinfo  {journal} {Phys. Rev. D}\ }\textbf {\bibinfo
  {volume} {109}},\ \bibinfo {pages} {044032} (\bibinfo {year} {2024})},\
  \Eprint {http://arxiv.org/abs/2303.05419} {arXiv:2303.05419 [gr-qc]}
  \BibitemShut {NoStop}%
\bibitem [{\citenamefont {Flanagan}\ and\ \citenamefont
  {Hughes}(1998)}]{Flanagan:1997kp}%
  \BibitemOpen
  \bibfield  {author} {\bibinfo {author} {\bibfnamefont {E.~E.}\ \bibnamefont
  {Flanagan}}\ and\ \bibinfo {author} {\bibfnamefont {S.~A.}\ \bibnamefont
  {Hughes}},\ }\href {\doibase 10.1103/PhysRevD.57.4566} {\bibfield  {journal}
  {\bibinfo  {journal} {Phys. Rev. D}\ }\textbf {\bibinfo {volume} {57}},\
  \bibinfo {pages} {4566} (\bibinfo {year} {1998})},\ \Eprint
  {http://arxiv.org/abs/gr-qc/9710129} {arXiv:gr-qc/9710129} \BibitemShut
  {NoStop}%
\bibitem [{\citenamefont {Ajith}\ \emph {et~al.}(2011)\citenamefont {Ajith}
  \emph {et~al.}}]{Ajith:2009bn}%
  \BibitemOpen
  \bibfield  {author} {\bibinfo {author} {\bibfnamefont {P.}~\bibnamefont
  {Ajith}} \emph {et~al.},\ }\href {\doibase 10.1103/PhysRevLett.106.241101}
  {\bibfield  {journal} {\bibinfo  {journal} {Phys. Rev. Lett.}\ }\textbf
  {\bibinfo {volume} {106}},\ \bibinfo {pages} {241101} (\bibinfo {year}
  {2011})},\ \Eprint {http://arxiv.org/abs/0909.2867} {arXiv:0909.2867 [gr-qc]}
  \BibitemShut {NoStop}%
\bibitem [{\citenamefont {Ohme}(2012)}]{Ohme:2012cba}%
  \BibitemOpen
  \bibfield  {author} {\bibinfo {author} {\bibfnamefont {F.}~\bibnamefont
  {Ohme}},\ }\emph {\bibinfo {title} {{Bridging the Gap between Post-Newtonian
  Theory and Numerical Relativity in Gravitational-Wave Data Analysis}}},\
  \href@noop {} {Ph.D. thesis},\ \bibinfo  {school} {Potsdam U.} (\bibinfo
  {year} {2012})\BibitemShut {NoStop}%
\bibitem [{\citenamefont {Chua}\ and\ \citenamefont
  {Cutler}(2022)}]{Chua:2021aah}%
  \BibitemOpen
  \bibfield  {author} {\bibinfo {author} {\bibfnamefont {A.~J.~K.}\
  \bibnamefont {Chua}}\ and\ \bibinfo {author} {\bibfnamefont {C.~J.}\
  \bibnamefont {Cutler}},\ }\href {\doibase 10.1103/PhysRevD.106.124046}
  {\bibfield  {journal} {\bibinfo  {journal} {Phys. Rev. D}\ }\textbf {\bibinfo
  {volume} {106}},\ \bibinfo {pages} {124046} (\bibinfo {year} {2022})},\
  \Eprint {http://arxiv.org/abs/2109.14254} {arXiv:2109.14254 [gr-qc]}
  \BibitemShut {NoStop}%
\bibitem [{\citenamefont {Roulet}\ \emph {et~al.}(2022)\citenamefont {Roulet},
  \citenamefont {Olsen}, \citenamefont {Mushkin}, \citenamefont {Islam},
  \citenamefont {Venumadhav}, \citenamefont {Zackay},\ and\ \citenamefont
  {Zaldarriaga}}]{Roulet:2022kot}%
  \BibitemOpen
  \bibfield  {author} {\bibinfo {author} {\bibfnamefont {J.}~\bibnamefont
  {Roulet}}, \bibinfo {author} {\bibfnamefont {S.}~\bibnamefont {Olsen}},
  \bibinfo {author} {\bibfnamefont {J.}~\bibnamefont {Mushkin}}, \bibinfo
  {author} {\bibfnamefont {T.}~\bibnamefont {Islam}}, \bibinfo {author}
  {\bibfnamefont {T.}~\bibnamefont {Venumadhav}}, \bibinfo {author}
  {\bibfnamefont {B.}~\bibnamefont {Zackay}}, \ and\ \bibinfo {author}
  {\bibfnamefont {M.}~\bibnamefont {Zaldarriaga}},\ }\href {\doibase
  10.1103/PhysRevD.106.123015} {\bibfield  {journal} {\bibinfo  {journal}
  {Phys. Rev. D}\ }\textbf {\bibinfo {volume} {106}},\ \bibinfo {pages}
  {123015} (\bibinfo {year} {2022})},\ \Eprint
  {http://arxiv.org/abs/2207.03508} {arXiv:2207.03508 [gr-qc]} \BibitemShut
  {NoStop}%
\bibitem [{\citenamefont {Bruegmann}\ \emph {et~al.}(2008)\citenamefont
  {Bruegmann}, \citenamefont {Gonzalez}, \citenamefont {Hannam}, \citenamefont
  {Husa}, \citenamefont {Sperhake},\ and\ \citenamefont
  {Tichy}}]{Bruegmann:2006ulg}%
  \BibitemOpen
  \bibfield  {author} {\bibinfo {author} {\bibfnamefont {B.}~\bibnamefont
  {Bruegmann}}, \bibinfo {author} {\bibfnamefont {J.~A.}\ \bibnamefont
  {Gonzalez}}, \bibinfo {author} {\bibfnamefont {M.}~\bibnamefont {Hannam}},
  \bibinfo {author} {\bibfnamefont {S.}~\bibnamefont {Husa}}, \bibinfo {author}
  {\bibfnamefont {U.}~\bibnamefont {Sperhake}}, \ and\ \bibinfo {author}
  {\bibfnamefont {W.}~\bibnamefont {Tichy}},\ }\href {\doibase
  10.1103/PhysRevD.77.024027} {\bibfield  {journal} {\bibinfo  {journal} {Phys.
  Rev. D}\ }\textbf {\bibinfo {volume} {77}},\ \bibinfo {pages} {024027}
  (\bibinfo {year} {2008})},\ \Eprint {http://arxiv.org/abs/gr-qc/0610128}
  {arXiv:gr-qc/0610128} \BibitemShut {NoStop}%
\bibitem [{\citenamefont {Husa}\ \emph {et~al.}(2016)\citenamefont {Husa},
  \citenamefont {Khan}, \citenamefont {Hannam}, \citenamefont {P\"urrer},
  \citenamefont {Ohme}, \citenamefont {Jim\'enez~Forteza},\ and\ \citenamefont
  {Boh\'e}}]{Husa:2015iqa}%
  \BibitemOpen
  \bibfield  {author} {\bibinfo {author} {\bibfnamefont {S.}~\bibnamefont
  {Husa}}, \bibinfo {author} {\bibfnamefont {S.}~\bibnamefont {Khan}}, \bibinfo
  {author} {\bibfnamefont {M.}~\bibnamefont {Hannam}}, \bibinfo {author}
  {\bibfnamefont {M.}~\bibnamefont {P\"urrer}}, \bibinfo {author}
  {\bibfnamefont {F.}~\bibnamefont {Ohme}}, \bibinfo {author} {\bibfnamefont
  {X.}~\bibnamefont {Jim\'enez~Forteza}}, \ and\ \bibinfo {author}
  {\bibfnamefont {A.}~\bibnamefont {Boh\'e}},\ }\href {\doibase
  10.1103/PhysRevD.93.044006} {\bibfield  {journal} {\bibinfo  {journal} {Phys.
  Rev. D}\ }\textbf {\bibinfo {volume} {93}},\ \bibinfo {pages} {044006}
  (\bibinfo {year} {2016})},\ \Eprint {http://arxiv.org/abs/1508.07250}
  {arXiv:1508.07250 [gr-qc]} \BibitemShut {NoStop}%
\bibitem [{\citenamefont {Khan}\ \emph {et~al.}(2016)\citenamefont {Khan},
  \citenamefont {Husa}, \citenamefont {Hannam}, \citenamefont {Ohme},
  \citenamefont {P\"urrer}, \citenamefont {Jim\'enez~Forteza},\ and\
  \citenamefont {Boh\'e}}]{Khan:2015jqa}%
  \BibitemOpen
  \bibfield  {author} {\bibinfo {author} {\bibfnamefont {S.}~\bibnamefont
  {Khan}}, \bibinfo {author} {\bibfnamefont {S.}~\bibnamefont {Husa}}, \bibinfo
  {author} {\bibfnamefont {M.}~\bibnamefont {Hannam}}, \bibinfo {author}
  {\bibfnamefont {F.}~\bibnamefont {Ohme}}, \bibinfo {author} {\bibfnamefont
  {M.}~\bibnamefont {P\"urrer}}, \bibinfo {author} {\bibfnamefont
  {X.}~\bibnamefont {Jim\'enez~Forteza}}, \ and\ \bibinfo {author}
  {\bibfnamefont {A.}~\bibnamefont {Boh\'e}},\ }\href {\doibase
  10.1103/PhysRevD.93.044007} {\bibfield  {journal} {\bibinfo  {journal} {Phys.
  Rev. D}\ }\textbf {\bibinfo {volume} {93}},\ \bibinfo {pages} {044007}
  (\bibinfo {year} {2016})},\ \Eprint {http://arxiv.org/abs/1508.07253}
  {arXiv:1508.07253 [gr-qc]} \BibitemShut {NoStop}%
\bibitem [{\citenamefont {Kalaghatgi}\ \emph {et~al.}(2020)\citenamefont
  {Kalaghatgi}, \citenamefont {Hannam},\ and\ \citenamefont
  {Raymond}}]{Kalaghatgi:2019log}%
  \BibitemOpen
  \bibfield  {author} {\bibinfo {author} {\bibfnamefont {C.}~\bibnamefont
  {Kalaghatgi}}, \bibinfo {author} {\bibfnamefont {M.}~\bibnamefont {Hannam}},
  \ and\ \bibinfo {author} {\bibfnamefont {V.}~\bibnamefont {Raymond}},\ }\href
  {\doibase 10.1103/PhysRevD.101.103004} {\bibfield  {journal} {\bibinfo
  {journal} {Phys. Rev. D}\ }\textbf {\bibinfo {volume} {101}},\ \bibinfo
  {pages} {103004} (\bibinfo {year} {2020})},\ \Eprint
  {http://arxiv.org/abs/1909.10010} {arXiv:1909.10010 [gr-qc]} \BibitemShut
  {NoStop}%
\bibitem [{\citenamefont {Varma}\ \emph
  {et~al.}(2019{\natexlab{b}})\citenamefont {Varma}, \citenamefont {Field},
  \citenamefont {Scheel}, \citenamefont {Blackman}, \citenamefont {Kidder},\
  and\ \citenamefont {Pfeiffer}}]{Varma:2018mmi}%
  \BibitemOpen
  \bibfield  {author} {\bibinfo {author} {\bibfnamefont {V.}~\bibnamefont
  {Varma}}, \bibinfo {author} {\bibfnamefont {S.~E.}\ \bibnamefont {Field}},
  \bibinfo {author} {\bibfnamefont {M.~A.}\ \bibnamefont {Scheel}}, \bibinfo
  {author} {\bibfnamefont {J.}~\bibnamefont {Blackman}}, \bibinfo {author}
  {\bibfnamefont {L.~E.}\ \bibnamefont {Kidder}}, \ and\ \bibinfo {author}
  {\bibfnamefont {H.~P.}\ \bibnamefont {Pfeiffer}},\ }\href {\doibase
  10.1103/PhysRevD.99.064045} {\bibfield  {journal} {\bibinfo  {journal} {Phys.
  Rev. D}\ }\textbf {\bibinfo {volume} {99}},\ \bibinfo {pages} {064045}
  (\bibinfo {year} {2019}{\natexlab{b}})},\ \Eprint
  {http://arxiv.org/abs/1812.07865} {arXiv:1812.07865 [gr-qc]} \BibitemShut
  {NoStop}%
\bibitem [{\citenamefont {Balasubramanian}\ \emph {et~al.}(1996)\citenamefont
  {Balasubramanian}, \citenamefont {Sathyaprakash},\ and\ \citenamefont
  {Dhurandhar}}]{Balasubramanian:1995bm}%
  \BibitemOpen
  \bibfield  {author} {\bibinfo {author} {\bibfnamefont {R.}~\bibnamefont
  {Balasubramanian}}, \bibinfo {author} {\bibfnamefont {B.~S.}\ \bibnamefont
  {Sathyaprakash}}, \ and\ \bibinfo {author} {\bibfnamefont {S.~V.}\
  \bibnamefont {Dhurandhar}},\ }\href {\doibase 10.1103/PhysRevD.53.3033}
  {\bibfield  {journal} {\bibinfo  {journal} {Phys. Rev. D}\ }\textbf {\bibinfo
  {volume} {53}},\ \bibinfo {pages} {3033} (\bibinfo {year} {1996})},\ \bibinfo
  {note} {[Erratum: Phys.Rev.D 54, 1860 (1996)]},\ \Eprint
  {http://arxiv.org/abs/gr-qc/9508011} {arXiv:gr-qc/9508011} \BibitemShut
  {NoStop}%
\bibitem [{\citenamefont {Ajith}\ \emph {et~al.}(2012)\citenamefont {Ajith}
  \emph {et~al.}}]{Ajith:2012az}%
  \BibitemOpen
  \bibfield  {author} {\bibinfo {author} {\bibfnamefont {P.}~\bibnamefont
  {Ajith}} \emph {et~al.},\ }\href {\doibase 10.1088/0264-9381/29/12/124001}
  {\bibfield  {journal} {\bibinfo  {journal} {Class. Quant. Grav.}\ }\textbf
  {\bibinfo {volume} {29}},\ \bibinfo {pages} {124001} (\bibinfo {year}
  {2012})},\ \bibinfo {note} {[Erratum: Class.Quant.Grav. 30, 199401 (2013)]},\
  \Eprint {http://arxiv.org/abs/1201.5319} {arXiv:1201.5319 [gr-qc]}
  \BibitemShut {NoStop}%
\bibitem [{\citenamefont {Nelder}\ and\ \citenamefont
  {Mead}(1965)}]{10.1093/comjnl/7.4.308}%
  \BibitemOpen
  \bibfield  {author} {\bibinfo {author} {\bibfnamefont {J.~A.}\ \bibnamefont
  {Nelder}}\ and\ \bibinfo {author} {\bibfnamefont {R.}~\bibnamefont {Mead}},\
  }\href {\doibase 10.1093/comjnl/7.4.308} {\bibfield  {journal} {\bibinfo
  {journal} {The Computer Journal}\ }\textbf {\bibinfo {volume} {7}},\ \bibinfo
  {pages} {308} (\bibinfo {year} {1965})},\ \Eprint
  {http://arxiv.org/abs/https://academic.oup.com/comjnl/article-pdf/7/4/308/1013182/7-4-308.pdf}
  {https://academic.oup.com/comjnl/article-pdf/7/4/308/1013182/7-4-308.pdf}
  \BibitemShut {NoStop}%
\bibitem [{\citenamefont {Virtanen}\ \emph {et~al.}(2020)\citenamefont
  {Virtanen}, \citenamefont {Gommers}, \citenamefont {Oliphant}, \citenamefont
  {Haberland}, \citenamefont {Reddy}, \citenamefont {Cournapeau}, \citenamefont
  {Burovski}, \citenamefont {Peterson}, \citenamefont {Weckesser},
  \citenamefont {Bright}, \citenamefont {{van der Walt}}, \citenamefont
  {Brett}, \citenamefont {Wilson}, \citenamefont {Millman}, \citenamefont
  {Mayorov}, \citenamefont {Nelson}, \citenamefont {Jones}, \citenamefont
  {Kern}, \citenamefont {Larson}, \citenamefont {Carey}, \citenamefont {Polat},
  \citenamefont {Feng}, \citenamefont {Moore}, \citenamefont {{VanderPlas}},
  \citenamefont {Laxalde}, \citenamefont {Perktold}, \citenamefont {Cimrman},
  \citenamefont {Henriksen}, \citenamefont {Quintero}, \citenamefont {Harris},
  \citenamefont {Archibald}, \citenamefont {Ribeiro}, \citenamefont
  {Pedregosa}, \citenamefont {{van Mulbregt}},\ and\ \citenamefont {{SciPy 1.0
  Contributors}}}]{2020SciPy-NMeth}%
  \BibitemOpen
  \bibfield  {author} {\bibinfo {author} {\bibfnamefont {P.}~\bibnamefont
  {Virtanen}}, \bibinfo {author} {\bibfnamefont {R.}~\bibnamefont {Gommers}},
  \bibinfo {author} {\bibfnamefont {T.~E.}\ \bibnamefont {Oliphant}}, \bibinfo
  {author} {\bibfnamefont {M.}~\bibnamefont {Haberland}}, \bibinfo {author}
  {\bibfnamefont {T.}~\bibnamefont {Reddy}}, \bibinfo {author} {\bibfnamefont
  {D.}~\bibnamefont {Cournapeau}}, \bibinfo {author} {\bibfnamefont
  {E.}~\bibnamefont {Burovski}}, \bibinfo {author} {\bibfnamefont
  {P.}~\bibnamefont {Peterson}}, \bibinfo {author} {\bibfnamefont
  {W.}~\bibnamefont {Weckesser}}, \bibinfo {author} {\bibfnamefont
  {J.}~\bibnamefont {Bright}}, \bibinfo {author} {\bibfnamefont {S.~J.}\
  \bibnamefont {{van der Walt}}}, \bibinfo {author} {\bibfnamefont
  {M.}~\bibnamefont {Brett}}, \bibinfo {author} {\bibfnamefont
  {J.}~\bibnamefont {Wilson}}, \bibinfo {author} {\bibfnamefont {K.~J.}\
  \bibnamefont {Millman}}, \bibinfo {author} {\bibfnamefont {N.}~\bibnamefont
  {Mayorov}}, \bibinfo {author} {\bibfnamefont {A.~R.~J.}\ \bibnamefont
  {Nelson}}, \bibinfo {author} {\bibfnamefont {E.}~\bibnamefont {Jones}},
  \bibinfo {author} {\bibfnamefont {R.}~\bibnamefont {Kern}}, \bibinfo {author}
  {\bibfnamefont {E.}~\bibnamefont {Larson}}, \bibinfo {author} {\bibfnamefont
  {C.~J.}\ \bibnamefont {Carey}}, \bibinfo {author} {\bibfnamefont
  {{\.I}.}~\bibnamefont {Polat}}, \bibinfo {author} {\bibfnamefont
  {Y.}~\bibnamefont {Feng}}, \bibinfo {author} {\bibfnamefont {E.~W.}\
  \bibnamefont {Moore}}, \bibinfo {author} {\bibfnamefont {J.}~\bibnamefont
  {{VanderPlas}}}, \bibinfo {author} {\bibfnamefont {D.}~\bibnamefont
  {Laxalde}}, \bibinfo {author} {\bibfnamefont {J.}~\bibnamefont {Perktold}},
  \bibinfo {author} {\bibfnamefont {R.}~\bibnamefont {Cimrman}}, \bibinfo
  {author} {\bibfnamefont {I.}~\bibnamefont {Henriksen}}, \bibinfo {author}
  {\bibfnamefont {E.~A.}\ \bibnamefont {Quintero}}, \bibinfo {author}
  {\bibfnamefont {C.~R.}\ \bibnamefont {Harris}}, \bibinfo {author}
  {\bibfnamefont {A.~M.}\ \bibnamefont {Archibald}}, \bibinfo {author}
  {\bibfnamefont {A.~H.}\ \bibnamefont {Ribeiro}}, \bibinfo {author}
  {\bibfnamefont {F.}~\bibnamefont {Pedregosa}}, \bibinfo {author}
  {\bibfnamefont {P.}~\bibnamefont {{van Mulbregt}}}, \ and\ \bibinfo {author}
  {\bibnamefont {{SciPy 1.0 Contributors}}},\ }\href {\doibase
  10.1038/s41592-019-0686-2} {\bibfield  {journal} {\bibinfo  {journal} {Nature
  Methods}\ }\textbf {\bibinfo {volume} {17}},\ \bibinfo {pages} {261}
  (\bibinfo {year} {2020})}\BibitemShut {NoStop}%
\bibitem [{\citenamefont {Veitch}\ and\ \citenamefont
  {Del~Pozzo}(2013)}]{Veitch:2013aaa}%
  \BibitemOpen
  \bibfield  {author} {\bibinfo {author} {\bibfnamefont {J.}~\bibnamefont
  {Veitch}}\ and\ \bibinfo {author} {\bibfnamefont {W.}~\bibnamefont
  {Del~Pozzo}},\ }\href {https://dcc.ligo.org/LIGO-T1300326} {\emph {\bibinfo
  {title} {{Analytic Marginalisation of Phase Parameter}}}},\ \bibinfo {type}
  {Tech. Rep.}\ \bibinfo {number} {T1300326}\ (\bibinfo {year}
  {2013})\BibitemShut {NoStop}%
\bibitem [{\citenamefont {Farr}(2014)}]{Farr:2014aaa}%
  \BibitemOpen
  \bibfield  {author} {\bibinfo {author} {\bibfnamefont {W.~M.}\ \bibnamefont
  {Farr}},\ }\href {https://dcc.ligo.org/LIGO-T1400460} {\emph {\bibinfo
  {title} {{Marginalization of the time and phase parameters in CBC parameter
  estimation}}}},\ \bibinfo {type} {Tech. Rep.}\ \bibinfo {number} {T1400460}\
  (\bibinfo {year} {2014})\BibitemShut {NoStop}%
\bibitem [{\citenamefont {Singer}\ and\ \citenamefont
  {Price}(2016)}]{Singer:2015ema}%
  \BibitemOpen
  \bibfield  {author} {\bibinfo {author} {\bibfnamefont {L.~P.}\ \bibnamefont
  {Singer}}\ and\ \bibinfo {author} {\bibfnamefont {L.~R.}\ \bibnamefont
  {Price}},\ }\href {\doibase 10.1103/PhysRevD.93.024013} {\bibfield  {journal}
  {\bibinfo  {journal} {Phys. Rev. D}\ }\textbf {\bibinfo {volume} {93}},\
  \bibinfo {pages} {024013} (\bibinfo {year} {2016})},\ \Eprint
  {http://arxiv.org/abs/1508.03634} {arXiv:1508.03634 [gr-qc]} \BibitemShut
  {NoStop}%
\bibitem [{\citenamefont {Singer}\ \emph {et~al.}(2016)\citenamefont {Singer}
  \emph {et~al.}}]{Singer:2016eax}%
  \BibitemOpen
  \bibfield  {author} {\bibinfo {author} {\bibfnamefont {L.~P.}\ \bibnamefont
  {Singer}} \emph {et~al.},\ }\href {\doibase 10.3847/2041-8205/829/1/L15}
  {\bibfield  {journal} {\bibinfo  {journal} {Astrophys. J. Lett.}\ }\textbf
  {\bibinfo {volume} {829}},\ \bibinfo {pages} {L15} (\bibinfo {year}
  {2016})},\ \Eprint {http://arxiv.org/abs/1603.07333} {arXiv:1603.07333
  [astro-ph.HE]} \BibitemShut {NoStop}%
\bibitem [{\citenamefont {Thrane}\ and\ \citenamefont
  {Talbot}(2019)}]{Thrane:2018qnx}%
  \BibitemOpen
  \bibfield  {author} {\bibinfo {author} {\bibfnamefont {E.}~\bibnamefont
  {Thrane}}\ and\ \bibinfo {author} {\bibfnamefont {C.}~\bibnamefont
  {Talbot}},\ }\href {\doibase 10.1017/pasa.2019.2} {\bibfield  {journal}
  {\bibinfo  {journal} {Publ. Astron. Soc. Austral.}\ }\textbf {\bibinfo
  {volume} {36}},\ \bibinfo {pages} {e010} (\bibinfo {year} {2019})},\ \bibinfo
  {note} {[Erratum: Publ.Astron.Soc.Austral. 37, e036 (2020)]},\ \Eprint
  {http://arxiv.org/abs/1809.02293} {arXiv:1809.02293 [astro-ph.IM]}
  \BibitemShut {NoStop}%
\bibitem [{\citenamefont {Metropolis}\ and\ \citenamefont
  {Ulam}(1949)}]{metropolis1949monte}%
  \BibitemOpen
  \bibfield  {author} {\bibinfo {author} {\bibfnamefont {N.}~\bibnamefont
  {Metropolis}}\ and\ \bibinfo {author} {\bibfnamefont {S.}~\bibnamefont
  {Ulam}},\ }\href@noop {} {\bibfield  {journal} {\bibinfo  {journal} {Journal
  of the American statistical association}\ }\textbf {\bibinfo {volume} {44}},\
  \bibinfo {pages} {335} (\bibinfo {year} {1949})}\BibitemShut {NoStop}%
\bibitem [{\citenamefont {Skilling}(2004)}]{Skilling2004}%
  \BibitemOpen
  \bibfield  {author} {\bibinfo {author} {\bibfnamefont {J.}~\bibnamefont
  {Skilling}},\ }in\ \href {\doibase 10.1063/1.1835238} {\emph {\bibinfo
  {booktitle} {{AIP} Conference Proceedings}}}\ (\bibinfo  {publisher}
  {{AIP}},\ \bibinfo {year} {2004})\BibitemShut {NoStop}%
\bibitem [{\citenamefont {Skilling}(2006)}]{Skilling:2006}%
  \BibitemOpen
  \bibfield  {author} {\bibinfo {author} {\bibfnamefont {J.}~\bibnamefont
  {Skilling}},\ }\href {\doibase 10.1214/06-BA127} {\bibfield  {journal}
  {\bibinfo  {journal} {Bayesian Anal.}\ }\textbf {\bibinfo {volume} {1}},\
  \bibinfo {pages} {833} (\bibinfo {year} {2006})}\BibitemShut {NoStop}%
\bibitem [{\citenamefont {Veitch}\ \emph {et~al.}(2015)\citenamefont {Veitch}
  \emph {et~al.}}]{Veitch:2014wba}%
  \BibitemOpen
  \bibfield  {author} {\bibinfo {author} {\bibfnamefont {J.}~\bibnamefont
  {Veitch}} \emph {et~al.},\ }\href {\doibase 10.1103/PhysRevD.91.042003}
  {\bibfield  {journal} {\bibinfo  {journal} {Phys. Rev.}\ }\textbf {\bibinfo
  {volume} {D91}},\ \bibinfo {pages} {042003} (\bibinfo {year} {2015})},\
  \Eprint {http://arxiv.org/abs/1409.7215} {arXiv:1409.7215 [gr-qc]}
  \BibitemShut {NoStop}%
%%CITATION = ARXIV:1409.7215;%%
\bibitem [{\citenamefont {Lange}\ \emph {et~al.}(2018)\citenamefont {Lange},
  \citenamefont {O'Shaughnessy},\ and\ \citenamefont {Rizzo}}]{Lange:2018pyp}%
  \BibitemOpen
  \bibfield  {author} {\bibinfo {author} {\bibfnamefont {J.}~\bibnamefont
  {Lange}}, \bibinfo {author} {\bibfnamefont {R.}~\bibnamefont
  {O'Shaughnessy}}, \ and\ \bibinfo {author} {\bibfnamefont {M.}~\bibnamefont
  {Rizzo}},\ }\href@noop {} {\  (\bibinfo {year} {2018})},\ \Eprint
  {http://arxiv.org/abs/1805.10457} {arXiv:1805.10457 [gr-qc]} \BibitemShut
  {NoStop}%
\bibitem [{\citenamefont {Ashton}\ \emph {et~al.}(2019)\citenamefont {Ashton}
  \emph {et~al.}}]{Ashton:2018jfp}%
  \BibitemOpen
  \bibfield  {author} {\bibinfo {author} {\bibfnamefont {G.}~\bibnamefont
  {Ashton}} \emph {et~al.},\ }\href {\doibase 10.3847/1538-4365/ab06fc}
  {\bibfield  {journal} {\bibinfo  {journal} {Astrophys. J. Suppl.}\ }\textbf
  {\bibinfo {volume} {241}},\ \bibinfo {pages} {27} (\bibinfo {year} {2019})},\
  \Eprint {http://arxiv.org/abs/1811.02042} {arXiv:1811.02042 [astro-ph.IM]}
  \BibitemShut {NoStop}%
\bibitem [{\citenamefont {Biwer}\ \emph {et~al.}(2019)\citenamefont {Biwer},
  \citenamefont {Capano}, \citenamefont {De}, \citenamefont {Cabero},
  \citenamefont {Brown}, \citenamefont {Nitz},\ and\ \citenamefont
  {Raymond}}]{Biwer:2018osg}%
  \BibitemOpen
  \bibfield  {author} {\bibinfo {author} {\bibfnamefont {C.~M.}\ \bibnamefont
  {Biwer}}, \bibinfo {author} {\bibfnamefont {C.~D.}\ \bibnamefont {Capano}},
  \bibinfo {author} {\bibfnamefont {S.}~\bibnamefont {De}}, \bibinfo {author}
  {\bibfnamefont {M.}~\bibnamefont {Cabero}}, \bibinfo {author} {\bibfnamefont
  {D.~A.}\ \bibnamefont {Brown}}, \bibinfo {author} {\bibfnamefont {A.~H.}\
  \bibnamefont {Nitz}}, \ and\ \bibinfo {author} {\bibfnamefont
  {V.}~\bibnamefont {Raymond}},\ }\href {\doibase 10.1088/1538-3873/aaef0b}
  {\bibfield  {journal} {\bibinfo  {journal} {Publ. Astron. Soc. Pac.}\
  }\textbf {\bibinfo {volume} {131}},\ \bibinfo {pages} {024503} (\bibinfo
  {year} {2019})},\ \Eprint {http://arxiv.org/abs/1807.10312} {arXiv:1807.10312
  [astro-ph.IM]} \BibitemShut {NoStop}%
\bibitem [{\citenamefont {Smith}\ \emph {et~al.}(2020)\citenamefont {Smith},
  \citenamefont {Ashton}, \citenamefont {Vajpeyi},\ and\ \citenamefont
  {Talbot}}]{Smith:2019ucc}%
  \BibitemOpen
  \bibfield  {author} {\bibinfo {author} {\bibfnamefont {R.~J.~E.}\
  \bibnamefont {Smith}}, \bibinfo {author} {\bibfnamefont {G.}~\bibnamefont
  {Ashton}}, \bibinfo {author} {\bibfnamefont {A.}~\bibnamefont {Vajpeyi}}, \
  and\ \bibinfo {author} {\bibfnamefont {C.}~\bibnamefont {Talbot}},\ }\href
  {\doibase 10.1093/mnras/staa2483} {\bibfield  {journal} {\bibinfo  {journal}
  {Mon. Not. Roy. Astron. Soc.}\ }\textbf {\bibinfo {volume} {498}},\ \bibinfo
  {pages} {4492} (\bibinfo {year} {2020})},\ \Eprint
  {http://arxiv.org/abs/1909.11873} {arXiv:1909.11873 [gr-qc]} \BibitemShut
  {NoStop}%
\bibitem [{\citenamefont {Ashton}\ and\ \citenamefont
  {Talbot}(2021)}]{Ashton:2021anp}%
  \BibitemOpen
  \bibfield  {author} {\bibinfo {author} {\bibfnamefont {G.}~\bibnamefont
  {Ashton}}\ and\ \bibinfo {author} {\bibfnamefont {C.}~\bibnamefont
  {Talbot}},\ }\href {\doibase 10.1093/mnras/stab2236} {\bibfield  {journal}
  {\bibinfo  {journal} {Mon. Not. Roy. Astron. Soc.}\ }\textbf {\bibinfo
  {volume} {507}},\ \bibinfo {pages} {2037} (\bibinfo {year} {2021})},\ \Eprint
  {http://arxiv.org/abs/2106.08730} {arXiv:2106.08730 [gr-qc]} \BibitemShut
  {NoStop}%
\bibitem [{\citenamefont {Dax}\ \emph {et~al.}(2021)\citenamefont {Dax},
  \citenamefont {Green}, \citenamefont {Gair}, \citenamefont {Macke},
  \citenamefont {Buonanno},\ and\ \citenamefont {Sch\"olkopf}}]{Dax:2021tsq}%
  \BibitemOpen
  \bibfield  {author} {\bibinfo {author} {\bibfnamefont {M.}~\bibnamefont
  {Dax}}, \bibinfo {author} {\bibfnamefont {S.~R.}\ \bibnamefont {Green}},
  \bibinfo {author} {\bibfnamefont {J.}~\bibnamefont {Gair}}, \bibinfo {author}
  {\bibfnamefont {J.~H.}\ \bibnamefont {Macke}}, \bibinfo {author}
  {\bibfnamefont {A.}~\bibnamefont {Buonanno}}, \ and\ \bibinfo {author}
  {\bibfnamefont {B.}~\bibnamefont {Sch\"olkopf}},\ }\href {\doibase
  10.1103/PhysRevLett.127.241103} {\bibfield  {journal} {\bibinfo  {journal}
  {Phys. Rev. Lett.}\ }\textbf {\bibinfo {volume} {127}},\ \bibinfo {pages}
  {241103} (\bibinfo {year} {2021})},\ \Eprint
  {http://arxiv.org/abs/2106.12594} {arXiv:2106.12594 [gr-qc]} \BibitemShut
  {NoStop}%
\bibitem [{\citenamefont {Tiwari}\ \emph {et~al.}(2023)\citenamefont {Tiwari},
  \citenamefont {Hoy}, \citenamefont {Fairhurst},\ and\ \citenamefont
  {MacLeod}}]{Tiwari:2023mzf}%
  \BibitemOpen
  \bibfield  {author} {\bibinfo {author} {\bibfnamefont {V.}~\bibnamefont
  {Tiwari}}, \bibinfo {author} {\bibfnamefont {C.}~\bibnamefont {Hoy}},
  \bibinfo {author} {\bibfnamefont {S.}~\bibnamefont {Fairhurst}}, \ and\
  \bibinfo {author} {\bibfnamefont {D.}~\bibnamefont {MacLeod}},\ }\href
  {\doibase 10.1103/PhysRevD.108.023001} {\bibfield  {journal} {\bibinfo
  {journal} {Phys. Rev. D}\ }\textbf {\bibinfo {volume} {108}},\ \bibinfo
  {pages} {023001} (\bibinfo {year} {2023})},\ \Eprint
  {http://arxiv.org/abs/2303.01463} {arXiv:2303.01463 [astro-ph.HE]}
  \BibitemShut {NoStop}%
\bibitem [{\citenamefont {Speagle}(2020)}]{Speagle:2020}%
  \BibitemOpen
  \bibfield  {author} {\bibinfo {author} {\bibfnamefont {J.~S.}\ \bibnamefont
  {Speagle}},\ }\href {\doibase 10.1093/mnras/staa278} {\bibfield  {journal}
  {\bibinfo  {journal} {Monthly Notices of the Royal Astronomical Society}\
  }\textbf {\bibinfo {volume} {493}},\ \bibinfo {pages} {3132?3158} (\bibinfo
  {year} {2020})}\BibitemShut {NoStop}%
\bibitem [{\citenamefont {Branchesi}\ \emph {et~al.}(2023)\citenamefont
  {Branchesi} \emph {et~al.}}]{Branchesi:2023mws}%
  \BibitemOpen
  \bibfield  {author} {\bibinfo {author} {\bibfnamefont {M.}~\bibnamefont
  {Branchesi}} \emph {et~al.},\ }\href {\doibase 10.1088/1475-7516/2023/07/068}
  {\bibfield  {journal} {\bibinfo  {journal} {JCAP}\ }\textbf {\bibinfo
  {volume} {07}},\ \bibinfo {pages} {068} (\bibinfo {year} {2023})},\ \Eprint
  {http://arxiv.org/abs/2303.15923} {arXiv:2303.15923 [gr-qc]} \BibitemShut
  {NoStop}%
\bibitem [{\citenamefont {Henze}\ and\ \citenamefont
  {Zirkler}(1990)}]{henze1990class}%
  \BibitemOpen
  \bibfield  {author} {\bibinfo {author} {\bibfnamefont {N.}~\bibnamefont
  {Henze}}\ and\ \bibinfo {author} {\bibfnamefont {B.}~\bibnamefont
  {Zirkler}},\ }\href@noop {} {\bibfield  {journal} {\bibinfo  {journal}
  {Communications in statistics-Theory and Methods}\ }\textbf {\bibinfo
  {volume} {19}},\ \bibinfo {pages} {3595} (\bibinfo {year}
  {1990})}\BibitemShut {NoStop}%
\bibitem [{\citenamefont {Vallat}(2018)}]{Vallat2018}%
  \BibitemOpen
  \bibfield  {author} {\bibinfo {author} {\bibfnamefont {R.}~\bibnamefont
  {Vallat}},\ }\href {\doibase 10.21105/joss.01026} {\bibfield  {journal}
  {\bibinfo  {journal} {Journal of Open Source Software}\ }\textbf {\bibinfo
  {volume} {3}},\ \bibinfo {pages} {1026} (\bibinfo {year} {2018})}\BibitemShut
  {NoStop}%
\bibitem [{\citenamefont {{Lin}}(1991)}]{61115}%
  \BibitemOpen
  \bibfield  {author} {\bibinfo {author} {\bibfnamefont {J.}~\bibnamefont
  {{Lin}}},\ }\href {\doibase 10.1109/18.61115} {\bibfield  {journal} {\bibinfo
   {journal} {IEEE Transactions on Information Theory}\ }\textbf {\bibinfo
  {volume} {37}},\ \bibinfo {pages} {145} (\bibinfo {year} {1991})}\BibitemShut
  {NoStop}%
\bibitem [{\citenamefont {Birnbaum}\ and\ \citenamefont
  {Meyer}(1951)}]{birnbaum1951effect}%
  \BibitemOpen
  \bibfield  {author} {\bibinfo {author} {\bibfnamefont {Z.~W.}\ \bibnamefont
  {Birnbaum}}\ and\ \bibinfo {author} {\bibfnamefont {P.~L.}\ \bibnamefont
  {Meyer}},\ }\href@noop {} {\emph {\bibinfo {title} {On the effect of
  truncation in some or all coordinates of a multinormal population}}}\
  (\bibinfo  {publisher} {Laboratory of Statistical Research, Department of
  Mathematics, University of~…},\ \bibinfo {year} {1951})\BibitemShut
  {NoStop}%
\bibitem [{\citenamefont {Tallis}(1961)}]{tallis1961moment}%
  \BibitemOpen
  \bibfield  {author} {\bibinfo {author} {\bibfnamefont {G.~M.}\ \bibnamefont
  {Tallis}},\ }\href@noop {} {\bibfield  {journal} {\bibinfo  {journal}
  {Journal of the Royal Statistical Society Series B: Statistical Methodology}\
  }\textbf {\bibinfo {volume} {23}},\ \bibinfo {pages} {223} (\bibinfo {year}
  {1961})}\BibitemShut {NoStop}%
\bibitem [{\citenamefont {Ferguson}\ \emph {et~al.}(2021)\citenamefont
  {Ferguson}, \citenamefont {Jani}, \citenamefont {Laguna},\ and\ \citenamefont
  {Shoemaker}}]{Ferguson:2020xnm}%
  \BibitemOpen
  \bibfield  {author} {\bibinfo {author} {\bibfnamefont {D.}~\bibnamefont
  {Ferguson}}, \bibinfo {author} {\bibfnamefont {K.}~\bibnamefont {Jani}},
  \bibinfo {author} {\bibfnamefont {P.}~\bibnamefont {Laguna}}, \ and\ \bibinfo
  {author} {\bibfnamefont {D.}~\bibnamefont {Shoemaker}},\ }\href {\doibase
  10.1103/PhysRevD.104.044037} {\bibfield  {journal} {\bibinfo  {journal}
  {Phys. Rev. D}\ }\textbf {\bibinfo {volume} {104}},\ \bibinfo {pages}
  {044037} (\bibinfo {year} {2021})},\ \Eprint
  {http://arxiv.org/abs/2006.04272} {arXiv:2006.04272 [gr-qc]} \BibitemShut
  {NoStop}%
\bibitem [{\citenamefont {Jan}\ \emph {et~al.}(2024)\citenamefont {Jan},
  \citenamefont {Ferguson}, \citenamefont {Lange}, \citenamefont {Shoemaker},\
  and\ \citenamefont {Zimmerman}}]{Jan:2023raq}%
  \BibitemOpen
  \bibfield  {author} {\bibinfo {author} {\bibfnamefont {A.}~\bibnamefont
  {Jan}}, \bibinfo {author} {\bibfnamefont {D.}~\bibnamefont {Ferguson}},
  \bibinfo {author} {\bibfnamefont {J.}~\bibnamefont {Lange}}, \bibinfo
  {author} {\bibfnamefont {D.}~\bibnamefont {Shoemaker}}, \ and\ \bibinfo
  {author} {\bibfnamefont {A.}~\bibnamefont {Zimmerman}},\ }\href {\doibase
  10.1103/PhysRevD.110.024023} {\bibfield  {journal} {\bibinfo  {journal}
  {Phys. Rev. D}\ }\textbf {\bibinfo {volume} {110}},\ \bibinfo {pages}
  {024023} (\bibinfo {year} {2024})},\ \Eprint
  {http://arxiv.org/abs/2312.10241} {arXiv:2312.10241 [gr-qc]} \BibitemShut
  {NoStop}%
\bibitem [{\citenamefont {Thompson}\ \emph {et~al.}(2024)\citenamefont
  {Thompson}, \citenamefont {Hamilton}, \citenamefont {London}, \citenamefont
  {Ghosh}, \citenamefont {Kolitsidou}, \citenamefont {Hoy},\ and\ \citenamefont
  {Hannam}}]{Thompson:2023ase}%
  \BibitemOpen
  \bibfield  {author} {\bibinfo {author} {\bibfnamefont {J.~E.}\ \bibnamefont
  {Thompson}}, \bibinfo {author} {\bibfnamefont {E.}~\bibnamefont {Hamilton}},
  \bibinfo {author} {\bibfnamefont {L.}~\bibnamefont {London}}, \bibinfo
  {author} {\bibfnamefont {S.}~\bibnamefont {Ghosh}}, \bibinfo {author}
  {\bibfnamefont {P.}~\bibnamefont {Kolitsidou}}, \bibinfo {author}
  {\bibfnamefont {C.}~\bibnamefont {Hoy}}, \ and\ \bibinfo {author}
  {\bibfnamefont {M.}~\bibnamefont {Hannam}},\ }\href {\doibase
  10.1103/PhysRevD.109.063012} {\bibfield  {journal} {\bibinfo  {journal}
  {Phys. Rev. D}\ }\textbf {\bibinfo {volume} {109}},\ \bibinfo {pages}
  {063012} (\bibinfo {year} {2024})},\ \Eprint
  {http://arxiv.org/abs/2312.10025} {arXiv:2312.10025 [gr-qc]} \BibitemShut
  {NoStop}%
\bibitem [{\citenamefont {Ramos-Buades}\ \emph {et~al.}(2023)\citenamefont
  {Ramos-Buades}, \citenamefont {Buonanno}, \citenamefont {Estell\'es},
  \citenamefont {Khalil}, \citenamefont {Mihaylov}, \citenamefont {Ossokine},
  \citenamefont {Pompili},\ and\ \citenamefont
  {Shiferaw}}]{Ramos-Buades:2023ehm}%
  \BibitemOpen
  \bibfield  {author} {\bibinfo {author} {\bibfnamefont {A.}~\bibnamefont
  {Ramos-Buades}}, \bibinfo {author} {\bibfnamefont {A.}~\bibnamefont
  {Buonanno}}, \bibinfo {author} {\bibfnamefont {H.}~\bibnamefont
  {Estell\'es}}, \bibinfo {author} {\bibfnamefont {M.}~\bibnamefont {Khalil}},
  \bibinfo {author} {\bibfnamefont {D.~P.}\ \bibnamefont {Mihaylov}}, \bibinfo
  {author} {\bibfnamefont {S.}~\bibnamefont {Ossokine}}, \bibinfo {author}
  {\bibfnamefont {L.}~\bibnamefont {Pompili}}, \ and\ \bibinfo {author}
  {\bibfnamefont {M.}~\bibnamefont {Shiferaw}},\ }\href {\doibase
  10.1103/PhysRevD.108.124037} {\bibfield  {journal} {\bibinfo  {journal}
  {Phys. Rev. D}\ }\textbf {\bibinfo {volume} {108}},\ \bibinfo {pages}
  {124037} (\bibinfo {year} {2023})},\ \Eprint
  {http://arxiv.org/abs/2303.18046} {arXiv:2303.18046 [gr-qc]} \BibitemShut
  {NoStop}%
\bibitem [{\citenamefont {Chua}\ \emph {et~al.}(2020)\citenamefont {Chua},
  \citenamefont {Korsakova}, \citenamefont {Moore}, \citenamefont {Gair},\ and\
  \citenamefont {Babak}}]{Chua:2019wgs}%
  \BibitemOpen
  \bibfield  {author} {\bibinfo {author} {\bibfnamefont {A.~J.~K.}\
  \bibnamefont {Chua}}, \bibinfo {author} {\bibfnamefont {N.}~\bibnamefont
  {Korsakova}}, \bibinfo {author} {\bibfnamefont {C.~J.}\ \bibnamefont
  {Moore}}, \bibinfo {author} {\bibfnamefont {J.~R.}\ \bibnamefont {Gair}}, \
  and\ \bibinfo {author} {\bibfnamefont {S.}~\bibnamefont {Babak}},\ }\href
  {\doibase 10.1103/PhysRevD.101.044027} {\bibfield  {journal} {\bibinfo
  {journal} {Phys. Rev. D}\ }\textbf {\bibinfo {volume} {101}},\ \bibinfo
  {pages} {044027} (\bibinfo {year} {2020})},\ \Eprint
  {http://arxiv.org/abs/1912.11543} {arXiv:1912.11543 [astro-ph.IM]}
  \BibitemShut {NoStop}%
\bibitem [{\citenamefont {Liu}\ \emph {et~al.}(2023)\citenamefont {Liu},
  \citenamefont {Li},\ and\ \citenamefont {Chua}}]{Liu:2023oxw}%
  \BibitemOpen
  \bibfield  {author} {\bibinfo {author} {\bibfnamefont {M.}~\bibnamefont
  {Liu}}, \bibinfo {author} {\bibfnamefont {X.-D.}\ \bibnamefont {Li}}, \ and\
  \bibinfo {author} {\bibfnamefont {A.~J.~K.}\ \bibnamefont {Chua}},\ }\href
  {\doibase 10.1103/PhysRevD.108.103027} {\bibfield  {journal} {\bibinfo
  {journal} {Phys. Rev. D}\ }\textbf {\bibinfo {volume} {108}},\ \bibinfo
  {pages} {103027} (\bibinfo {year} {2023})},\ \Eprint
  {http://arxiv.org/abs/2307.07233} {arXiv:2307.07233 [astro-ph.IM]}
  \BibitemShut {NoStop}%
\bibitem [{\citenamefont {Afshordi}\ \emph {et~al.}(2023)\citenamefont
  {Afshordi} \emph {et~al.}}]{LISAConsortiumWaveformWorkingGroup:2023arg}%
  \BibitemOpen
  \bibfield  {author} {\bibinfo {author} {\bibfnamefont {N.}~\bibnamefont
  {Afshordi}} \emph {et~al.} (\bibinfo {collaboration} {LISA Consortium
  Waveform Working Group}),\ }\href@noop {} {\  (\bibinfo {year} {2023})},\
  \Eprint {http://arxiv.org/abs/2311.01300} {arXiv:2311.01300 [gr-qc]}
  \BibitemShut {NoStop}%
\bibitem [{\citenamefont {Hannam}\ \emph {et~al.}(2022)\citenamefont {Hannam}
  \emph {et~al.}}]{Hannam:2021pit}%
  \BibitemOpen
  \bibfield  {author} {\bibinfo {author} {\bibfnamefont {M.}~\bibnamefont
  {Hannam}} \emph {et~al.},\ }\href {\doibase 10.1038/s41586-022-05212-z}
  {\bibfield  {journal} {\bibinfo  {journal} {Nature}\ }\textbf {\bibinfo
  {volume} {610}},\ \bibinfo {pages} {652} (\bibinfo {year} {2022})},\ \Eprint
  {http://arxiv.org/abs/2112.11300} {arXiv:2112.11300 [gr-qc]} \BibitemShut
  {NoStop}%
\bibitem [{\citenamefont {Scheel}\ \emph {et~al.}(2025)\citenamefont {Scheel}
  \emph {et~al.}}]{Scheel:2025jct}%
  \BibitemOpen
  \bibfield  {author} {\bibinfo {author} {\bibfnamefont {M.~A.}\ \bibnamefont
  {Scheel}} \emph {et~al.},\ }\href@noop {} {\  (\bibinfo {year} {2025})},\
  \Eprint {http://arxiv.org/abs/2505.13378} {arXiv:2505.13378 [gr-qc]}
  \BibitemShut {NoStop}%
\bibitem [{\citenamefont {{LIGO Scientific Collaboration}}\ \emph
  {et~al.}(2018)\citenamefont {{LIGO Scientific Collaboration}}, \citenamefont
  {{Virgo Collaboration}},\ and\ \citenamefont {{KAGRA
  Collaboration}}}]{lalsuite}%
  \BibitemOpen
  \bibfield  {author} {\bibinfo {author} {\bibnamefont {{LIGO Scientific
  Collaboration}}}, \bibinfo {author} {\bibnamefont {{Virgo Collaboration}}}, \
  and\ \bibinfo {author} {\bibnamefont {{KAGRA Collaboration}}},\ }\href
  {\doibase 10.7935/GT1W-FZ16} {\enquote {\bibinfo {title} {{LVK} {A}lgorithm
  {L}ibrary - {LALS}uite},}\ }\bibinfo {howpublished} {Free software (GPL)}
  (\bibinfo {year} {2018})\BibitemShut {NoStop}%
\bibitem [{\citenamefont {Nitz}\ \emph {et~al.}(2024)\citenamefont {Nitz},
  \citenamefont {Harry}, \citenamefont {Brown}, \citenamefont {Biwer},
  \citenamefont {Willis}, \citenamefont {Canton}, \citenamefont {Capano},
  \citenamefont {Dent}, \citenamefont {Pekowsky}, \citenamefont {Davies},
  \citenamefont {De}, \citenamefont {Cabero}, \citenamefont {Wu}, \citenamefont
  {Williamson}, \citenamefont {Machenschalk}, \citenamefont {Macleod},
  \citenamefont {Pannarale}, \citenamefont {Kumar}, \citenamefont {Reyes},
  \citenamefont {dfinstad}, \citenamefont {Kumar}, \citenamefont {Tápai},
  \citenamefont {Singer}, \citenamefont {Kumar}, \citenamefont {veronica
  villa}, \citenamefont {maxtrevor}, \citenamefont {Gadre}, \citenamefont
  {Khan}, \citenamefont {Fairhurst},\ and\ \citenamefont
  {Tolley}}]{alex_nitz_2024_10473621}%
  \BibitemOpen
  \bibfield  {author} {\bibinfo {author} {\bibfnamefont {A.}~\bibnamefont
  {Nitz}}, \bibinfo {author} {\bibfnamefont {I.}~\bibnamefont {Harry}},
  \bibinfo {author} {\bibfnamefont {D.}~\bibnamefont {Brown}}, \bibinfo
  {author} {\bibfnamefont {C.~M.}\ \bibnamefont {Biwer}}, \bibinfo {author}
  {\bibfnamefont {J.}~\bibnamefont {Willis}}, \bibinfo {author} {\bibfnamefont
  {T.~D.}\ \bibnamefont {Canton}}, \bibinfo {author} {\bibfnamefont
  {C.}~\bibnamefont {Capano}}, \bibinfo {author} {\bibfnamefont
  {T.}~\bibnamefont {Dent}}, \bibinfo {author} {\bibfnamefont {L.}~\bibnamefont
  {Pekowsky}}, \bibinfo {author} {\bibfnamefont {G.~S.~C.}\ \bibnamefont
  {Davies}}, \bibinfo {author} {\bibfnamefont {S.}~\bibnamefont {De}}, \bibinfo
  {author} {\bibfnamefont {M.}~\bibnamefont {Cabero}}, \bibinfo {author}
  {\bibfnamefont {S.}~\bibnamefont {Wu}}, \bibinfo {author} {\bibfnamefont
  {A.~R.}\ \bibnamefont {Williamson}}, \bibinfo {author} {\bibfnamefont
  {B.}~\bibnamefont {Machenschalk}}, \bibinfo {author} {\bibfnamefont
  {D.}~\bibnamefont {Macleod}}, \bibinfo {author} {\bibfnamefont
  {F.}~\bibnamefont {Pannarale}}, \bibinfo {author} {\bibfnamefont
  {P.}~\bibnamefont {Kumar}}, \bibinfo {author} {\bibfnamefont
  {S.}~\bibnamefont {Reyes}}, \bibinfo {author} {\bibnamefont {dfinstad}},
  \bibinfo {author} {\bibfnamefont {S.}~\bibnamefont {Kumar}}, \bibinfo
  {author} {\bibfnamefont {M.}~\bibnamefont {Tápai}}, \bibinfo {author}
  {\bibfnamefont {L.}~\bibnamefont {Singer}}, \bibinfo {author} {\bibfnamefont
  {P.}~\bibnamefont {Kumar}}, \bibinfo {author} {\bibnamefont {veronica
  villa}}, \bibinfo {author} {\bibnamefont {maxtrevor}}, \bibinfo {author}
  {\bibfnamefont {B.~U.~V.}\ \bibnamefont {Gadre}}, \bibinfo {author}
  {\bibfnamefont {S.}~\bibnamefont {Khan}}, \bibinfo {author} {\bibfnamefont
  {S.}~\bibnamefont {Fairhurst}}, \ and\ \bibinfo {author} {\bibfnamefont
  {A.}~\bibnamefont {Tolley}},\ }\href {\doibase 10.5281/zenodo.10473621}
  {\enquote {\bibinfo {title} {gwastro/pycbc: v2.3.3 release of pycbc},}\ }
  (\bibinfo {year} {2024})\BibitemShut {NoStop}%
\bibitem [{\citenamefont {Hoy}\ and\ \citenamefont
  {Raymond}(2021)}]{Hoy:2020vys}%
  \BibitemOpen
  \bibfield  {author} {\bibinfo {author} {\bibfnamefont {C.}~\bibnamefont
  {Hoy}}\ and\ \bibinfo {author} {\bibfnamefont {V.}~\bibnamefont {Raymond}},\
  }\href {\doibase 10.1016/j.softx.2021.100765} {\bibfield  {journal} {\bibinfo
   {journal} {SoftwareX}\ }\textbf {\bibinfo {volume} {15}},\ \bibinfo {pages}
  {100765} (\bibinfo {year} {2021})},\ \Eprint
  {http://arxiv.org/abs/2006.06639} {arXiv:2006.06639 [astro-ph.IM]}
  \BibitemShut {NoStop}%
\bibitem [{\citenamefont {Hunter}(2007)}]{Hunter:2007}%
  \BibitemOpen
  \bibfield  {author} {\bibinfo {author} {\bibfnamefont {J.~D.}\ \bibnamefont
  {Hunter}},\ }\href {\doibase 10.1109/MCSE.2007.55} {\bibfield  {journal}
  {\bibinfo  {journal} {Computing in Science \& Engineering}\ }\textbf
  {\bibinfo {volume} {9}},\ \bibinfo {pages} {90} (\bibinfo {year}
  {2007})}\BibitemShut {NoStop}%
\bibitem [{\citenamefont {Harris}\ \emph {et~al.}(2020)\citenamefont {Harris},
  \citenamefont {Millman}, \citenamefont {van~der Walt}, \citenamefont
  {Gommers}, \citenamefont {Virtanen}, \citenamefont {Cournapeau},
  \citenamefont {Wieser}, \citenamefont {Taylor}, \citenamefont {Berg},
  \citenamefont {Smith}, \citenamefont {Kern}, \citenamefont {Picus},
  \citenamefont {Hoyer}, \citenamefont {van Kerkwijk}, \citenamefont {Brett},
  \citenamefont {Haldane}, \citenamefont {del R{\'{i}}o}, \citenamefont
  {Wiebe}, \citenamefont {Peterson}, \citenamefont {G{\'{e}}rard-Marchant},
  \citenamefont {Sheppard}, \citenamefont {Reddy}, \citenamefont {Weckesser},
  \citenamefont {Abbasi}, \citenamefont {Gohlke},\ and\ \citenamefont
  {Oliphant}}]{harris2020array}%
  \BibitemOpen
  \bibfield  {author} {\bibinfo {author} {\bibfnamefont {C.~R.}\ \bibnamefont
  {Harris}}, \bibinfo {author} {\bibfnamefont {K.~J.}\ \bibnamefont {Millman}},
  \bibinfo {author} {\bibfnamefont {S.~J.}\ \bibnamefont {van~der Walt}},
  \bibinfo {author} {\bibfnamefont {R.}~\bibnamefont {Gommers}}, \bibinfo
  {author} {\bibfnamefont {P.}~\bibnamefont {Virtanen}}, \bibinfo {author}
  {\bibfnamefont {D.}~\bibnamefont {Cournapeau}}, \bibinfo {author}
  {\bibfnamefont {E.}~\bibnamefont {Wieser}}, \bibinfo {author} {\bibfnamefont
  {J.}~\bibnamefont {Taylor}}, \bibinfo {author} {\bibfnamefont
  {S.}~\bibnamefont {Berg}}, \bibinfo {author} {\bibfnamefont {N.~J.}\
  \bibnamefont {Smith}}, \bibinfo {author} {\bibfnamefont {R.}~\bibnamefont
  {Kern}}, \bibinfo {author} {\bibfnamefont {M.}~\bibnamefont {Picus}},
  \bibinfo {author} {\bibfnamefont {S.}~\bibnamefont {Hoyer}}, \bibinfo
  {author} {\bibfnamefont {M.~H.}\ \bibnamefont {van Kerkwijk}}, \bibinfo
  {author} {\bibfnamefont {M.}~\bibnamefont {Brett}}, \bibinfo {author}
  {\bibfnamefont {A.}~\bibnamefont {Haldane}}, \bibinfo {author} {\bibfnamefont
  {J.~F.}\ \bibnamefont {del R{\'{i}}o}}, \bibinfo {author} {\bibfnamefont
  {M.}~\bibnamefont {Wiebe}}, \bibinfo {author} {\bibfnamefont
  {P.}~\bibnamefont {Peterson}}, \bibinfo {author} {\bibfnamefont
  {P.}~\bibnamefont {G{\'{e}}rard-Marchant}}, \bibinfo {author} {\bibfnamefont
  {K.}~\bibnamefont {Sheppard}}, \bibinfo {author} {\bibfnamefont
  {T.}~\bibnamefont {Reddy}}, \bibinfo {author} {\bibfnamefont
  {W.}~\bibnamefont {Weckesser}}, \bibinfo {author} {\bibfnamefont
  {H.}~\bibnamefont {Abbasi}}, \bibinfo {author} {\bibfnamefont
  {C.}~\bibnamefont {Gohlke}}, \ and\ \bibinfo {author} {\bibfnamefont {T.~E.}\
  \bibnamefont {Oliphant}},\ }\href {\doibase 10.1038/s41586-020-2649-2}
  {\bibfield  {journal} {\bibinfo  {journal} {Nature}\ }\textbf {\bibinfo
  {volume} {585}},\ \bibinfo {pages} {357} (\bibinfo {year}
  {2020})}\BibitemShut {NoStop}%
\bibitem [{\citenamefont {Synge}(1960)}]{Synge:1960ueh}%
  \BibitemOpen
  \bibinfo {editor} {\bibfnamefont {J.~L.}\ \bibnamefont {Synge}},\ ed.,\
  \href@noop {} {\emph {\bibinfo {title} {{Relativity: The General theory}}}}\
  (\bibinfo {year} {1960})\BibitemShut {NoStop}%
\bibitem [{\citenamefont {Mahalanobis}(2018)}]{Mahalanobis}%
  \BibitemOpen
  \bibfield  {author} {\bibinfo {author} {\bibfnamefont {P.}~\bibnamefont
  {Mahalanobis}},\ }\href {\doibase 10.1007/s13171-019-00164-5} {\bibfield
  {journal} {\bibinfo  {journal} {Sankhya A}\ }\textbf {\bibinfo {volume}
  {80}},\ \bibinfo {pages} {1} (\bibinfo {year} {2018})}\BibitemShut {NoStop}%
\bibitem [{\citenamefont {Dhani}\ \emph {et~al.}(2024)\citenamefont {Dhani},
  \citenamefont {V\"olkel}, \citenamefont {Buonanno}, \citenamefont {Estelles},
  \citenamefont {Gair}, \citenamefont {Pfeiffer}, \citenamefont {Pompili},\
  and\ \citenamefont {Toubiana}}]{Dhani:2024jja}%
  \BibitemOpen
  \bibfield  {author} {\bibinfo {author} {\bibfnamefont {A.}~\bibnamefont
  {Dhani}}, \bibinfo {author} {\bibfnamefont {S.}~\bibnamefont {V\"olkel}},
  \bibinfo {author} {\bibfnamefont {A.}~\bibnamefont {Buonanno}}, \bibinfo
  {author} {\bibfnamefont {H.}~\bibnamefont {Estelles}}, \bibinfo {author}
  {\bibfnamefont {J.}~\bibnamefont {Gair}}, \bibinfo {author} {\bibfnamefont
  {H.~P.}\ \bibnamefont {Pfeiffer}}, \bibinfo {author} {\bibfnamefont
  {L.}~\bibnamefont {Pompili}}, \ and\ \bibinfo {author} {\bibfnamefont
  {A.}~\bibnamefont {Toubiana}},\ }\href@noop {} {\  (\bibinfo {year}
  {2024})},\ \Eprint {http://arxiv.org/abs/2404.05811} {arXiv:2404.05811
  [gr-qc]} \BibitemShut {NoStop}%
\bibitem [{\citenamefont {Vallisneri}(2008)}]{Vallisneri:2007ev}%
  \BibitemOpen
  \bibfield  {author} {\bibinfo {author} {\bibfnamefont {M.}~\bibnamefont
  {Vallisneri}},\ }\href {\doibase 10.1103/PhysRevD.77.042001} {\bibfield
  {journal} {\bibinfo  {journal} {Phys. Rev. D}\ }\textbf {\bibinfo {volume}
  {77}},\ \bibinfo {pages} {042001} (\bibinfo {year} {2008})},\ \Eprint
  {http://arxiv.org/abs/gr-qc/0703086} {arXiv:gr-qc/0703086} \BibitemShut
  {NoStop}%
\bibitem [{\citenamefont {Neidinger}(2010)}]{doi:10.1137/080743627}%
  \BibitemOpen
  \bibfield  {author} {\bibinfo {author} {\bibfnamefont {R.~D.}\ \bibnamefont
  {Neidinger}},\ }\href {\doibase 10.1137/080743627} {\bibfield  {journal}
  {\bibinfo  {journal} {SIAM Review}\ }\textbf {\bibinfo {volume} {52}},\
  \bibinfo {pages} {545} (\bibinfo {year} {2010})},\ \Eprint
  {http://arxiv.org/abs/https://doi.org/10.1137/080743627}
  {https://doi.org/10.1137/080743627} \BibitemShut {NoStop}%
\bibitem [{\citenamefont {Edwards}\ \emph {et~al.}(2024)\citenamefont
  {Edwards}, \citenamefont {Wong}, \citenamefont {Lam}, \citenamefont {Coogan},
  \citenamefont {Foreman-Mackey}, \citenamefont {Isi},\ and\ \citenamefont
  {Zimmerman}}]{Edwards:2023sak}%
  \BibitemOpen
  \bibfield  {author} {\bibinfo {author} {\bibfnamefont {T.~D.~P.}\
  \bibnamefont {Edwards}}, \bibinfo {author} {\bibfnamefont {K.~W.~K.}\
  \bibnamefont {Wong}}, \bibinfo {author} {\bibfnamefont {K.~K.~H.}\
  \bibnamefont {Lam}}, \bibinfo {author} {\bibfnamefont {A.}~\bibnamefont
  {Coogan}}, \bibinfo {author} {\bibfnamefont {D.}~\bibnamefont
  {Foreman-Mackey}}, \bibinfo {author} {\bibfnamefont {M.}~\bibnamefont {Isi}},
  \ and\ \bibinfo {author} {\bibfnamefont {A.}~\bibnamefont {Zimmerman}},\
  }\href {\doibase 10.1103/PhysRevD.110.064028} {\bibfield  {journal} {\bibinfo
   {journal} {Phys. Rev. D}\ }\textbf {\bibinfo {volume} {110}},\ \bibinfo
  {pages} {064028} (\bibinfo {year} {2024})},\ \Eprint
  {http://arxiv.org/abs/2302.05329} {arXiv:2302.05329 [astro-ph.IM]}
  \BibitemShut {NoStop}%
\end{thebibliography}%

\end{document}